\UseRawInputEncoding
\documentclass[aps, prl, superscriptaddress, citeautoscript]{revtex4-2}
\setcitestyle{super}

\usepackage{setspace}
\usepackage{graphicx}
\usepackage{amsmath,amssymb,amsfonts}
\usepackage{xcolor}
\usepackage{textcomp}
\usepackage{booktabs}
\usepackage{listings}
\usepackage[english]{babel}

\usepackage{multirow}
\usepackage{amsthm}
\usepackage{mathrsfs}
\usepackage[title]{appendix}
\usepackage{algorithm}
\usepackage{algorithmicx}
\usepackage{algpseudocode}
\usepackage[version=3]{mhchem} 
\usepackage{csquotes}
\usepackage{array}
\usepackage{cancel}
\usepackage[normalem]{ulem}
\usepackage{etoolbox} 
\usepackage{lineno}

\makeatletter
\let\l@en\l@english
\makeatother
\usepackage{hyperref}

\newcommand*\bifi[0]{di([1,1'-biphenyl]-4-yl)carbene}
\newcommand*\bike[0]{di([1,1'-biphenyl]-4-yl)methanone}
\newcommand*\ciso[0]{\ensuremath{^{13}\text{C}}}

\newcommand*\cis[0]{\ensuremath{^{12}\mathrm{C}}}

\newcommand*\Tg[0]{\ensuremath{{T}_{0}}}
\newcommand*\Tgx[0]{\ensuremath{{T}_{0\mathrm{x}}}}
\newcommand*\Tgy[0]{\ensuremath{{T}_{0\mathrm{y}}}}
\newcommand*\Tgz[0]{\ensuremath{{T}_{0\mathrm{z}}}}
\newcommand*\Te[0]{\ensuremath{{T}_{1}}}
\newcommand*\Tex[0]{\ensuremath{{T}_{1\mathrm{x}}}}
\newcommand*\Tey[0]{\ensuremath{{T}_{1\mathrm{y}}}}
\newcommand*\Tez[0]{\ensuremath{{T}_{1\mathrm{z}}}}
\newcommand*\T[1]{\ensuremath{{T}_{#1}}}
\newcommand*\Sg[0]{\ensuremath{{S}_{0}}}
\newcommand*\qm[0]{qubit molecule}
\newcommand*\qms[0]{qubit molecules}

\newcommand*\Des[0]{\ensuremath{D_\mathrm{es}}}
\newcommand*\Egs[0]{\ensuremath{E_\mathrm{gs}}}
\newcommand*\Ees[0]{\ensuremath{E_\mathrm{es}}}

\begin{document}

\title[A Single-Molecule Spin-Photon Interface]{A Single-Molecule Spin-Photon Interface}
	
\author{Simon Roggors}
\affiliation{NVision Imaging Technologies GmbH, Wolfgang-Paul-Str. 2, 89081 Ulm, Germany}
\affiliation{Institute for Quantum Optics (IQO), Ulm University, Meyerhofstra\ss e M26, 89069 Ulm, Germany}

\author{Thomas Unden}
\affiliation{NVision Imaging Technologies GmbH, Wolfgang-Paul-Str. 2, 89081 Ulm, Germany}

\author{Anna Aubele}
\affiliation{NVision Imaging Technologies GmbH, Wolfgang-Paul-Str. 2, 89081 Ulm, Germany}

\author{Paul Mentzel}
\affiliation{NVision Imaging Technologies GmbH, Wolfgang-Paul-Str. 2, 89081 Ulm, Germany}

\author{Gregor Bayer}
\affiliation{NVision Imaging Technologies GmbH, Wolfgang-Paul-Str. 2, 89081 Ulm, Germany}

\author{Alon Salhov}
\affiliation{NVision Imaging Technologies GmbH, Wolfgang-Paul-Str. 2, 89081 Ulm, Germany}
\affiliation{Racah Institute of Physics, The Hebrew University of Jerusalem, Jerusalem 9190401, Israel}

\author{Jochen Scharpf}
\affiliation{NVision Imaging Technologies GmbH, Wolfgang-Paul-Str. 2, 89081 Ulm, Germany}

\author{Martin B. Plenio}
\affiliation{Institute of Theoretical Physics, Ulm University, Albert Einstein Allee 11, 89069 Ulm, Germany}
\affiliation{Center for Integrated Quantum Science and Technology (IQST), Ulm University, 89069 Ulm, Germany}

\author{Alex Retzker}
\affiliation{Racah Institute of Physics, The Hebrew University of Jerusalem, Jerusalem 9190401, Israel}

\author{Fedor Jelezko}
\affiliation{Institute for Quantum Optics (IQO), Ulm University, Meyerhofstra\ss e 4, 89069 Ulm, Germany}
\affiliation{Center for Integrated Quantum Science and Technology (IQST), Ulm University, 89069 Ulm, Germany}

\author{Tim R. Eichhorn}
\affiliation{NVision Imaging Technologies GmbH, Wolfgang-Paul-Str. 2, 89081 Ulm, Germany}

\author{Tobias A. Schaub}
\affiliation{NVision Imaging Technologies GmbH, Wolfgang-Paul-Str. 2, 89081 Ulm, Germany}
\affiliation{Institute for Quantum Optics (IQO), Ulm University, Meyerhofstra\ss e 4, 89069 Ulm, Germany}

\author{Matthias Pfender}
\email{matthias@nvision-imaging.com}
\affiliation{NVision Imaging Technologies GmbH, Wolfgang-Paul-Str. 2, 89081 Ulm, Germany}

\author{Philipp Neumann}
\affiliation{NVision Imaging Technologies GmbH, Wolfgang-Paul-Str. 2, 89081 Ulm, Germany}

\author{Ilai Schwartz}
\email{ilai@nvision-imaging.com}
\affiliation{NVision Imaging Technologies GmbH, Wolfgang-Paul-Str. 2, 89081 Ulm, Germany}
	
\begin{abstract}
Optical interfaces that connect long-lived spin qubits to photons are a central requirement for quantum networking and distributed quantum information processing.
Currently, solid-state atomic defects are leading candidates due to their inherent spin and optical coherence\cite{awschalomQuantumTechnologiesOptically2018,heppElectronicStructureSilicon2014}.
Building on these advancements, synthetically tailored molecular systems represent a fundamental change in the field, utilizing precise atomic control and consistent bottom-up assembly\cite{wasielewskiExploitingChemistryMolecular2020b, toninelliSingleOrganicMolecules2021}.
However, the lack of a robust spin-photon interface combining bright fluorescence, high spectral stability, and the persistent spin lifetimes inherent to ground-state systems has prohibited the detection of individual molecular qubits.
Here we show that a triplet ground state carbene molecule, embedded within a structurally matched host crystal, functions as a robust spin-photon interface with single-molecule addressability.
The system exhibits narrow zero-phonon lines, spectral stability over more than an hour, spin-selective optical transitions and single-molecule optically detected magnetic resonance.
Coherent control yields millisecond-scale dynamical-decoupling coherence and tens-of-milliseconds spin relaxation at a temperature of $4.5$\,K.
These results establish molecular qubits as a viable platform for single-emitter quantum optics\cite{wriggeEfficientCouplingPhotons2008,wangCoherentCouplingSingle2017} while preserving the advantages of bottom-up chemical design~\cite{baylissEnhancingSpinCoherence2022} and processable materials~\cite{lombardiPhotostableMoleculesChip2018,hailNanoprintingOrganicMolecules2019,huangOnchipQuantumInterference2025,langeCavityQEDMolecular2025}.
\end{abstract}

\maketitle

\makeatletter
\let\oldaddcontentsline\addcontentsline
\renewcommand{\addcontentsline}[3]{}
\makeatother

Currently, color centers in wide-bandgap semiconductors, spearheaded by the nitrogen-vacancy center in diamond, have driven landmark achievements from loophole-free Bell inequality tests~\cite{hensenLoopholefreeBellInequality2015} and multi-node quantum networks~\cite{hermansQubitTeleportationNonneighbouring2022a}, to fault-tolerant quantum error correction~\cite{abobeihFaulttolerantOperationLogical2022a}.
Optical spin readout has been successfully demonstrated across a variety of molecular platforms.
For instance, organometallic chromium(IV) complexes~\cite{baylissOpticallyAddressableMolecular2020} have proven to be a highly versatile platform, with the ability to tune spin properties by chemically changing the qubit or the matrix~\cite{baylissEnhancingSpinCoherence2022}.
Meanwhile, purely organic qubits such as diradicals~\cite{gorgonReversibleSpinopticalInterface2023,koppLuminescentOrganicTriplet2024,chowdhuryBrightTripletBright2025} and carbenes~\cite{roggorsOpticallyDetectedMagnetic2025a} offer a distinct advantage; the absence of heavy atoms minimizes spin-orbit coupling, leading to significantly extended spin lifetimes and cyclical, coherent optical transitions.
Additionally, molecular rare-earth complexes leverage the inherently narrow optical linewidths of rare-earth ions within molecular crystal environments~\cite{serranoUltranarrowOpticalLinewidths2022}, facilitating the optical readout of both electron~\cite{weissHighresolutionMolecularSpinphoton2025} and nuclear~\cite{vasilenkoOpticallyDetectedNuclear2026} spin states.
To date, experimental progress, however, has been confined to bulk ensembles.
To transcend the constraints of ensemble-averaged platforms and enable isolated single-molecule addressability, we designed a robust molecular qubit, namely \bifi{}, hosted within an isosteric, crystalline \bike{} matrix, providing high fluorescence emission, spectral stability, and favorable inter-system crossing (ISC) dynamics.

\subsection*{Qubit Design}
\begin{figure}[htbp]
	\centering
	\includegraphics[width=18cm]{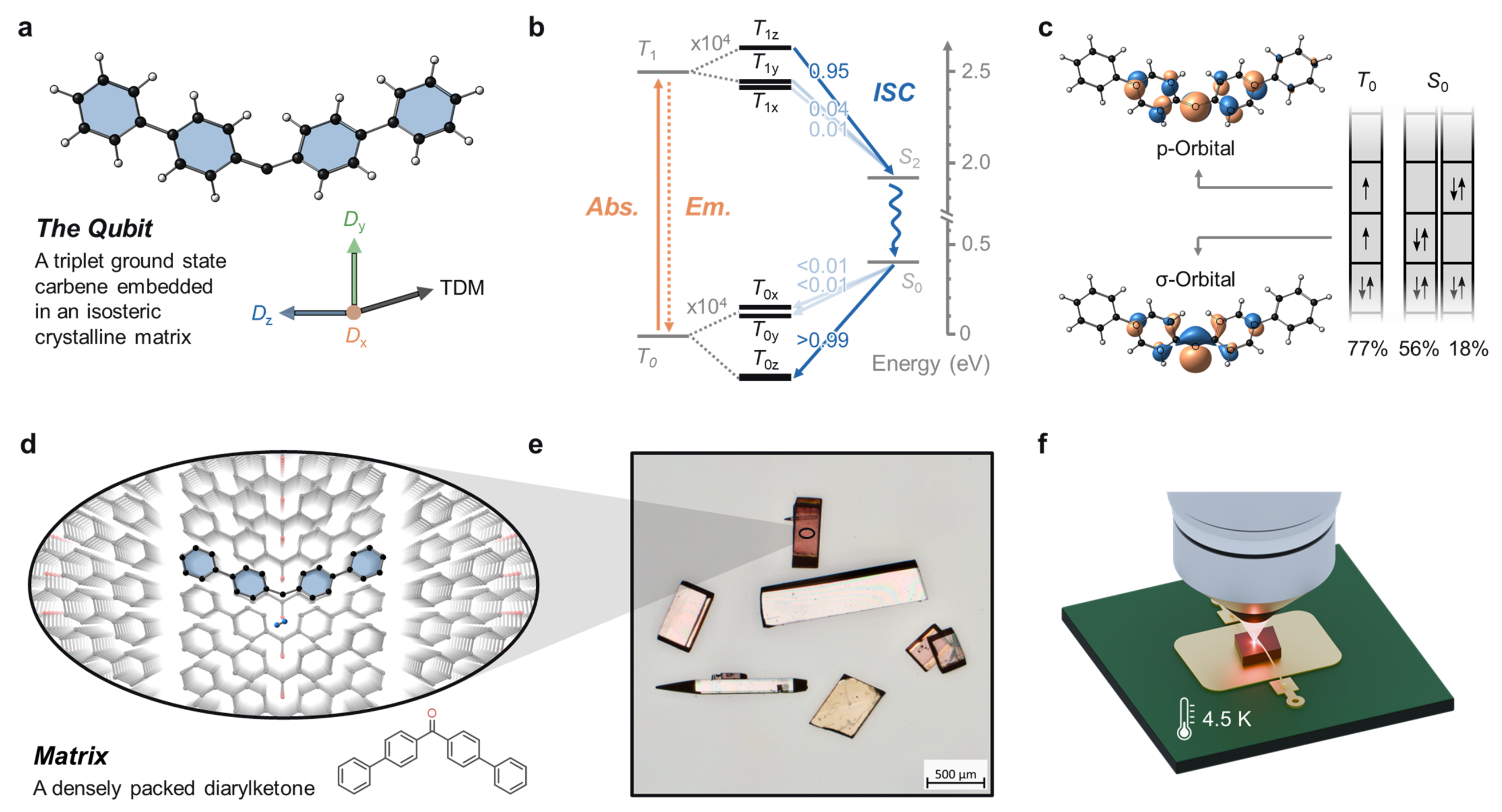}
	\caption{\textbf{A molecular color-center with a triplet ground state doped into a crystalline matrix.} 
		\textbf{a}, Molecular structure of the qubit molecule and nearby nitrogen molecule together with the ZFS tensor and the transition dipole moment (TDM) from SA8-CASSCF(12,12)/QD-SC-NEVPT2/def2-TZVPP calculations (see Methods).
		\textbf{b}, Scaled energy levels (ZFS enlarged for clarity) showing the ZFS change between ground and photoexcited triplet states. Note the sign change of the ZFS parameter in $\Tg{}$ and $\Te{}$ indicated by reordered x,y,z substates. The blue arrows illustrate optical spin pumping via ISC through the singlet manifold to the \Tgz{} ground state, numbers denote relative ISC rates.
		\textbf{c}, Spin-bearing $\sigma$ and p-orbitals (isosurface=0.04) and major spin configurations of the $\Tg{}$ and $\Sg{}$ states.
		\textbf{d}, The carbene and the nitrogen molecule embedded in the crystalline lattice of ketone matrix.
		\textbf{e}, Representative crystals with a uniform 2.5 mol\% doping rate shown in multiple orientations. Different colors stem from variations of the effective optical density along the viewing axis, scaling with crystal thickness and orientation.
		\textbf{f}, Schematic of the low-temperature confocal microscopy system. A MW wire is shown positioned over the doped crystal sample for local spin manipulation.
	}
	\label{fig1}
\end{figure}
Our \qm{} is a bent, pseudo \textit{C}$_{2}$-symmetric diarylcarbene (carbene angle approx. 142$^\circ$) with two \textit{para}-connected biphenyls and derives its unique characteristics from the ferromagnetic exchange of two strongly correlated, unpaired electrons which are co-localized on the divalent carbon center and the proximal aryl units (Fig. \ref{fig1}a).
To gain a deeper understanding of the compound \textit{in situ}, we model the geometry of the carbene embedded in a crystalline cluster of 18 matrix molecules derived from single crystal X-ray diffraction data (see Methods) \cite{neeseSoftwareUpdateORCA2025,Bannwarth_GFN2_2019}. 
Exploratory time-dependent density functional theory calculations of the carbene qubit extracted from the cluster model predict the \T{0}$\rightarrow$\T{1} electronic transition to possess locally excited state character --- a key requirement for strong emission into the zero-phonon line (ZPL, see SI IV. E.).
Based on multiconfigurational quantum chemical calculations followed by perturbational treatment to account for dynamic correlation (CASSCF/NEVPT2, see Methods) we find this transition to involve a formal redistribution of spin density from the central carbene p-orbital to an adjacent phenyl ring (Fig. \ref{fig1}b, see SI, IV. F.).
This shift induces a substantial change in the magnitude and sign of the zero-field splitting (ZFS) parameters from an oblate ZFS tensor in the ground state (\textit{D}$_{T_0,\mathrm{calc}} = 11797$\,MHz, \textit{E}$_{T_0,\mathrm{calc}} = -516$\,MHz) to a prolate tensor in the excited state (\textit{D}$_{T_1,\mathrm{calc}} = -6161$\,MHz, \textit{E}$_{T_1,\mathrm{calc}} = 276$\,MHz).
The resulting large frequency separation of spin-conserving transitions of 17166\,MHz (18749\,MHz) between $T_{0z} \rightarrow T_{1z}$ and $T_{0y} \rightarrow T_{1y}$ ($T_{0x} \rightarrow  T_{1x}$) is facilitating the optical addressability of individual spin states.
The predictions also indicate that lowest lying singlet state $\Sg$ has a closed-shell character (Fig. \ref{fig1}c), which is destabilized to the ground state triplet by approximately $\Delta\textit{E}_{ST}$ = 0.47 eV in the $T_{0}$ geometry. 
In line with previous theoretical studies~\cite{pohElectronSextetsOptically2025e, pohQuantifyingSpinOpticalProperties2026a} on the spin-optical interfaces of bent carbenes, the electronic structure renders intersystem crossing (ISC) between the singlet and the triplet ground state El-Sayed~\cite{el-sayedSpinOrbitCoupling1963}-allowed, promoting a highly spin-selective population of the \Tgz{} state due to symmetry (for a discussion on the excited state \Tez{}-selectivity see SI). 

Motivated by early reports on paradigmatic diphenylcarbene in diphenylketone~\cite{Anderson_Electron_1976, Graham_Magnetic_1986}, we mitigate the inherent reactivity of carbenes in our experiments,~\cite{bertrandCarbeneChemistryFleeting2002,hiraiPersistentTripletCarbenes2009b} using a precursor strategy in which a photoactivatable diaryldiazomethane precursor is embedded in an isosteric, rigid crystalline matrix, namely \bike{} (Fig. \ref{fig1}d; see the SI). Two key advantages emerge from this specific crystalline matrix that merit emphasis:
\textit{i}) It imposes a rigid constraint that planarizes the biphenyl units (Ph--Ph dihedral angle ≈6$^\circ$ \textit{vs.} 36$^\circ$ in the gas phase), thereby minimizing structural reorganization in the electronically excited $\Te{}$ state and maximizing the branching ratio into the ZPL (see SI, IV. D), and ii) it diminishes wavefunction overlap of the qubit with the surrounding matrix via a dominant edge-to-face packing motif~\cite{hestandExpandedTheoryJMolecular2018}. 
Homogeneously doped crystals with different morphologies and doping concentrations were prepared via standard solution-phase methods.
For instance, vapor diffusion crystallization within few days yielded tabular or prismatic millimeter-sized bulk crystals (Fig. \ref{fig1}e), while drop-casting produced parallelogramic submicron thin plates, all of which exhibited high optical quality (see the SI and the Methods section). 
Beyond its aforementioned primary function, the photoactivation method offers a powerful feature in that it enables the activation of carbenes with sub-micrometer precision.
This level of control, achieved via confocal microscopy, allows the \qm{} density to be regulated independently of the overall doping level.

\subsection*{Optical Spectroscopy and Optically Detected Magnetic Resonance}
The precursor-doped crystal (25\,ppm doping ratio) was oriented within a cryo-confocal microscope (see Fig.~\ref{fig1}f and Methods) to maximize coupling efficiency for both incident laser light and emitted photons.
After cool-down to 4.5\,K, the diffraction-limited laser focus was kept for about two minutes at several positions on the crystal surface at a wavelength of 580\,nm and $5\,$\textmu W power. Photoelimination of nitrogen\cite{pitesaPhotoeliminationNitrogenDiazoalkanes2020c} then locally converted all precursor molecules into dense ensembles of qubit molecules.
Subsequently, we performed thermal annealing to 200\,K, relieving local matrix strain and stabilizing the crystalline environment, which was further corroborated by EPR measurements (see Methods and SI, VI. C.).
Upon continuing laser excitation the \qms{} fluoresce and were thus observable in a confocal microscopy image (see inset in Fig.~\ref{fig1,5}a).
\begin{figure}[htbp]
	\centering
	\includegraphics[width=18cm]{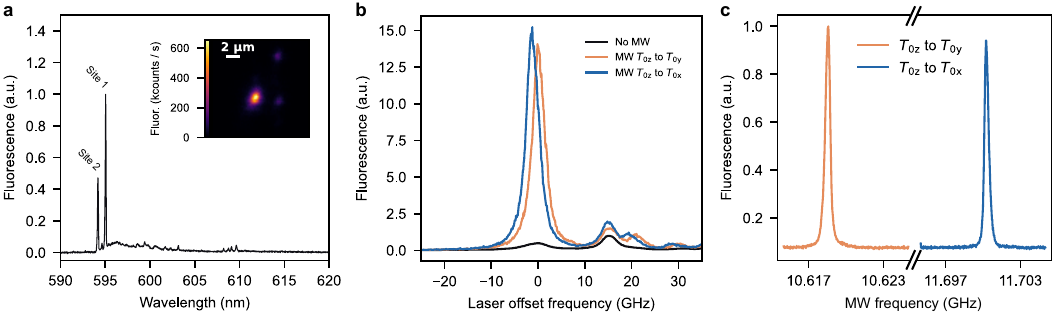}
	\caption{\textbf{Ensemble optical spectroscopy and optically detected magnetic resonance.} 
		\textbf{a}, The fluorescence spectrum of a carbene ensemble under 580\,nm excitation exhibits two dominant ZPLs. The confocal microscopy scan in the inset shows several locally created ensembles.
		\textbf{b}, The excitation spectra show accumulated fluorescence photons (detection wavelength $>605\,$nm) when scanning the laser frequency across the longer wavelength ZPL (black). Additional MW radiation resonant to spin transitions $\Tgz{} \leftrightarrow \Tgy{}$ (orange) or $\Tgz{} \leftrightarrow \Tgx{}$ (blue) significantly alters the fluorescence response.
		\textbf{c}, The optically detected magnetic resonance (ODMR) spectrum revealing the spin resonances exploited in \textbf{b} with same color coding.
	}
	\label{fig1,5}
\end{figure}
The ensemble fluorescence spectrum exhibits two sharp ZPLs (at $594.18$\,nm and $595.04$\,nm Fig.~\ref{fig1,5}a, limited by the spectrometer's resolution to about 0.1\,nm full-width at half-maximum (FWHM), amounting to about 19\% of total fluorescence), accompanied by a lower-energy vibronic and phononic sideband.
We attribute the two ZPLs to the presence of two nearly identical, yet structurally distinguishable sites for \qms{} within the host crystal (comparable, for example, to terrylene molecules in \textit{p}-terphenyl crystals \cite{kummerTerrylenePterphenylNovel1994a}).
This assignment is further corroborated by detailed ODMR and EPR measurements (see the SI, VI. A. and VII. B.)
Subsequent resonant optical excitation experiments focused on the higher-wavelength molecular site ($595.04$\,nm) to prevent unwanted off-resonant excitation of the lower-wavelength species.
Scanning the laser frequency across the ZPL and detecting the phonon sideband emission ($>605$\,nm) reveals two inhomogeneously broadened excitation lines with narrow FWHM of $\sim3\,$GHz (see black curve in Fig.~\ref{fig1,5}b).
Comparably narrow lines are known for example for pentacene or terrylene in \textit{p}-terphenyl~\cite{ambroseDetectionSpectroscopySingle1991a,kummerTerrylenePterphenylNovel1994a}.
The spectral splitting between the lines in Fig.~\ref{fig1,5}b matches the expected energy difference for spin-conserving optical transitions in related carbene molecules~\cite{roggorsOpticallyDetectedMagnetic2025a}.
To unambiguously assign these optical transitions to specific spin states, in addition to laser excitation, we applied resonant microwave (MW) fields corresponding to either the $\Tgz{} \leftrightarrow \Tgy{}$ or $\Tgz{} \leftrightarrow \Tgx{}$ transition.
As shown in Fig.~\ref{fig1,5}b, these MW fields greatly increase the intensity of the lower-frequency excitation line and, in addition, reveal a sub-structure.
Indeed, the lower-energy peak is found to be a composite of two overlapping optical transitions, namely $\Tgx{} \rightarrow \Tex{}$ and $\Tgy{} \rightarrow \Tey{}$.
Optical excitation of the latter transitions induces parasitic ISC to the closest singlet state ($\Te{} \rightarrow S_2$), followed by an ISC decay mainly into the triplet ground state \Tgz{}, eventually removing the molecules from the optical excitation loop (compare also the calculated ISC rates in Fig.~\ref{fig1}b).
The application of resonant microwaves replenishes \Tgx{} or \Tgy{} populations from the \Tgz{} state.
This selectively enhances the fluorescence intensity of the corresponding optical excitation line, allowing us to directly map specific optical transitions to their distinct triplet spin states.
The excitation line corresponding to the $\Tgz{} \rightarrow \Tez{}$ transition is low in amplitude due to a higher ISC rate for $\Tez \rightarrow S_2$ compared to $\Tex{} \rightarrow S_2$ and $\Tey{} \rightarrow S_2$, in accordance with the calculated values and excited state lifetime measurements (see Fig.~\ref{fig1}b and SI III. C.).
For the sake of consistency, all laser excitation spectra are plotted against the laser frequency offset relative to the center of the lower frequency optical transition, which includes both the \Tgx{} to \Tex{} and \Tgy{} to \Tey{} transitions.
Note that the excitation spectrum in Fig.~\ref{fig1,5}b contains additional features beyond $\sim 20$\,GHz, that are also modulated when applying microwaves.
These spectral features are consistent with isotope effects due to natural abundance \ciso{} shifting the ZPL of the \qms{} \cite{doberer13CIsotopeShifts1983,brouwer13CIsotopeEffects1996,kummerAbsorptionExcitationEmission1997} (see SI VII. D.).

The ODMR spectra on ensembles of \qms{} reveals two sharp and strong fluorescence maxima when the swept MW frequency corresponds to transition $\Tgz{} \leftrightarrow \Tgy{}$ at 10.618\,GHz or $\Tgz{} \leftrightarrow \Tgx{}$ at 11.700\,GHz, while the laser is kept at laser offset frequency around $0\,$GHz (see Fig.~\ref{fig1,5}c).
From these values, the ground-state ZFS parameters are deduced to be $|D_{T_0}| = 11.1597(2)$\,GHz and $|E_{T_0}|=540.9(2)$\,MHz (the sign of the ZFS parameters is obtained by quantum chemical simulations, as well as by EPR measurements in the SI VI. A.).
Notably, the ensemble ODMR resonances show a full width at half maximum of about 400\,kHz, providing an upper bound of the distribution of ZFS parameters $D_{T_0}$ and $E_{T_0}$ in the probed ensemble.

\subsection*{Single Molecule Spectroscopy}
\begin{figure}[htbp]
	\centering
	\includegraphics[width=18cm]{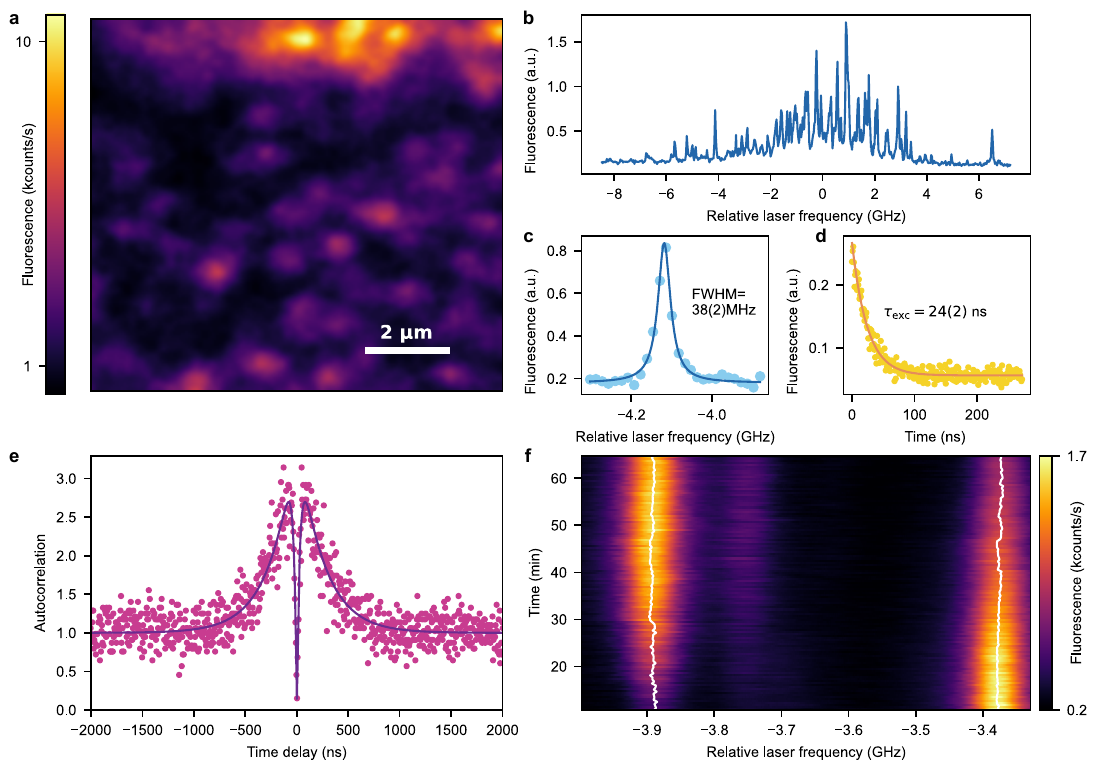}
	\caption{\textbf{Single molecule microscopy, spectroscopy and spectral stability.}
		\textbf{a}, Confocal fluorescence microscopy showing a multitude of diffraction-limited bright spots in a sparsely activated region with 3\,nW of resonant laser excitation. Fluorescence was enhanced by simultaneous microwave excitation (see Methods). A Gaussian filter with $\sigma\approx100\,$nm was applied to smooth the image.
		\textbf{b}, The laser excitation spectrum (with MW as in \textbf{a}) reveals multiple sharp resonances (\textit{c.f.} the previous inhomogeneously broadened spectrum of Fig.~\ref{fig1,5}b).
		\textbf{c}, Excerpt from a spectral region of \textbf{b} showing an isolated resonance with a linewidth of 38(2)\,MHz.
		\textbf{d}, Pulsed laser excitation on this resonance revealing a fluorescence lifetime of 24(2)\,ns.
		\textbf{e}, Fluorescence autocorrelation measurement confirms the detection of a single emitter ($g^{(2)} (0)=0.142^{+0.169}_{-0.023}$).
		\textbf{f}, Long-term, repeated acquisition of fluorescence excitation spectra demonstrates spectral diffusion of the individual emitters with a standard deviation of about 2.6\,MHz during one hour.
	}\label{fig2}
\end{figure}
Using a gradual photoactivation procedure, we obtain \textit{in situ} spatial control over the \qm{} density (see Methods). 
Within a weakly activated region, the inhomogeneously broadened ensemble ZPL resolves into distinct, potentially single-molecule excitation lines (see Fig.~\ref{fig2}b, microwave radiation targeting the $\Tgz{} \leftrightarrow \Tgx{}$ and $\Tgz{} \leftrightarrow \Tgy{}$ spin transitions was applied to enhance the fluorescence).
Resonant optical excitation further isolates a subset of these activated molecules, yielding a sparse spatial distribution characteristic of individual emitters (see Fig.~\ref{fig2}a)~\cite{moernerOpticalDetectionSpectroscopy1989,orritSinglePentaceneMolecules1990a,ambroseDetectionSpectroscopySingle1991a}.
To confirm the observation of a single molecule, we focused the laser onto an isolated spot and directed the fluorescence into a Hanbury Brown and Twiss setup.
The resulting photon autocorrelation function exhibits a zero-delay coincidence dip to $g^{(2)} (0)=0.142^{+0.169}_{-0.023}$, well below the threshold of 0.5 indicative of a single emitter \cite{kimblePhotonAntibunchingResonance1977} (see Fig.~\ref{fig2}e and Methods).
Scanning the excitation laser across a single-molecule resonance revealed a Lorentzian lineshape with a FWHM of $38(2)$\,MHz (see Fig.~\ref{fig2}c).
We determined the corresponding excited-state lifetime via rapid modulation of the excitation light.
Given the measured decay time of 24(2)\,ns (see Fig.~\ref{fig2}d), the lifetime-limited linewidth of the optical transition is calculated to be $\Delta \nu_{\mathrm{lifetime}} = \frac{1}{2\pi \tau_{\mathrm{exc}}}=6.6(5)$\,MHz.
Finally, repeated acquisition of excitation spectra over more than an hour (Fig.~\ref{fig2}f) demonstrates exceptional spectral stability, with the fitted center frequencies of two distinct molecular lines exhibiting standard deviations of just 2.4\,MHz and 2.6\,MHz.
By removing one of the two applied microwave signals, and thus modulating the excitation spectrum as in Fig.~\ref{fig1,5}b, the spectral separation of the $\Tgx{} \rightarrow \Tex{}$ and $\Tgy{} \rightarrow \Tey{}$ transitions was determined to be $1.555(1)\,$GHz (see SI VII. E.).
The latter frequency separation is the smallest of the spin-selective optical transitions, while still allowing for highly spin-selective optical excitation due to the narrow homogeneous linewidth of 38\,MHz. 

\subsection*{Single Spin Readout}
\begin{figure}[htbp]
	\centering
	\includegraphics[width=18cm]{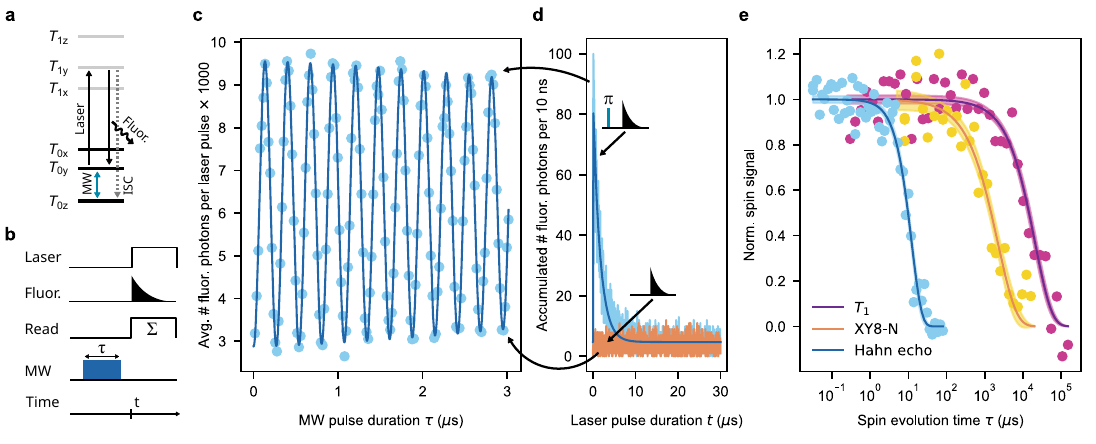}
	\caption{\textbf{Single spin initialization, manipulation and readout.}
		\textbf{a}, Energy level diagram with most important transitions: resonant laser excitation from $\Tgy{} \rightarrow \Tey{}$, excited state decay either via fluorescence to \Tgy{} or via ISC to \Tgz{} and MW field resonant to transition $\Tgz{} \leftrightarrow \Tgy{}$.
		\textbf{b}, Pulse sequence for spin Rabi oscillations: A MW pulse of duration $\tau$ coherently drives the initialized spin from \Tgz{} to \Tgy{} and back.
		The laser pulse invokes fluorescence only if the spin is in state \Tgy{}, and after prolonged exposure of a few \textmu s initializes the system in \Tgz{} via ISC. The amount of fluorescence photons constitutes the readout signal. The sequence is repeated for different $\tau$.
		\textbf{c}, Resulting fluorescence signal showing Rabi oscillations of a single carbene spin with a frequency of 3.7\, MHz. 
		\textbf{d}, Exemplary fluorescence responses upon laser excitation for the electron spin in state \Tgz{} (orange, no MW pulse after initialization) or \Tgy{} (blue, MW $\pi$-pulse after initialization).
		\textbf{e}, Lifetimes of the single spin for various MW pulses sequences. During a Hahn-echo sequence ($\pi/2-\tau/2-\pi-\tau/2-\pi/2$) the spin signal decays after $12.2(6)\,$\textmu s to 1/e, during an XY8-$N$ sequence ($\pi/2-\left[ \mathrm{XY8}\right]^N-\pi/2$, with $1\,$\textmu s $\pi$-pulse separation and $\tau=N\cdot8\cdot1\,$\textmu s for $N=1..2000$) after $2.2(3)\,$ms and during a $T_1$-sequence ($\pi-\tau$ or $\tau$) after $21(2)\,$ms.
	}\label{fig3}
\end{figure}

Optically detected single spin manipulations were performed in a similar fashion as the ensemble measurements shown in Fig.~\ref{fig1,5}c.
A laser pulse resonant to the optical single molecule transition $\Tgx{} \rightarrow \Tex{}$ or $\Tgy{} \rightarrow \Tey{}$ was used to measure the spin state, yielding a high fluorescence count rate when the spin of that molecule is in state \Tgx{} or \Tgy{}, respectively, and only background photons otherwise (see Fig.~\ref{fig3}a, d).
Furthermore, during laser illumination, the ISC eventually leads to the initialization of the electron spin in \Tgz{}.
In between laser pulses, a MW pulse sequence coherently manipulates the electron spin in the triplet ground state and thus affects the fluorescence response to the subsequent laser pulse (see Fig.~\ref{fig3}b).
As a prime example, Fig.~\ref{fig3}c presents the Rabi oscillation of a single \qm{} due to a MW pulse resonant to spin transition $\Tgz{} \leftrightarrow \Tgy{}$.
More complex MW sequences were applied to measure spin properties like the Hahn-echo spin coherence time $T_2=12.2(6)\,$\textmu s, the spin coherence time $T_2^\mathrm{(XY8)}=2.2(3)\,$ms during an XY8 dynamical decoupling sequence and the spin lattice relaxation time $T_1=21(2)\,$ms (see. Fig.~\ref{fig3}e and Methods).
Given the abundance of proton nuclear spins in the environment, these long spin lifetimes are maintained due to the suppressive effect of the large ZFS in \T{0} compared to common hyperfine couplings~\cite{baylissEnhancingSpinCoherence2022,roggorsOpticallyDetectedMagnetic2025a}.

\subsection*{Conclusion and Outlook}\label{sec3}
The demonstration of optical initialization, coherent control, and readout of individual molecular qubits establishes a novel and chemically versatile spin-photon interface.
By substitutionally embedding photoactivated triplet ground state carbenes within an isosteric, structurally matched crystalline matrix, we achieved single-emitter addressability characterized by narrow optical linewidths and exceptional long-term spectral stability.
This pristine interface not only facilitated highly efficient spin-selective excitation but also enabled the coherent control and measurement of the molecule's magnetic substructure, yielding a $T_2^\mathrm{(XY8)}$ time of 2.2(3)\,ms and $T_1$ time of 21(2)\,ms at 4.5\,K.
These single-molecule values represent a substantial leap for molecular qubits, extending previously reported values on ensembles by more than one order of magnitude \cite{baylissEnhancingSpinCoherence2022,roggorsOpticallyDetectedMagnetic2025a}.
Crucially, this performance simultaneously bridges the gap to established inorganic color centers.
While the nitrogen-vacancy center remains an exceptional benchmark for solid-state spin lifetimes \cite{jarmolaTemperatureMagneticFieldDependentLongitudinal2012a,bar-gillSolidstateElectronicSpin2013a}, our molecular platform already rivals the phonon-limited coherence of inversion-symmetric defects, as e.g. the silicon-vacancy center in diamond\cite{pingaultCoherentControlSiliconvacancy2017,sukachevSiliconVacancySpinQubit2017a}.

Looking ahead, the inherent versatility of this molecular system is expected to unlock a multitude of experiments in fundamental quantum science.
The absence of extraneous electron defects within the host crystal, a common limitation in systems such as diamond or silicon carbide, provides a significantly lower-noise environment for exploring complex interactions between multiple electron spins.
In addition, the large transverse term of the electron spin ZFS ($E_{T_0} = 540.9(2)$\,MHz) decouples the electron spin from the surrounding hydrogen nuclear spins~\cite{miaoUniversalCoherenceProtection2020} which are present in the molecule.
These experiments are further facilitated by the ease of doping the crystals with varying concentrations of qubit molecules, coupled with the selective spatial activation of the carbenes, allowing for the creation of electron spin concentrations ranging from very sparse to extremely dense.
Furthermore, the bottom-up nature of chemical synthesis allows for the deterministic creation of localized nuclear spin registers through site-selective isotopic labeling, offering a degree of rational design and uniformity currently unattainable in many solid-state platforms.
This capacity for rational design extends also to the targeted optimization of photophysical parameters, such as excited-state lifetimes and inter-system crossing dynamics, which can be tuned to meet the specific requirements of various experimental protocols.

Beyond its utility as a research tool, the platform offers immediate pathways for advanced technological applications.
The intrinsic processability of these molecular systems, including their application in very thin films, facilitates their direct integration with mature photonic integrated circuits based on materials such as silicon nitride or lithium niobate~\cite{lombardiPhotostableMoleculesChip2018,huangOnchipQuantumInterference2025,langeCavityQEDMolecular2025}.
Such integration provides a viable route for the on-chip routing of photons entangled with single qubit molecules, serving as a fundamental building block for quantum repeater nodes and distributed quantum computing.
Crucially, the substantial energy separation between spin-selective transitions of at least $1.555(1)$\,GHz is more than two orders of magnitude greater than the lifetime-limited linewidth of $6.6$\,MHz.
This ensures that high-fidelity spin addressability will persist even when excited-state lifetimes are deliberately shortened via coupling to high-finesse optical cavities.
Ultimately, this work introduces a structurally precise and chemically tunable interface that promises a scalable framework for the next generation of quantum technologies. 

\section*{Methods}
\subsection*{Computational Methods}
Multiscale ONIOM (QM/QM2) optimization\cite{svenssonONIOMMultilayeredIntegrated1996, chung_oniom_2015} as implemented in ORCA 6.1.0\cite{neeseORCAQuantumChemistry2020, neeseSoftwareUpdateORCA2025} was used to simulate the carbene with a nearby nitrogen \textit{in situ}, \textit{i.e.} specifically taking into account the matrix environment.
ONIOM calculations were performed by dividing a 837-atom system derived from the experimental single crystal structure of the matrix (CCDC 2542406) into a 309-atom high-level QM subsystem (carbene+N$_2$ and first shell of six surrounding matrix molecules) and a fixed-coordinate, 528-atom, low-level QM2 subsystem (second shell of 12 surrounding matrix molecules), utilizing a subtractive coupling scheme with electrostatic embedding. We note, that, it is important to use nonspherical atomic form factors within the NoSpherA2 extension\cite{kleemissAccurateCrystalStructures2021} of Olex2 during structure refinement to ensure the crystallographic data is suitable for generating model clusters without computational post-treatment of hydrogens.\cite{dolomanovOLEX2CompleteStructure2009a} The central, high level QM model part was treated by unrestricted Kohn-Sham (UKS) calculations with density functional theory (DFT) using the B97-3c composite approach~\cite{brandenburgB973cRevisedLowcost2018}. This method natively incorporates the modified Ahlrichs’ triple-zeta split-valence basis set (def2-mTZVP)~\cite{weigend_balanced_2005} alongside the atom-pairwise dispersion correction with the Becke-Johnson damping scheme (D3BJ)~\cite{grimme_consistent_2010, grimme_effect_2011}. The Coulomb integrals were evaluated using the Split-RI-J approximation with the def2-mTZVP/J auxiliary basis set~\cite{weigend_ri_2006}. Matrix molecules in the QM2 region were computed using the semiempirical extended tight-binding method GFN2-xTB (XTB2)~\cite{Bannwarth_GFN2_2019}. The relaxed cluster was used to extract the final coordinates of the carbene qubit in the triplet ground state (see Fig.~\ref{fig1}). 
Time-dependent density functional theory (TD-DFT) calculations employed the B3LYP functional\cite{becke_densityfunctional_1993, lee_development_1988, stephens_ab_1994} in combination with the def2-TZVPP basis set\cite{weigend_balanced_2005} and the D4 dispersion correction.\cite{caldeweyher_extension_2019} In this calculation, the Coulomb and exchange contributions were evaluated using the resolution-of-identity approximation \cite{neese_soc_2005} in conjunction with the RIJCOSX scheme.\cite{helmich-Improved_2021} 
For prediction of spin dynamics we use the state-averaged complete active space self-consistent field (SA-CASSCF) method~\cite{roosCompleteActiveSpace1980,lawleyInitioMethodsQuantum1987} corrected by the van Vleck quasi-degenerate (QD) extension to the strongly contracted second-order N-electron valence state perturbation theory (SC-NEVPT2) to account for dynamic correlation (spin-orbit and spin-spin coupling)~\cite{roemeltExcitedStatesLarge2013,angeliIntroductionNelectronValence2001, angeliNelectronValenceState2002}. The active space included twelve electrons in twelve orbitals and the trust-region augmented hessian (TRAH) algorithmwas employed for robust wavefunction convergence\cite{helmich-parisTrustregionAugmentedHessian2022} . The wavefunction is state-averaged over all states that are close in energy to the region of interest, in our case the first three triplet and five singlet states (SA8). The def2-TZVPP basis set~\cite{weigend_balanced_2005} was employed for all atoms, alongside the def2-TZVPP/C~\cite{hellweg_aux_2007} auxiliary basis set to accelerate the evaluation of Coulomb and exchange integrals via the RIJCOSX approximation~\cite{neese_rijcosx_2009}.

\subsection*{Sample Preparation}
All crystals of the \qm{} doped into the matrix were grown by solution based crystallization methods and under exclusion of daylight, \textit{i.e.} under red light $\geq$600\,nm to avoid slow decomposition of the diazomethane precursor. In a typical procedure, \bike{} is dissolved in hot THF (stabilizer free, HPLC grade, 0.014 M) and cooled to room temperature before the diazomethane precursor dopant is added as a solid and dissolved using an ultrasonication bath for a few seconds.
The corresponding solution is filtered through a syringe filter (PTFE $0.2\,$\textmu m) and added to vials (2-5 mL) standing in a screw cap jar. The jar is filled with \textit{n}-hexane (fill height ca 2-3 cm) and sealed.
Upon standing in the dark at room temperature for several days doped crystals form at the bottom of the vials.
The supernatant is removed using a Pasteur pipette to avoid mechanical stress on the crystals.
The residue is then washed twice with small volumes of THF/\textit{n}-hexane (1:2) without allowing the crystals to dry out.
This approach preserves the surface quality and prevents secondary nucleation or island-like growth. The resulting, air-dried crystals remain stable in a freezer ($T = -22\,^\circ$C) for at least six months.
A doped crystal of $\sim1$\,mm$^3$ volume (25 ppm doping ratio) with optically flawless appearance was selected and mounted onto a cryogenic cold finger using vacuum grease to ensure optimal thermal contact. The crystal was oriented such that its transition dipole moment vector of the $T_0 \rightarrow T_1$ transition was within the plane of polarization of the laser beam. A 25\,\textmu m diameter copper wire was positioned across the sample to facilitate MW delivery for spin manipulation.

\subsection*{Cryogenic Microscopy and Spectroscopy Experiment}
Optical excitation and fluorescence experiments were performed using a custom-built cryogenic confocal microscope operating at a temperature of approximately $4.5$\,K at the sample mount.
The molecular crystal sample was attached to a sample platform (4.5\,K) of a closed-cycle, optical cryostat.
The microscope objective (NA=0.9, WD=1\,mm) is within the vacuum shroud of the cryostat but at room temperature and separated from the cold sample by a thin metal plate thermalized at $50\,$K with a small aperture ($\approx 4\,$mm).
The sample platform can be coarsely moved w.r.t. the objective by slip-stick positioners, and a galvo-scanner outside the cryostat allows fine scanning of the laser beam across the sample via a telecentric lens system.
Optical excitation is performed using a tunable ($570$\,nm to $620$\,nm) dye laser using Rhodamine-6G as laser medium.
Initial photoactivation and off-resonant optical excitation of the \qms{} is performed at a wavelength of $580$\,nm.
For full activation $5\,$\textmu W of laser light was focused on the crystal surface for about 2 minutes; and for mild activation about $1\,$\textmu W of focused laser light was scanned across the crystal surface.
For annealing, the crystal with as-activated carbenes was heated up to 200\,K for about one hour.
Fluorescence spectra were recorded under off-resonant laser illumination (580\,nm, $5\,$\textmu W, 545/70\,nm bandpass filter for laser clean up) using a 590\,nm longpass filter in front of the spectrometer.
Resonant laser excitation for excitation spectroscopy, microscopy or spin state initialization and readout typically requires powers of $\approx 3\,$nW, a 600\,nm shortpass filter for clean up and a 605\,nm longpass for fluorescence detection.
The angle of linear laser polarization was adjusted to maximize \qm{} fluorescence.
For some experiments, continuous-wave (CW) resonant optical excitation was accompanied by CW resonant spin excitation in the ground state spin triplet to increase fluorescence intensity.
In detail, when resonant optical excitation on the optical transitions $\Tgx{} \rightarrow \Tex{}$ or $\Tgy{} \rightarrow \Tey{}$ shelves the molecule into the dark state \Tgz{}, resonant MW spin excitation on transitions $\Tgz{} \leftrightarrow \Tgx$ or $\Tgz{} \leftrightarrow \Tgy{}$, respectively, retrieves this molecule from \Tgz{} and thus enables continuous optical excitation.
The shelving process is the effect causing the fluorescence decay depicted in Fig.~\ref{fig3}d.
This technique was used for experiments with results depicted in Figs. \ref{fig1,5}b, c and \ref{fig2}.
The laser intensity can be switched on and off within a few tens of nanoseconds using an acousto-optical modulator (AOM) that yields an on/off ratio of the laser intensity of more than $10^4$.
An electro-optical modulator (EOM) with a bandwidth of $5\,$GHz is used to provide faster switching times required for excited state lifetime measurements.
The strength of the applied microwave field was varied between experiments.
It was strongest for the Rabi measurement in Fig.~\ref{fig3}c.
For microwave-assisted fluorescence excitation spectra as shown in Fig.~\ref{fig1,5}b the microwave field was reduced by more than a factor of ten, corresponding to a Rabi frequency of about 250\,kHz.
For CW-ODMR measurements as shown in Fig.~\ref{fig1,5}c, the MW driving was reduced further to a Rabi frequency of about 40\,kHz.

\subsection*{Single spin initialization, coherent control and readout}
The procedure for conducting basic single spin manipulation and readout measurements is described in the main text for the example of Rabi oscillations (see Fig.~\ref{fig3}b,c,d).
In the latter case, the electron spin was manipulated during laser-off periods and the readout signal comprised the sum of all fluorescence photons during laser-on periods (see Fig.~\ref{fig3}d) divided by the number of laser pulse repetitions.
For the pulsed spin experiments depicted in Fig.~\ref{fig3}e, namely Hahn-echo, XY8-$N$ and $T_1$ measurements, the spin signal is derived slightly differently.
In the latter cases each data point is the difference of two complementary measurement results.
For example, in the $T_1$ measurement two complementary measurements for a certain data point are i) a bare waiting time $\tau$ before the next readout laser pulse and ii) a MW $\pi$-pulse followed by the same waiting time prior to the next readout laser pulse.
Here, i) yields a low and ii) a high number of fluorescence photons.
When the spin states of the complementary measurements have decayed to the equilibrium state, the number of fluorescence photons is supposed to be equal and thus the difference zero.
In the case of the Hahn-echo and XY8-$N$ measurements, two complementary measurements differ only by the phase of the last $\pi/2$-pulse within the MW sequence; it is either $0$ or $\pi$.
Finally, a stretched exponential decay, $f(\tau) = A \cdot \exp{\left\{ -\left( \tau / T\right)^\beta \right\}} + y_0$, was fitted to the data; for the $T_1$ measurement $\beta\overset{!}{=}1$.
The displayed data is then normalized to the respective value of $A$.
The stretch-factor $\beta$ for Hahn-echo and XY8-$N$ measurements are $1.7(2)$ and $0.9(2)$, respectively.
Deviations from $\beta=1$ for Hahn-echo and XY8-$N$ measurements are known for decoherence of single spins and hint towards non-Markovian processes, as expected for proximal electron and nuclear spins such as protons in the present case \cite{mimsPhaseMemoryElectron1968,chinQuantumMetrologyNonMarkovian2012,delangeUniversalDynamicalDecoupling2010}.

\subsection*{Fit of the second-order photon autocorrelation}
The measured second-order autocorrelation function $g^{(2)}(\tau)$ is analyzed using a three-level system model to capture both the single-photon antibunching, as well as the bunching dynamics of the singlet state.
The ideal correlation function is parameterized as
\begin{equation*}
	g^{(2)}_\mathrm{ideal}(\tau)=1-(1-g_0+A)e^{-|\tau-\tau_0|/\tau_\mathrm{anti}}+A\cdot e^{-|\tau-\tau_0|/\tau_\mathrm{bunch}}
\end{equation*}
with the relative amplitude $A$ of the bunching behavior, the depth of the antibunching $g_0$, the timescales of the (anti-)~bunching $\tau_\mathrm{anti}$ and $\tau_\mathrm{bunch}$ and the electronic coincidence shift $\tau_0$.
To account for finite timing resolution of the measurement, $g^{2}_\mathrm{ideal}(\tau)$ is convolved with a Gaussian instrument response function.
The total standard deviation of the temporal jitter $\sigma_\mathrm{total}=1.56$\,ns was calculated by taking the root-mean-square of two times the variance of the detector jitter ($\sigma_\mathrm{SPAD}=0.425$\,ns) and the uniform discretization error introduced by the 5\,ns temporal binning of the correlation data ($\sigma_\mathrm{bin}=5/\sqrt{12}$\,ns).
Finally, the convolved model is fitted to the raw correlation data using a non-linear least-squares algorithm, incorporating a baseline scaling parameter to ensure the fitted $g^{(2)}(\tau)$ correctly normalizes to 1 as $\tau \rightarrow \infty$.
The fitted values in Fig.\ref{fig2}e are $\tau_\mathrm{bunch}=245(9)$\,ns, $\tau_\mathrm{anti}=28(2)$\,ns and $g^{(2)} (0)=0.142^{+0.169}_{-0.023}$.
To estimate the uncertainty of the $g^{(2)}(0)$ value, a Monte Carlo simulation is performed by generating synthetic data sets from the covariance matrix of the fit.

\section*{Data availability}
Crystallographic X-Ray data that support the findings of this study have been deposited in the Cambridge Structural Database with the CCDC accession number 2542406.
The data that support the findings of this study are available from the corresponding authors upon reasonable request.

\section*{Code availability}
All analysis codes related to the current study are available from the corresponding author upon reasonable request.

\section*{Supplementary information}
Supplementary Information containing details on chemical synthesis and material fabrication, bulk optical spectroscopy, details on computational chemistry methods and results, EPR spectroscopy and additional cryogenic confocal spectroscopy measurements is attached to this document.

\section*{Acknowledgements}
Mass spectrometry and X-ray crystallography was conducted at the Service Center mass spectrometry and X-ray crystallography facility at the University of Ulm.
Photoluminescent quantum yields were recorded at the Institute for Organic Chemistry III, University of Ulm.
The authors would further like to thank Dr. P. Hautle, Paul Scherrer Institute, Switzerland, for loaning of the EPR cryostat and Dr. Mathias Hermann (Institute of Organic Chemistry II and Advanced Materials, Ulm University) for X-ray crystallographic analysis.

\section*{Funding}
The authors acknowledge support by the state of Baden--Württemberg through bwHPC and the German Research Foundation (DFG) through grant no INST 40/575-1 FUGG (JUSTUS 2 cluster).
The work was furthermore made possible by the DLR Quantum Computing Initiative and the Federal Ministry for Economic Affairs and Climate Action (COMIQC project, https://qci.dlr.de/projects/COMIQC).
A.S. gratefully acknowledges the support of the Clore Israel Foundation Scholars Programme, the Israeli Council for Higher Education, and the Milner Foundation.
A.R. acknowledges the support of ERC grant QRES, project number 770929, Quantera grant MfQDS, the Israeli Science foundation (ISF), the Schwartzmann university chair and the Israeli Innovation Authority under the project “Quantum Computing Infrastructures”.

\section*{Author contributions}
T.A.S. and T.R.E. performed computational chemistry calculations and their analysis.
A.A. and P.M. performed chemical synthesis, under supervision from T.A.S.
J.S. and T.R.E. performed crystal growth and sample preparation, with support from T.A.S.
S.R. and T.U. performed cryogenic spectroscopic measurements, with support from G.B. and under supervision from F.J., M.P., P.N.
P.M. and G.B. performed time-correlated single photon-counting measurements and analysis, under supervision from P.N. and T.R.E.
T.R.E. performed and analysed microscopy and EPR measurements.
A.S., M.B.P., A.R., T.R.E, T.A.S., M.P. and P.N. analysed the data and interpreted results.
A.S., M.B.P. and A.R. provided theoretical support to the measurements.
M.P. and I.S. supervised the project.
T.A.S., T.R.E., M.P. and I.S. wrote the manuscript with input from all co-authors.

\section*{Materials and Correspondence}
Requests for material and correspondence may be addressed to Matthias Pfender or Ilai Schwartz.


\clearpage
\pagebreak

\makeatletter
\let\addcontentsline\oldaddcontentsline
\makeatother

\setcounter{equation}{0}
\setcounter{figure}{0}
\setcounter{table}{0}
\setcounter{section}{0}

\makeatletter
\renewcommand{\theequation}{S\arabic{equation}}
\renewcommand{\thefigure}{S\arabic{figure}}
\renewcommand{\thetable}{S\arabic{table}}
\makeatother

\begin{center}
	\textbf{\LARGE Supplementary Information: "A Single-Molecule Spin-Photon Interface"}
\end{center}
\vspace{10pt}

\tableofcontents
\vspace{20pt}

\section{Chemistry and material preparation}
\subsection{Synthesis and characterization}
\subsubsection{Experimental methods}
All commercially available chemicals were purchased from Sigma Aldrich, TCI Germany, VWR International, Fischer Scientific, Acros Organics, ABCR, ChemPur, Strem, BLDpharm, and Alfa Aesar. Unless otherwise noted, commercially available materials were used without purification. Moisture and oxygen sensitive reactions were carried out in flame-dried glassware and under an inert atmosphere of purified argon using syringe/septa technique. Commercially available anhydrous tetrahydrofuran ($\ge$99.9\%, THF) was stored over molecular sieve (3\,\AA) and used without purification. 

Thin-layer chromatography (TLC) was performed on the aluminium plates coated with 0.20 mm thickness of Silica gel 60 F$_{254}$ (Macherey-Nagel). Developing plates were visualized using UV light at wavelength of 254 and 365 nm. $^1$H and $^{13}$C NMR spectra were recorded on a Bruker Avance III HD 400 ($^1$H: 400.1\,MHz, $^{13}$C: 100.6\,MHz) spectrometer at 25\,°C. The chemical shifts ($\delta$) are reported in parts per million (ppm) and are referenced to solvent signal: $\delta = 5.32$\,ppm ($^1$H) and $\delta = 53.8$\,ppm ($^{13}$C) for CD$_2$Cl$_2$ and $\delta$ = 3.58\,ppm ($^1$H) and $\delta$ = 67.2\,ppm ($^{13}$C) for THF-d$_8$. Coupling constants ($J$) are given in Hz and the apparent resonance multiplicity is reported as singlet (s), doublet (d), triplet (t), or multiplet (m). 2D Correlation NMR spectra were recorded for signal assignment of all target compounds. 

FT-IR spectra were recorded on pure powders using a Spectrum Two FT-IR spectrometer (PerkinElmer, USA) equipped with a GladiATR™ ATR accessory (Pike Technologies, USA). 

UV/Vis absorption spectra were recorded on a Thermo Scientific™ GENESYS™ 150 spectrophotometer. All measurements were performed in methylcyclohexane using quartz cuvettes with a path length of 1\,cm at room temperature. The extinction coefficient $\epsilon$ was determined by measuring the absorption at different concentrations. 

High resolution MALDI mass spectra were recorded on a Bruker SolariX using Dithranol as matrix. Chemical ionisation (CI) mass spectra were recorded on a Thermo Scientific ISQ LT single quadrupole mass spectrometer by direct vaporization of samples into the MS ionization chamber (DIP) using methane as reagent gas. For sample \textbf{S3} the molecular mass was not observed which we ascribe to the delicate combination of thermal, light, and acid sensitivity of the diaryldiazomethanes. Instead, the MALDI mass spectrum of \textbf{S3} shows fragment ions at $m/z$ 319.14902 (100\%) resulting from N$_2$-cleavage and protonation together with a larger cluster at 665.29577 (74\%) [2M–N$_2$+H]$^+$.

Automated flash chromatography was performed on a Biotage® Selekt system using Biotage® Sfär silica gel cartridges (20\,\textmu m). Product purification was monitored by UV absorbance at 250\,nm and 300\,nm. Matrix compound \textbf{S1} was purified by vacuum sublimation in the last step to afford sufficiently pure material. Elemental analysis was carried out on a Elementar vario Micro cube at the Core Facility Elemental, Molecular and Materials Analysis (University Ulm). Melting points were determined with a Büchi M-565 using 2\,°C/min heating rate.

\subsubsection{Synthesis of di([1,1'-biphenyl]-4-yl)diazomethane}

\begin{figure}[htpb]
	\centering
	\includegraphics[width=\textwidth]{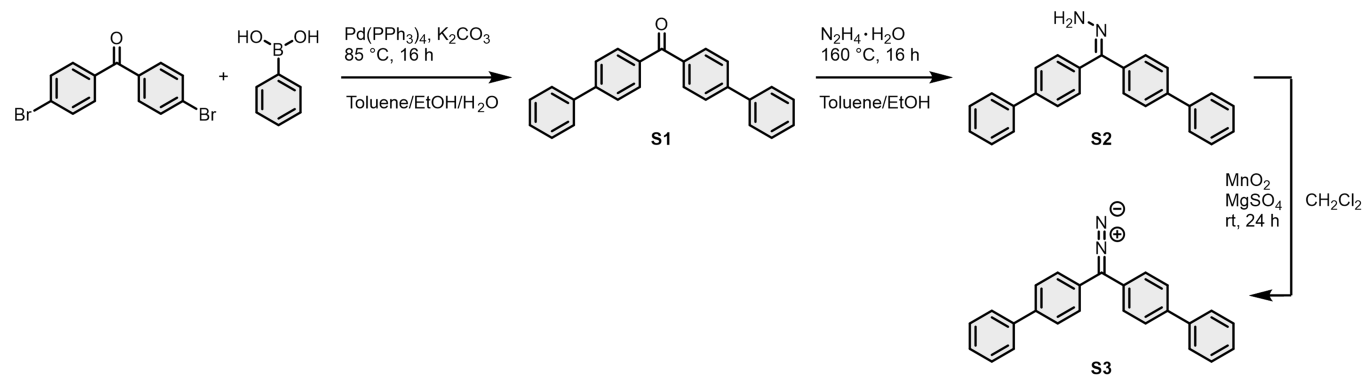}
	\caption{Synthetic route to matrix S1 and dopant \textbf{S3}.}
	\label{fig:SI_Syn_1}
\end{figure}

\textbf{Di([1,1'-biphenyl]-4-yl)methanone (S1)}: 
The compound was prepared according to a modified literature procedure.\cite{kalomenopoulosEnantioselectiveSynthesisAAryl2025} 4,4'-Dibromobenzophenone (5.00\,g, 14.7\,mmol) and phenylboronic acid (4.12\,g, 33.8\,mmol) were dissolved in a degassed mixture of toluene (123\,mL, 0.12\,M), ethanol (63\,mL, 0.23\,M) and water (63\,mL, 0.23\,M). K$_2$CO$_3$ (12.2\,g, 88.2\,mmol) and Pd(PPh$_3$)$_4$ (0.34\,g, 0.29\,mmol) were added, and the reaction mixture was stirred for 16\,h at 85\,°C. After cooling to room temperature, water was added and the layers were separated. The aqueous phase was extracted with ethyl acetate (2 $\times$ 200\,mL). The combined organic phases were washed with water (300\,mL) and brine (200\,mL), dried over anhydrous Na$_2$SO$_4$, filtered, and concentrated under reduced pressure. 
The crude product was purified by column chromatography (flash-silica gel; CH$_2$Cl$_2$/cyclohexane 1:1) to obtain di([1,1'-biphenyl]-4-yl)methanone (\textbf{S1}) as a colourless solid (4.5\,g, 13.5\,mmol, 92\% yield). Analytically pure material was obtained by vacuum sublimation (3 $\times$ 10$^{-2}$\,mbar, 195\,°C). 
$R_f$\,(SiO$_2$, CH$_2$Cl$_2$/$n$-hexane 3:2)\,=\,0.33.
Mp. 238 -- 240\,°C (lit.\cite{kalomenopoulosEnantioselectiveSynthesisAAryl2025}: 238 -- 240\,°C);
$^1$H NMR (400\,MHz, CD$_2$Cl$_2$) $\delta$ 7.92 (d, $J\,=\,8.7$\,Hz, 4H, H1/H3), 7.76 (d, $J\,=\,8.7$\,Hz, 4H, H4/H12), 7.70 (d, $J\,=\,8.2$\,Hz, 4H, H7/H11), 7.50 (m, 4H, H8/H10), 7.42 (t, $J\,=\,7.3$\,Hz, 2H, H9) ppm;
$^{13}$C NMR (101\,MHz, CD$_2$Cl$_2$) $\delta$ 195.9 (C13), 145.4 (C5), 140.3 (C6), 136.9 (C2), 131.0 (C1/3), 129.4 (C8/10), 128.6 (C9), 127.7 (C7/11), 127.3 (C4/12) ppm;
IR (ATR): $\tilde{\nu}\,=\,1639$\,(C=O)\,cm$^{-1}$;
UV/Vis (methylcyclohexane): $\lambda_{\text{max}}\,=\,295$\,nm ($\epsilon\,=\,30\,400$\,M$^{-1}$\,cm$^{-1}$); 
HRMS (APCI, positive mode) $m/z$ calcd for C$_{25}$H$_{18}$O: 335.14304 [M+H]$^+$, found 335.14300;
Elemental analysis (\%calcd, \%found for C$_{25}$H$_{18}$O): C (89.79, 90.12), H (5.43, 5.67). 
The analytical data is in accordance with literature data.\cite{kalomenopoulosEnantioselectiveSynthesisAAryl2025} \\

\textbf{(Di([1,1'-biphenyl]-4-yl)methylene)hydrazine (S2):} The compound was prepared according to a modified literature procedure.\cite{schmittGeneralApproachNHPyrazoles2015a} A mixture of di([1,1'-biphenyl]-4-yl)methanone  (300\,mg, 0.90\,mmol), hydrazine monohydrate (N$_2$H$_4$ 64-65\% 0.14\,mL, 2.70\,mmol), toluene (anhydrous, 4.50\,mL, 0.20\,M) and EtOH (anhydrous, 2.25\,mL, 0.40\,M) was heated at 160\,°C for 16\,h in a high-pressure vessel sealed with a PTFE-lined screw cap. The reaction mixture was allowed to cool to room temperature over 1\,h, and the solvents were removed under reduced pressure. Purification by flash column chromatography on silica gel (CH$_2$Cl$_2$) afforded di([1,1'-biphenyl]-4-yl)methylene)hydrazine (\textbf{S2}) as a colourless solid (250\,mg, 0.72\,mmol, 80\% yield). 
$R_f$\,(SiO$_2$, CH$_2$Cl$_2$)\,=\,0.16; 
Mp. 178 -- 181\,°C;
$^1$H NMR (400\,MHz, CD$_2$Cl$_2$) $\delta$ 7.80 (d, $J\,=\,8.60$\,Hz, 2H), 7.69 (d, $J\,=\,7.0$ Hz, 2H), 7.62 (d, $J\,=\,7.1$\,Hz, 2H), 7.59 -- 7.54 (m, 4H), 7.50 (t, $J\,=\,7.5$\,Hz, 2H), 7.45 -- 7.39 (m, 5H), 7.36 -- 7.32 (m, 1H), 5.55 (s, 2H) ppm;
$^{13}$C NMR (101\,MHz, CD$_2$Cl$_2$) $\delta$ 148.2, 142.1, 141.0, 140.9, 140.8, 138.1, 132.3, 129.8, 129.3, 129.2, 128.4, 128.1, 127.8, 127.5, 127.3, 127.2, 127.1 ppm;
IR (ATR): $\tilde{\nu}$\,=\,3200, 3363\,(N--H), 1486\,(C=N)\,cm$^{-1}$;
HRMS (APCI, positive mode) $m/z$ calcd for C$_{25}$H$_{20}$N$_2$: 349.16993 [M+H]$^+$, found 349.16974;
Elemental analysis (\%calcd, \%found for C$_{25}$H$_{20}$N$_2$): C (86.17, 86.23), H (5.79, 6.01), N (8.04, 7.97).\\

\textbf{Di([1,1'-biphenyl]-4-yl)diazomethane (S3):} The compound was handled under exclusion of daylight; all manipulations were carried out under red light $\ge$600\,nm. A mixture of (di([1,1'-biphenyl]-4-yl)methylene)hydrazine (160\,mg, 0.46\,mmol), anhydrous MgSO$_4$ (66\,mg, 0.55\,mmol), and CH$_2$Cl$_2$ (anhydrous, 9.2\,mL, 0.05\,M) was cooled to 0\,°C under vigorous stirring. Activated MnO$_2$ (150\,mg, 1.72\,mmol) was added in one portion, resulting in the immediate formation of a deep purple suspension. The suspension was allowed to warm to room temperature and stirred for 24\,h. The resulting mixture was filtered through a plug of activated basic alumina (Brockmann activity I) and the solids were rinsed with a mixture of $n$-hexane/CH$_2$Cl$_2$ 1:1 (80\,mL) until the filtrate was colorless. The filtrate was concentrated at room temperature by rotary evaporation and stored at 7\,°C for a few hours. The formed dark purple needles were filtered off and dried under vacuum to give the analytically pure product (140\,mg, 0.40\,mmol, 88\%).
$R_f$\,(Al$_2$O$_3$ basic, CH$_2$Cl$_2$/$n$-hexane 3:2)\,=\,0.88; 
Mp. 238 -- 241\,°C;
$^1$H NMR (400\,MHz, THF-$d_8$) $\delta$ 7.71 (d, $J\,=\,8.7$ Hz, 4H, H4/12), 7.65 (d, $J\,=\,8.3$ Hz, 4H, H7/11), 7.43 -- 7.40 (m, 8H, H1/3/8/10), 7.30 (t, $J\,=\,7.3$\,Hz, 2H, H9) ppm;
$^{13}$C NMR (101\,MHz, THF-$d_8$) $\delta$ 141.1 (C2), 139.3 (C6), 129.4 (C1/3), 129.1 (C5), 128.3 (C4/12), 127.8 (C9), 127.2 (C7/11), 126.1 (C8/10) ppm (consistent with previous reports, the diazomethane carbon gives only a weak signal and was not observed\cite{roggorsOpticallyDetectedMagnetic2025a});
IR (ATR): $\tilde{\nu}\,=\,2024$\,(N$\equiv$N)\,cm$^{-1}$;
UV/Vis (methylcyclohexane): $\lambda_{\text{max}}\,=\,544$ nm ($\epsilon\,=\,130$ M$^{-1}$\,cm$^{-1}$);
MS (MALDI, positive mode) $m/z$ 319.14902 ([M--N$_2$+H]$^+$, 100\%), 665.29577 ([2M–N$_2$+H]$^+$, 74\%);
Elemental analysis (\%calcd, \%found for C$_{25}$H$_{18}$N$_2$): C (86.68, 86.79), H (5.24, 5.29), N (8.09, 8.03).

\newpage
\subsection{NMR spectra}
\begin{figure}[htpb]
	\centering
	\includegraphics[width=0.70\textwidth]{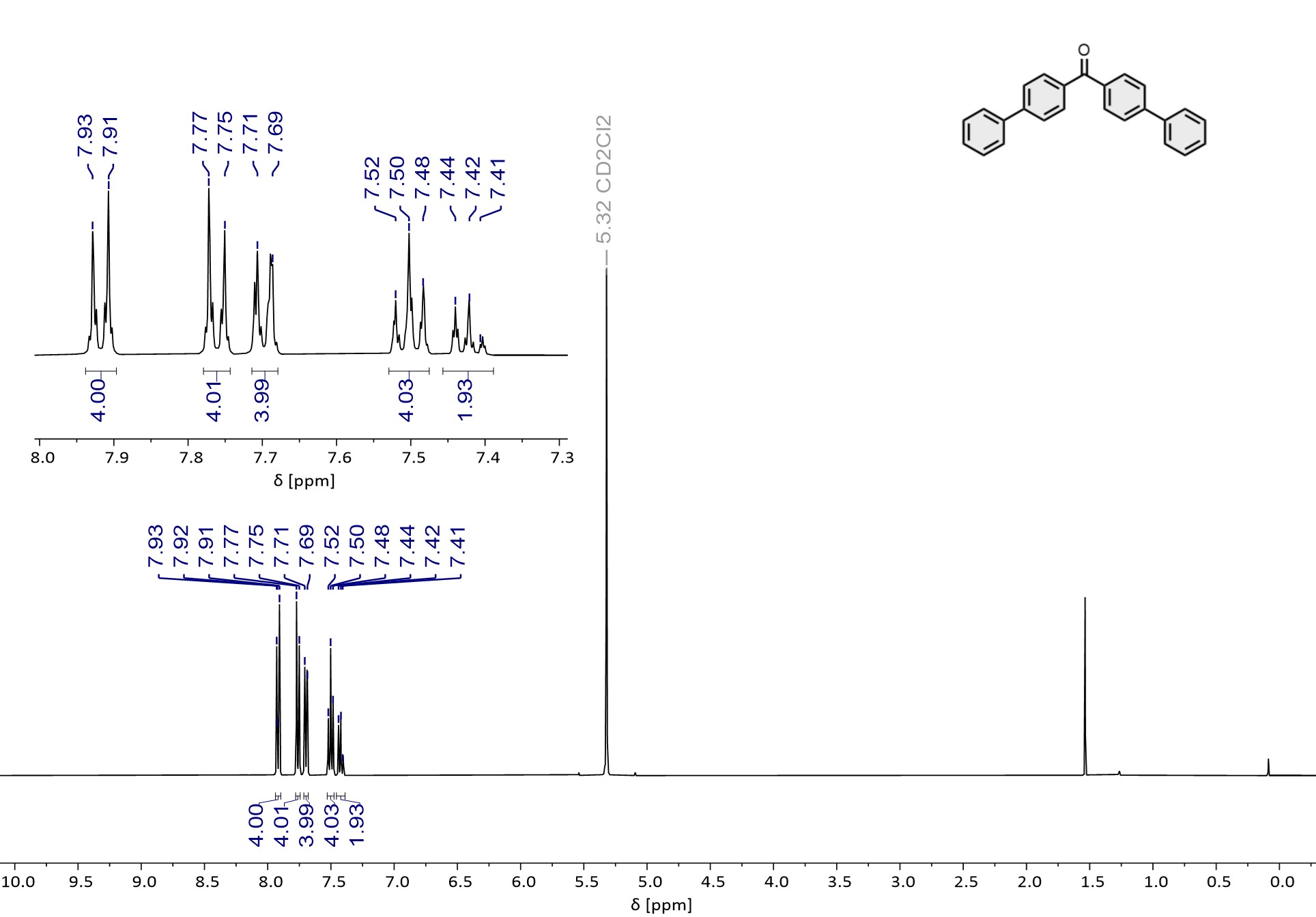}
	\caption{$^1$H NMR spectrum (400.1 MHz, CD$_2$Cl$_2$) of compound \textbf{S1} recorded at rt.}
	\label{fig:Ketone_1HNMR}
\end{figure}
\begin{figure}[htpb]
	\centering
	\includegraphics[width=0.70\textwidth]{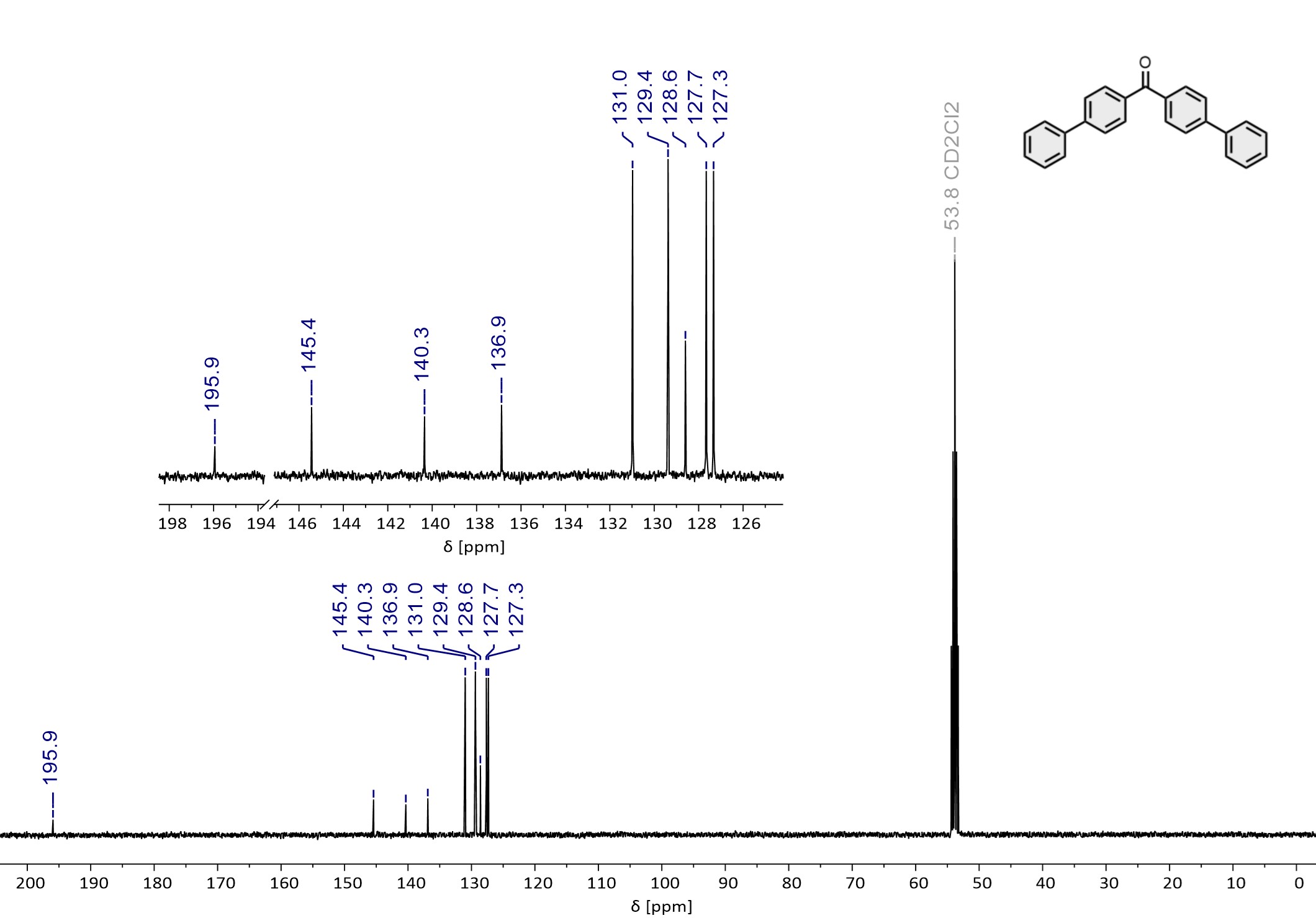}
	\caption{$^{13}$C NMR spectrum (100.6 MHz, CD$_2$Cl$_2$) of compound \textbf{S1} recorded at rt.}
	\label{fig:Ketone_13CNMR}
\end{figure}
\begin{figure}[htpb]
	\centering
	\includegraphics[width=0.8\textwidth]{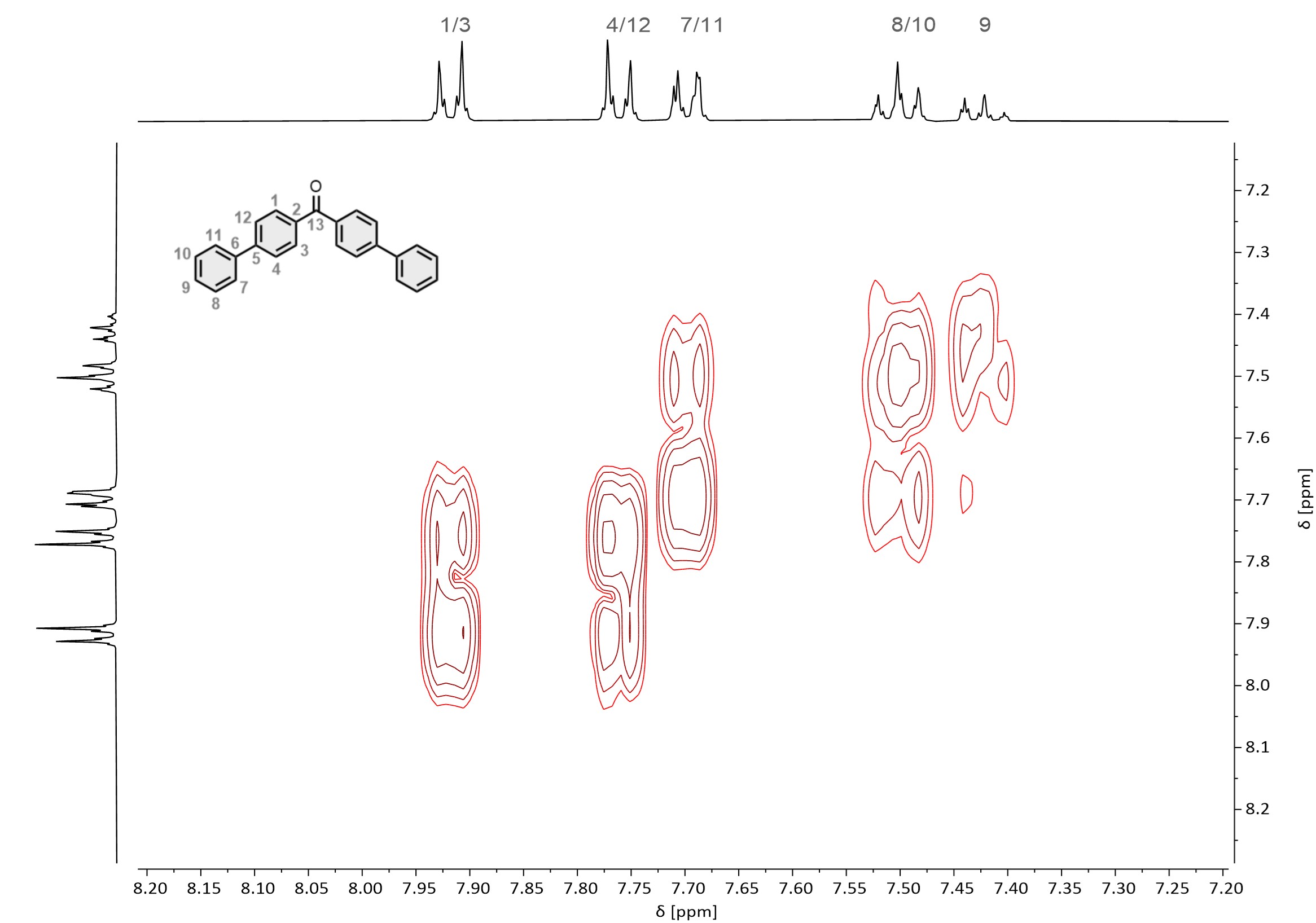}
	\caption{Two-dimensional $^1$H,$^1$H-COSY NMR spectrum (400.1 MHz, CD$_2$Cl$_2$) of \textbf{S1} at rt.}
	\label{fig:Ketone_COSY}
\end{figure}
\begin{figure}[htpb]
	\centering
	\includegraphics[width=0.8\textwidth]{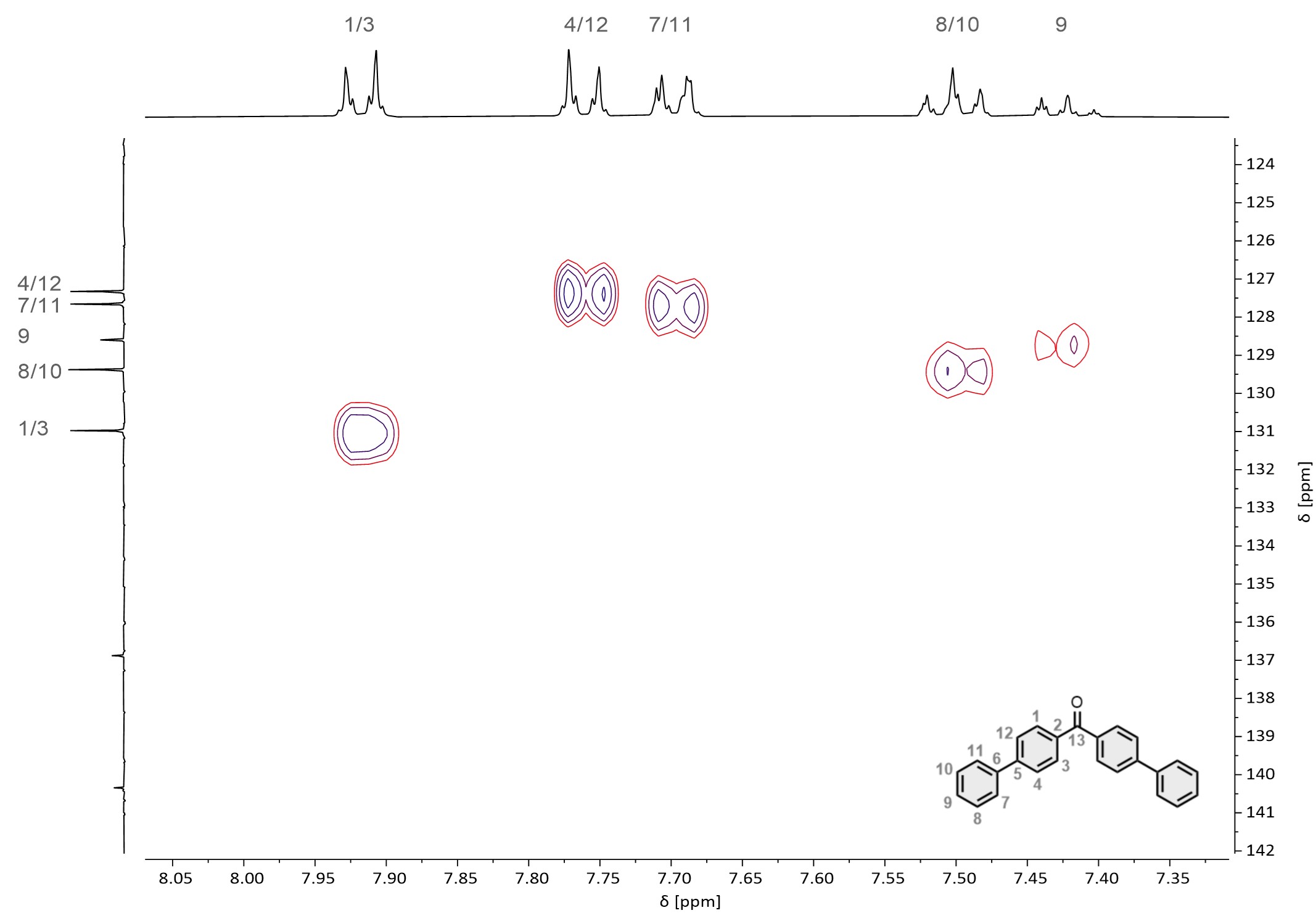}
	\caption{2D $^1$H,$^{13}$C-HSQC NMR spectrum ($^1$H: 400.1 MHz, $^{13}$C: 100.6 MHz, CD$_2$Cl$_2$) of \textbf{S1} rt.}
	\label{fig:Ketone_HSQC}
\end{figure}
\begin{figure}[htpb]
	\centering
	\includegraphics[width=0.8\textwidth]{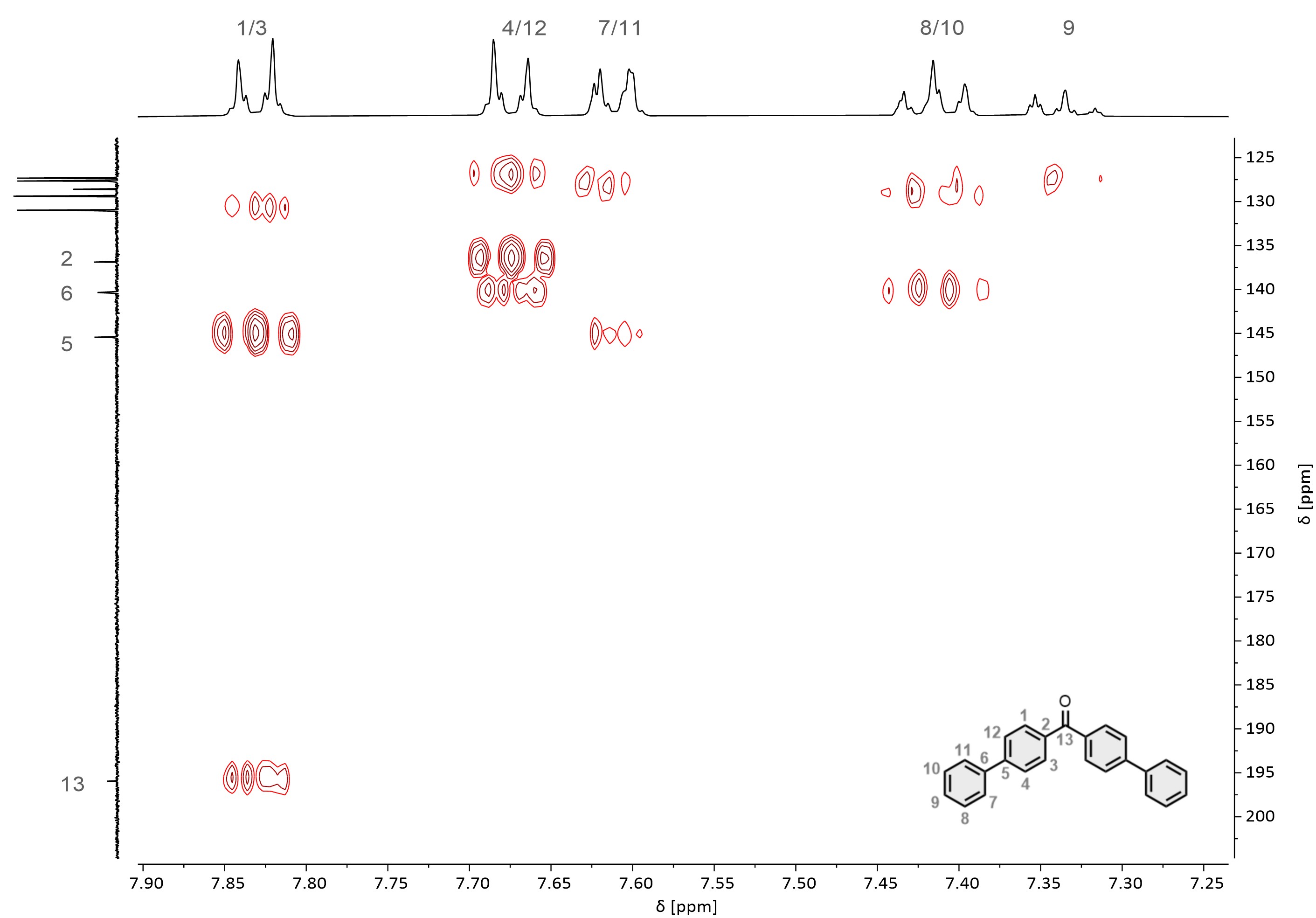}
	\caption{2D $^1$H,$^{13}$C-HMBC NMR spectrum ($^1$H: 400.1 MHz, $^{13}$C: 100.6 MHz, CD$_2$Cl$_2$) of \textbf{S1} at rt.}
	\label{fig:Ketone_HMBC}
\end{figure}
\begin{figure}[htpb]
	\centering
	\includegraphics[width=0.8\textwidth]{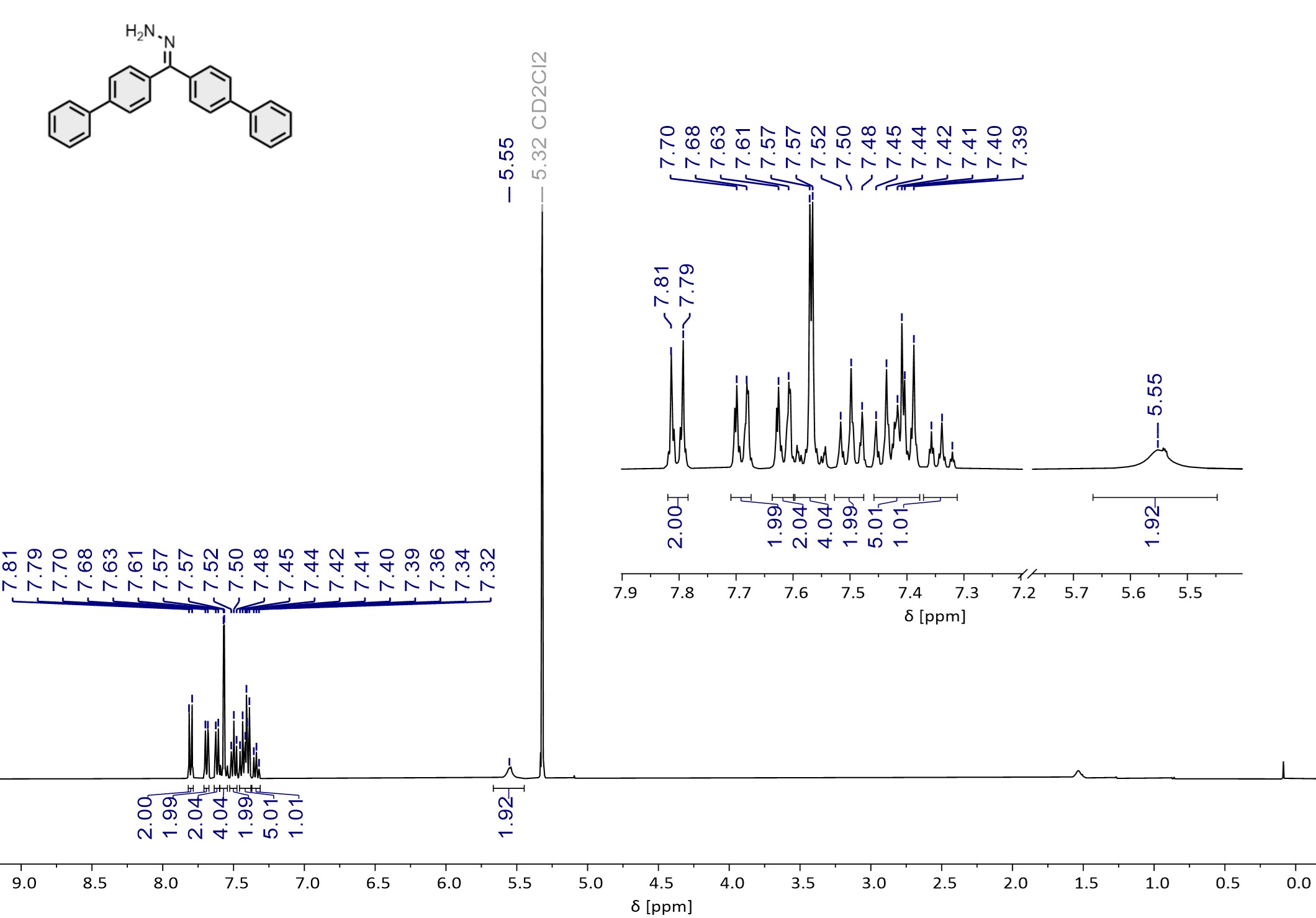}
	\caption{$^1$H NMR spectrum (400.1 MHz, CD$_2$Cl$_2$) of compound \textbf{S2} recorded at rt.}
	\label{fig:Hydrazone_1HNMR}
\end{figure}
\begin{figure}[htpb]
	\centering
	\includegraphics[width=0.8\textwidth]{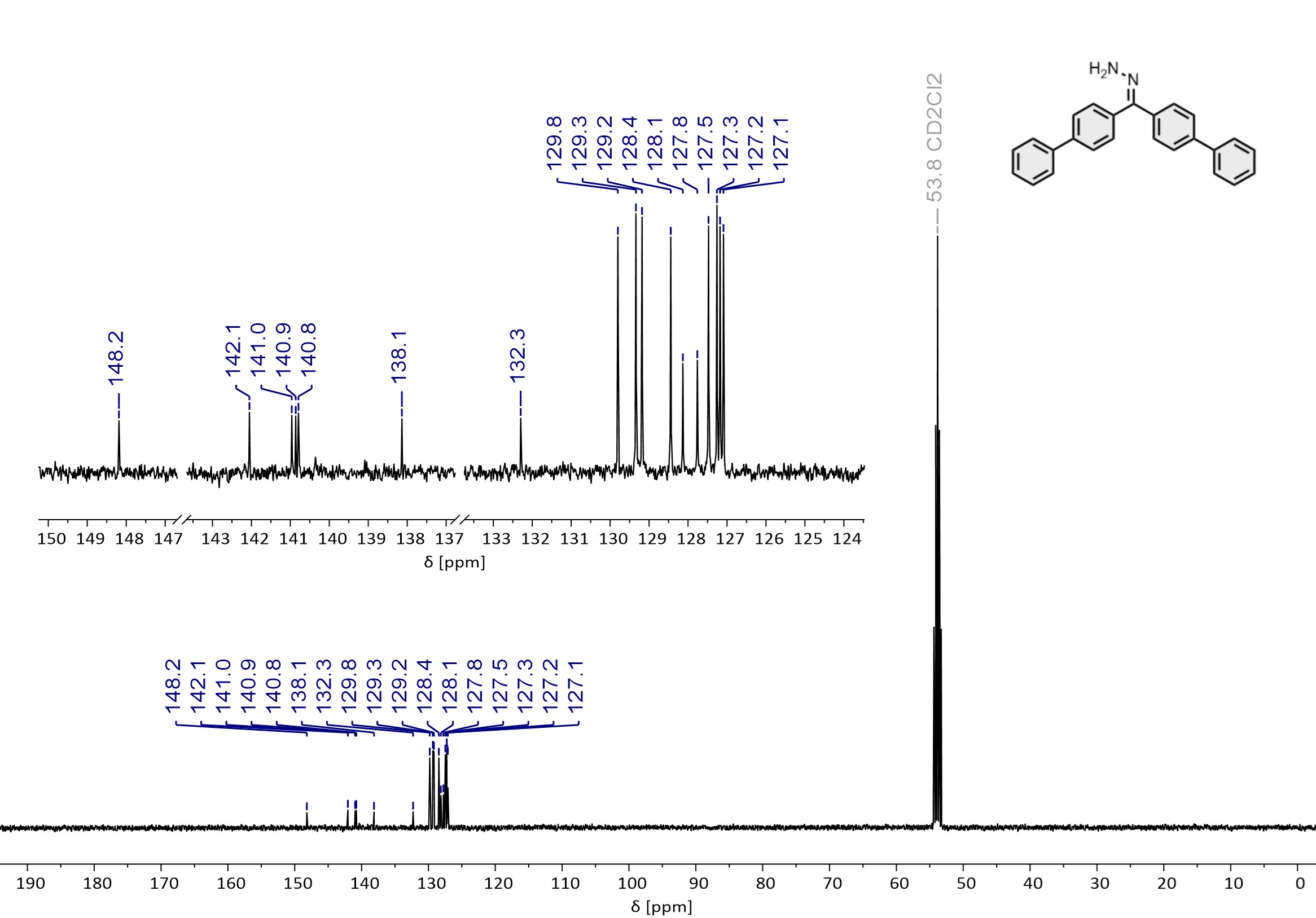}
	\caption{$^{13}$C NMR spectrum (100.6 MHz, CD$_2$Cl$_2$) of compound \textbf{S2} recorded at rt.}
	\label{fig:Hydrazone_13CNMR}
\end{figure}
\begin{figure}[htpb]
	\centering
	\includegraphics[width=0.8\textwidth]{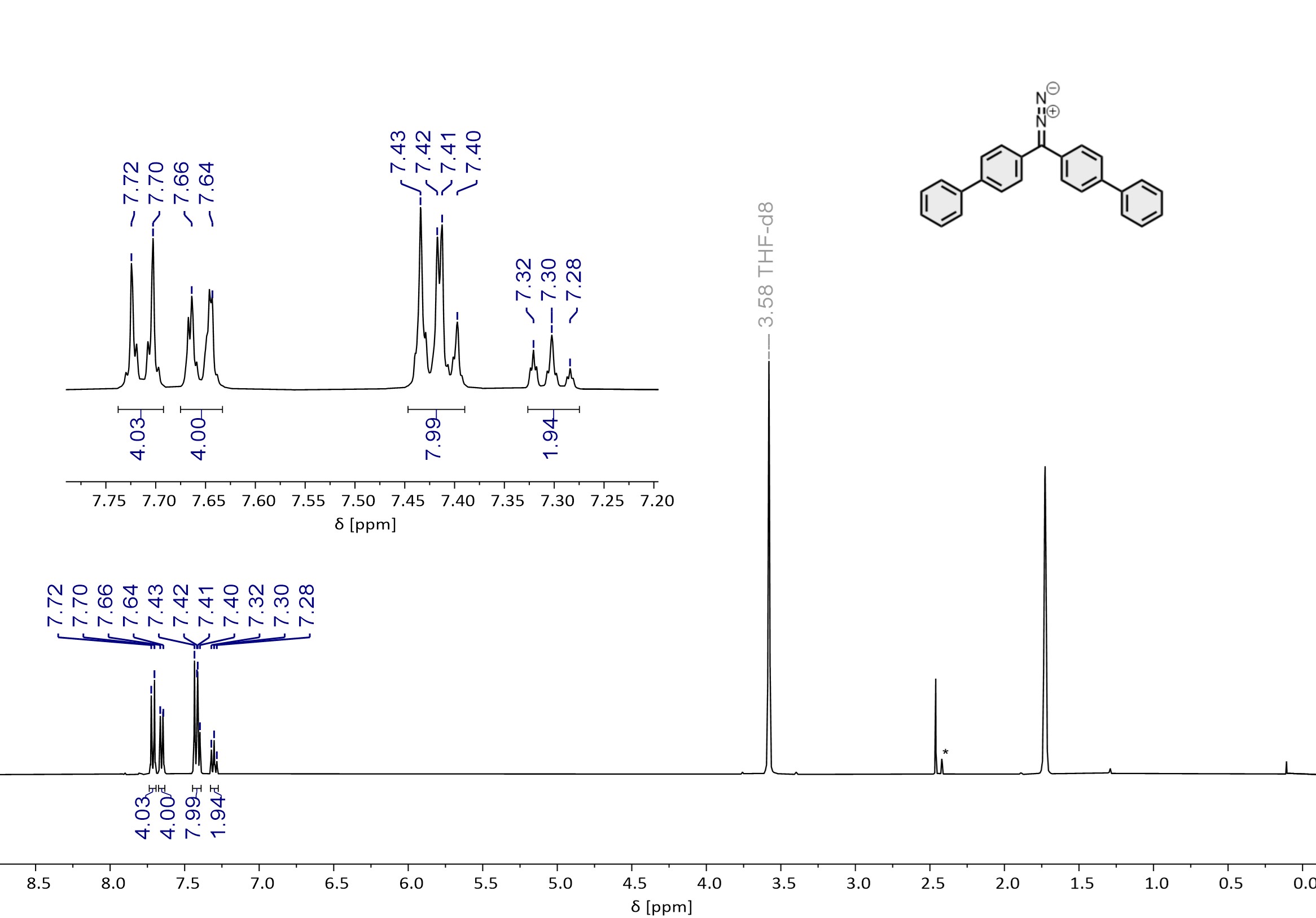}
	\caption{$^1$H NMR spectrum (400.1 MHz, THF-\textit{d}$_8$) of compound \textbf{S3} recorded at rt;  *denotes impurity from solvent.}
	\label{fig:Diazo_1HNMR}
\end{figure}
\begin{figure}[htpb]
	\centering
	\includegraphics[width=0.8\textwidth]{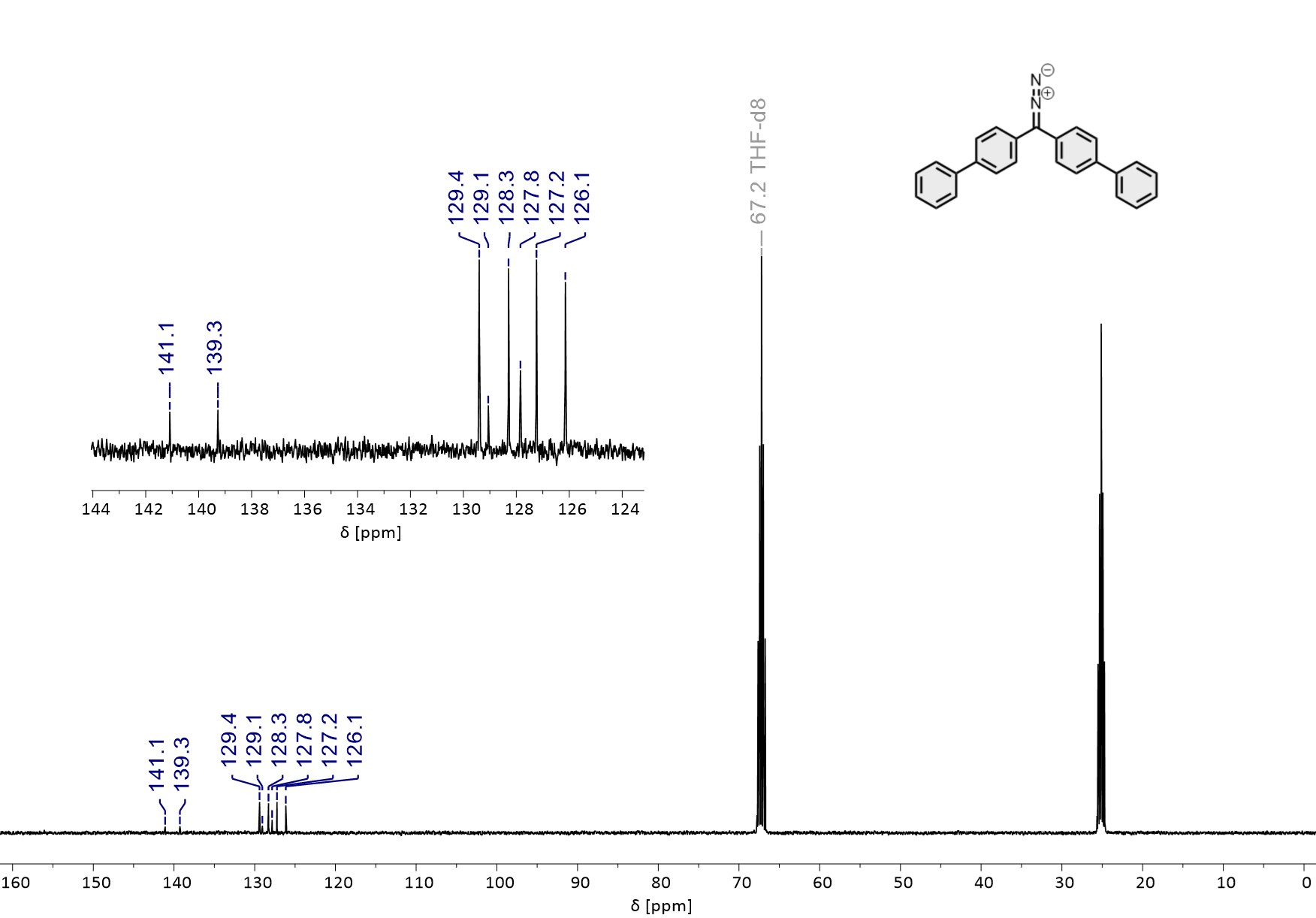}
	\caption{$^13$C NMR spectrum (100.6 MHz, THF-\textit{d}$_8$) of compound \textbf{S3} recorded at rt.}
	\label{fig:Diazo_13CNMR}
\end{figure}
\begin{figure}[htpb]
	\centering
	\includegraphics[width=0.8\textwidth]{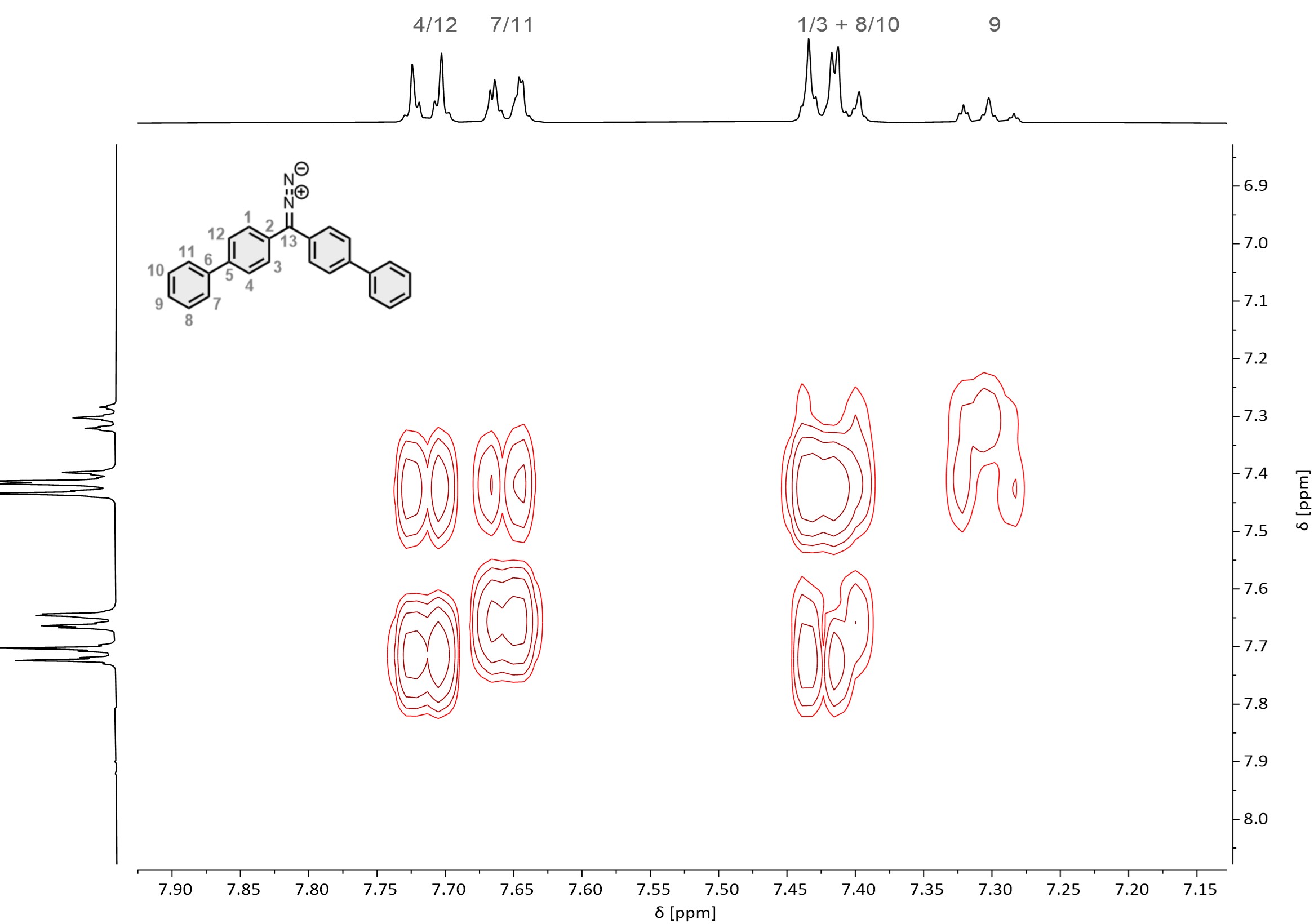}
	\caption{Two-dimensional $^1$H,$^1$H-COSY NMR spectrum (400.1 MHz, THF-\textit{d}$_8$) of \textbf{S3} at rt.}
	\label{fig:Diazo_COSY}
\end{figure}
\begin{figure}[htpb]
	\centering
	\includegraphics[width=0.8\textwidth]{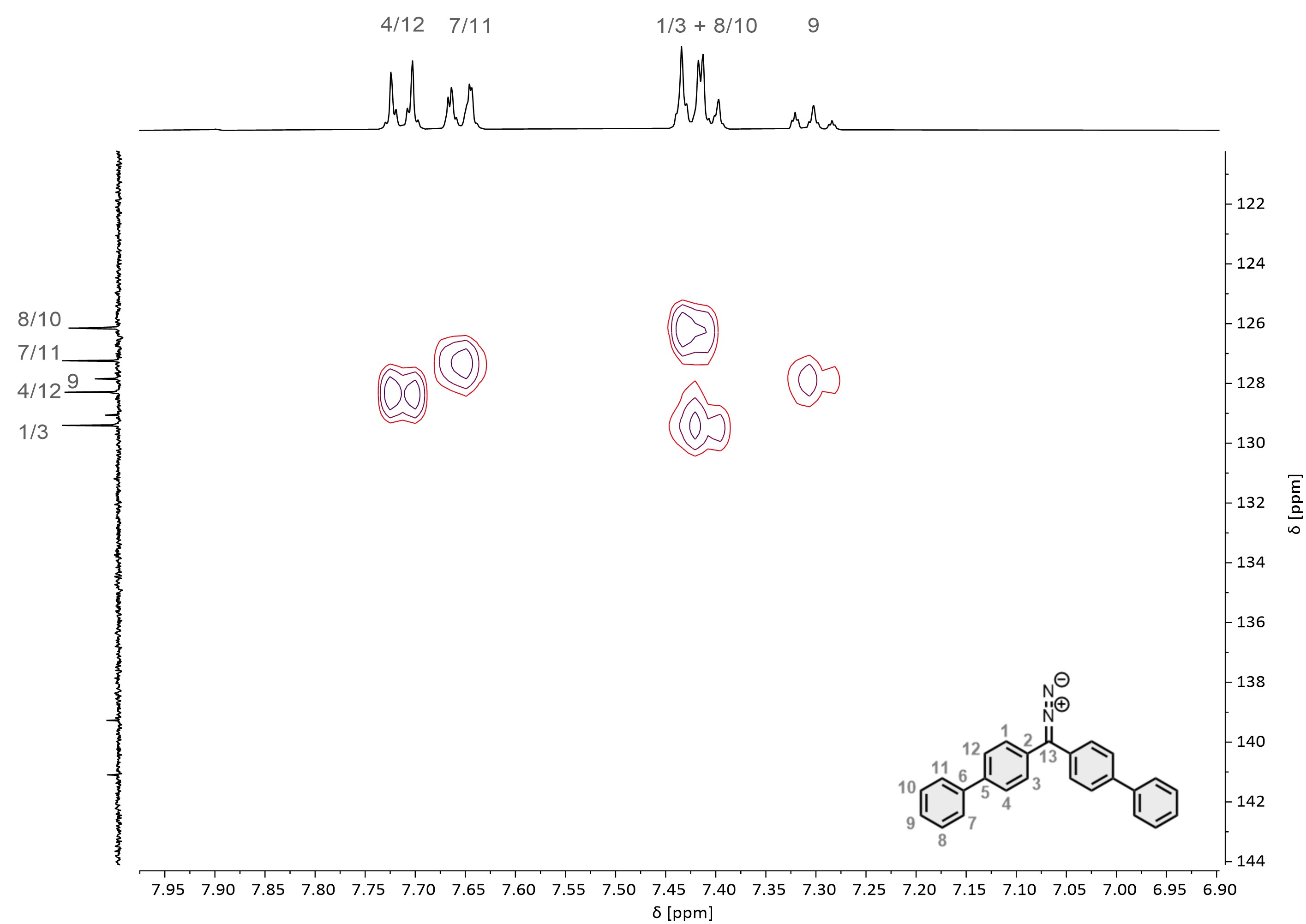}
	\caption{2D $^1$H,$^{13}$C-HSQC NMR spectrum ($^1$H: 400.1 MHz, $^{13}$C: 100.6 MHz, THF-\textit{d}$_8$) of \textbf{S3} rt.}
	\label{fig:Diazo_HSQC}
\end{figure}
\begin{figure}[htpb]
	\centering
	\includegraphics[width=0.8\textwidth]{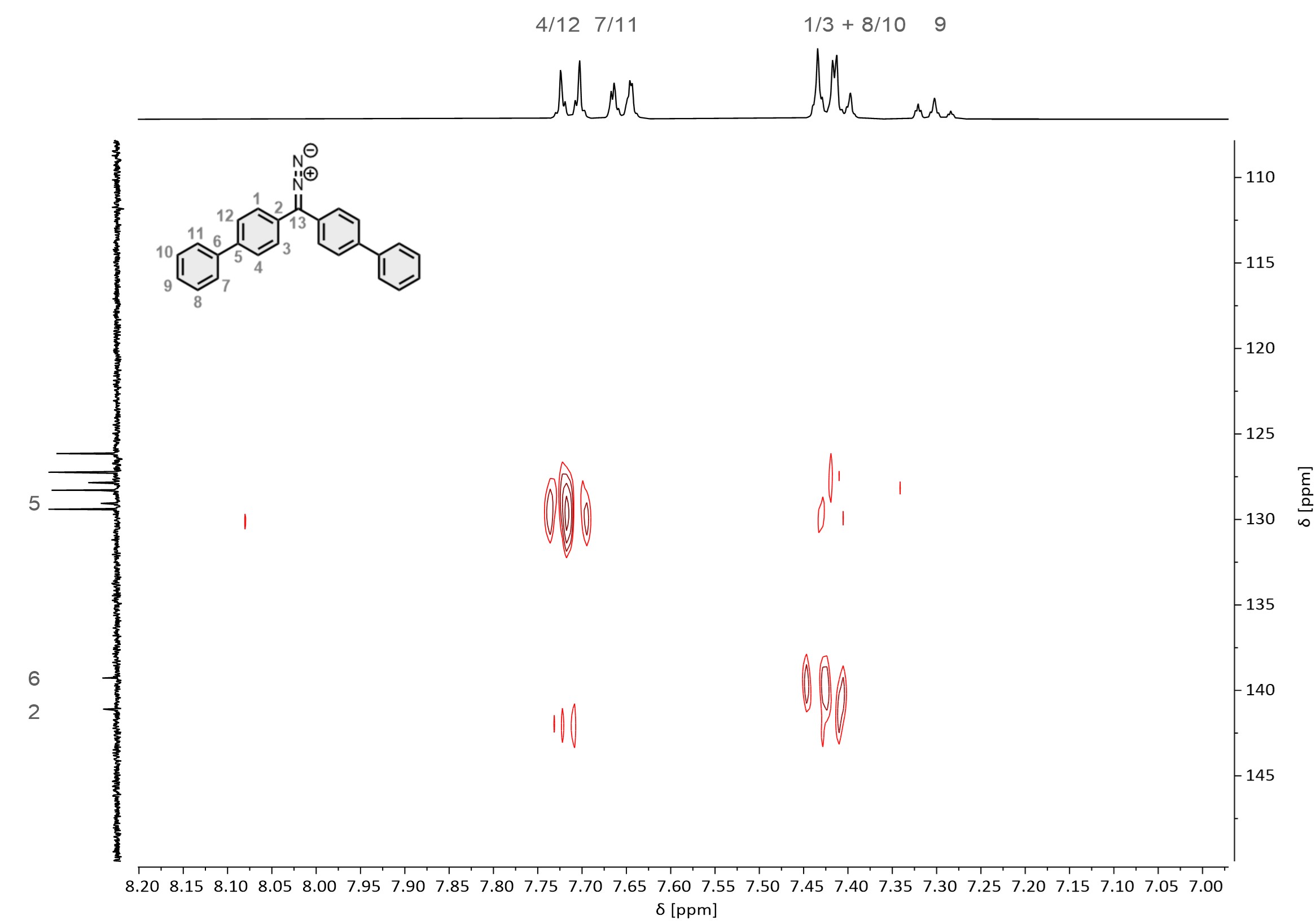}
	\caption{2D $^1$H,$^{13}$C-HMBC NMR spectrum ($^1$H: 400.1 MHz, $^{13}$C: 100.6 MHz, THF-\textit{d}$_8$) of \textbf{S3} at rt.}
	\label{fig:Diazo_HMBC}
\end{figure}

\newpage
\section{Material fabrication and sample preparation}

\subsection{Growth of doped molecular crystals}
Details on the crystal growth can be found in the \textit{Methods} section. 
The doping ratio $\xi$, defined as the molar ratio of dopant/matrix was estimated based on the single crystal structure data of the matrix (see section: X-ray crystallographic analysis). Using the cell volume of 1670 \AA$^3$, the molecular volume calculates to 417.5 \AA$^3$. The average distance \textit{d} (nm) was then estimated as $\xi=0.4175/d^3$ with $\xi$ being the doping ratio. 
\begin{table}[H]
	\centering
	\caption{Exemplary doping ratios and the resulting average distance.}
	\label{tab:doping_ratio_distance}
	\begin{tabular}{cc}
		\toprule
		Ratio $\xi$ (dopant/matrix)  &  Avg. distance (nm) \\ 
		\midrule
		0.025        &  2.6 \\
		0.0025       &  5.5 \\
		0.00025      &  12  \\
		0.000025     &  25  \\
		\bottomrule
	\end{tabular}
\end{table}

\subsection{X-ray crystallographic analysis}
\begin{table}[b!]
	\centering
	\setlength{\tabcolsep}{6pt}
	\caption{Crystallographic data for CCDC~2542406 (\textbf{S1}).}
	\begin{tabular}{@{}ll@{}}
		\toprule
		\textbf{Parameter}  &  \textbf{Value}\\
		\midrule
		CCDC number                               &  2542406\\
		Empirical formula                         &  C$_{25}$H$_{18}$O\\
		Formula weight                            &  334.42\\
		Temperature (K)                           &  150.00(10)\\
		Crystal system                            &  Orthorhombic\\
		Space group (No.)                         &  P b c n (60)\\
		$a$ (\AA)                                 &  7.3846(1)\\
		$b$ (\AA)                                 &  6.1526(1)\\
		$c$ (\AA)                                 &  36.7595(4)\\
		$\alpha$ (°)                              &  90\\
		$\beta$ (°)                               &  90\\
		$\gamma$ (°)                              &  90\\
		Volume (\AA$^{3}$)                        &  1670.15(4)\\
		$Z$                                       &  4\\
		$\rho_{\text{calc}}$ (g cm$^{-3}$)        &  1.330 \\
		$\mu$ (mm$^{-1}$)                         &  0.614\\
		$F(000)$                                  &  706\\
		Crystal size (mm$^{3}$)                   &  0.10 $\times$ 0.19 $\times$ 0.89\\
		Crystal colour                            &  clear light colourless\\
		Crystal shape                             &  needle\\
		Radiation                                 &  CuK$\alpha$ ($\lambda = 1.54184$ \AA)\\
		h,k,lmax                                  &  9, 7, 46\\
		Reflections collected                     &  29204\\
		Independent reflections                   &  1740 [$R_{\text{int}} = 0.0616$]\\
		Completeness to $\theta = 76.67^\circ$    &  99.4 \%\\
		Abs. corr. $T_\text{min}$/$T_\text{max}$  &  0.695, 1.000 \\
		Goodness-of-fit on $F^{2}$                &  1.0536\\
		Final $R$ indexes [$I\!\ge\!2\sigma(I)$]  &  $R_{1}=0.0222$, $wR_{2}=0.0535$\\
		Final $R$ indexes [all data]              &  $R_{1}=0.0247$, $wR_{2}=0.0546$\\
		Largest peak/hole (e \AA$^{-3}$)          &  0.1179 / -0.1701\\
		\bottomrule
	\end{tabular}
	\label{tab:Xray_DNK}
\end{table}
Crystallographic data for the structure reported in this paper have been deposited with the Cambridge Crystallographic Data Centre.\cite{groomCambridgeStructuralDatabase2016} CCDC 2542406 contains the supplementary crystallographic data for this paper. These data can be obtained free of charge from The Cambridge Crystallographic Data Centre via \href{https://www.ccdc.cam.ac.uk/structures}{https://www.ccdc.cam.ac.uk/structures}. This report and the CIF file were generated using FinalCif.\cite{kratzert} 

Clear, colorless single crystals suitable for crystallographic analysis were grown by slow gas-phase diffusion of \textit{n}-hexane into a solution of \textbf{S1} in THF (0.014\,M) at room temperature over several days. A rod-like prismatic crystal was prepared in perfluoroalkylether (viscosity 1800 cSt, CAS: 69991-67-9, ABCR) and mounted on a nylon loop. Data were collected on a Rigaku/Oxford Diffraction SuperNova diffractometer, utilizing Cu K($\alpha$ radiation ($\alpha$) = 1.54184 Å) from a microfocus source and an Atlas CCD detector under a stream of nitrogen at 150 K. All data were integrated with CrysAlisPro and a multi-scan absorption correction using SCALE3 ABSPACK was applied.\cite{CrysalisPro} The structure was solved by direct methods with SHELXT 2018/2 and refined by full-matrix least-squares methods against $F^{2}$ using SHELXL-2018/3.\cite{sheldrickCrystalStructureRefinement2015,sheldrickSHELXTIntegratedSpacegroup2015}
The crystal structure was refined against $F^{2}$ using the NoSpherA2\cite{kleemissAccurateCrystalStructures2021} implementation within the Olex2 software suite.\cite{dolomanovOLEX2CompleteStructure2009a} Unlike conventional independent atom models, in this approach nonspherical atomic form factors were calculated via tailored aspherical electron densities. These densities were generated using a Hirshfeld Atom Refinement (HAR) approach, with the underlying wavefunction calculated at the B3LYP/cc-pVTZ level of theory using the ORCA 5.0 interface. Hydrogen atom positions and anisotropic displacement parameters were refined freely without the use of a riding model, providing a more physically accurate representation of the nuclear positions and bonding density.

Compound \textbf{S1} crystallizes in the orthorhombic space group Pbcn (No. 60). The unit cell parameters at 150.0(1)\,K are a = 7.3846(1)\,\AA, b = 6.1526(1)\,\AA, c = 36.7595(4)\,\AA, with $\alpha$ = $\beta$ = $\gamma$ = 90° and a total cell volume of $V = 1670.15(4)\,$\AA$^3$ (Tab. \ref{tab:Xray_DNK}). The unit cell contains Z = 4 molecules, while the asymmetric unit contains half of a molecule (Z' = 0.5). This reduction in the asymmetric unit arises because the molecule sits on a special position, specifically a twofold rotation axis (Wyckoff position 4c) passing through the carbonyl group (O001 and C005), as indicated by their site symmetry order of 2.

\begin{figure}[b]
	\centering
	\includegraphics[width=0.8\columnwidth]{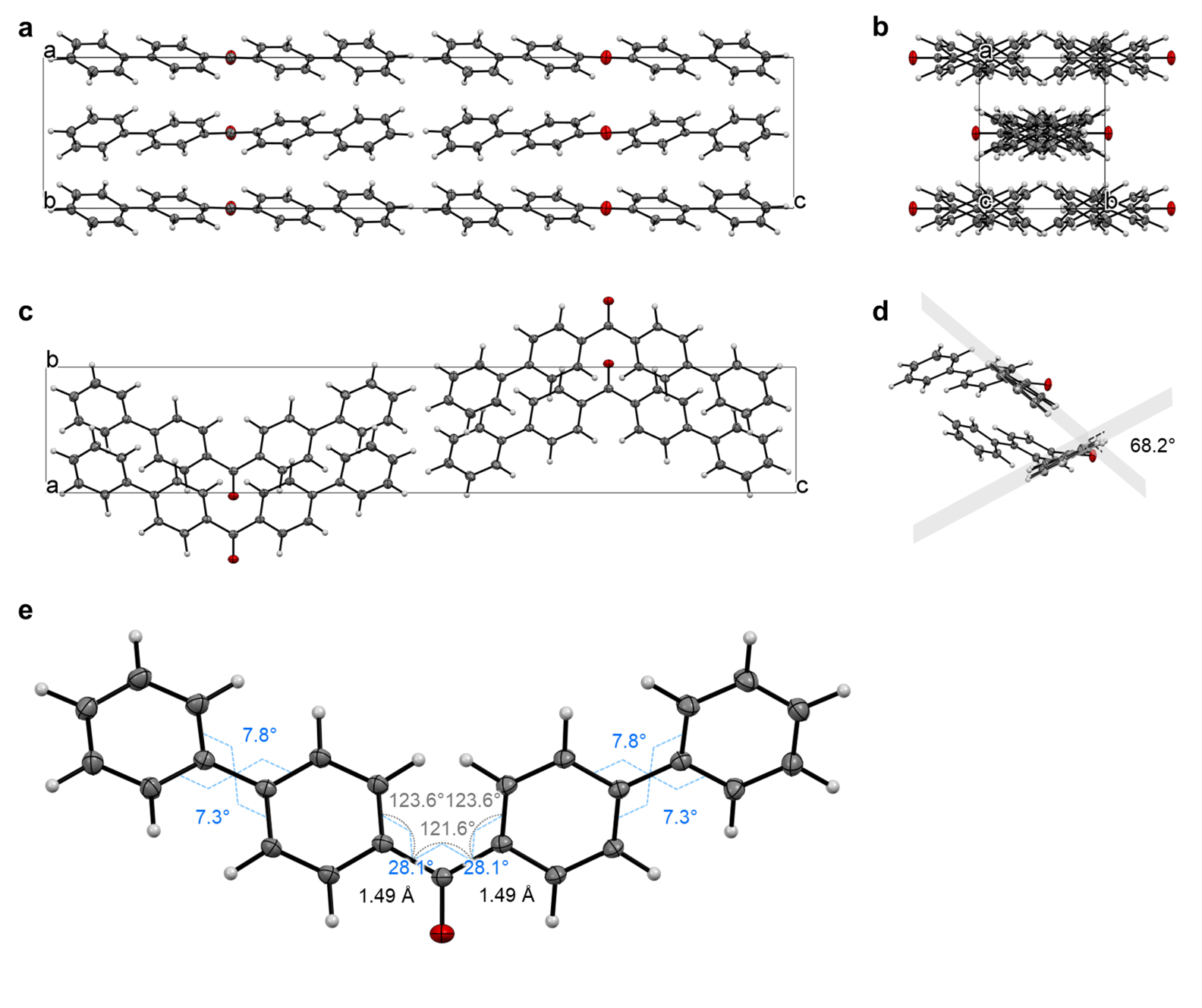}
	\caption{Unit cell of \textbf{S1} along the b- (a), c- (b), and a-axis (c) at 50\% probability level; A pair of neighboring molecules with view along the biphenyl long axis illustrating the dihedral tilt angle of herringbone stacked biphenyl units (d); Molecular structure with selected dihedral angles and the central ketone angle (e).}
	\label{Fig.XRay}
\end{figure}
The unit cell of \textbf{S1} shows a tightly packed arrangement in which no $\pi$-$\pi$ interactions between aryl units are present (Fig. \ref{Fig.XRay}a-c). Instead, we find a herringbone-type arrangement of biphenyls of adjacent molecules with a dihedral tilt angle $\phi$ of 68.2°(Fig. \ref{Fig.XRay}d) The biphenyl side arms were significantly co-planarized (dihedral angle of 7.3°/7.8°), which we ascribe to crystal packing forces - a feature well known for biphenyl (see e.g. CCDC 1111363 \cite{charbonneauBiphenylThreedimensionalData1977}, Fig. \ref{Fig.XRay}e).

\begin{figure}[H]
	\centering
	\includegraphics[width=0.7\columnwidth]{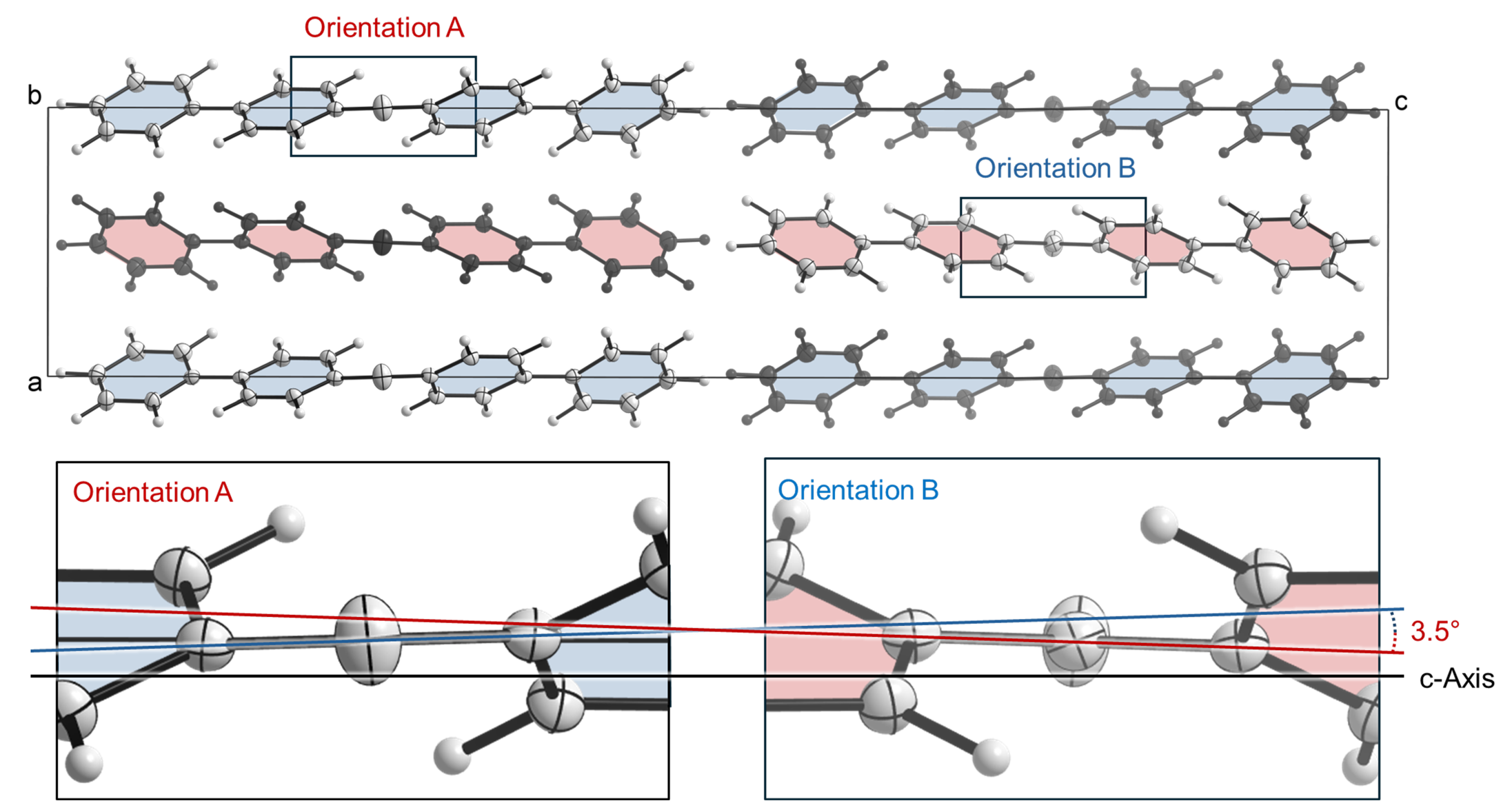}
	\caption{Illustration of the structural and orientational relationship of the four matrix molecules in the unit cell; Two pairs of enantiomers, \textit{P}- and \textit{M}-chirality with respect to the ketone aryls (dark and light gray atoms, respectively) and two unique orientations of the molecules (blue and red rings) deviating by an angle of 3.5° with respect to the C=O axis (b-axis of the unit cell, see the insets for a zoom-in).}
	\label{Fig.Enantiomers}
\end{figure}

The unit cell comprises two pairs of enantiomers (displaying \textit{P} and \textit{M} chirality, Fig. \ref{Fig.Enantiomers} denoted by dark and light gray colored atoms, respectively). The molecules C--C(=O)--C plane is slightly tilted relative to the unit cell c-axis by and angle of approximately $\pm$1.73° (blue and red colored rings). Within the crystal lattice, these represent two unique orientations, resulting in an overall difference in orientation of approximately 3.5°.

\begin{figure}[H]
	\centering
	\includegraphics[width=0.6\columnwidth]{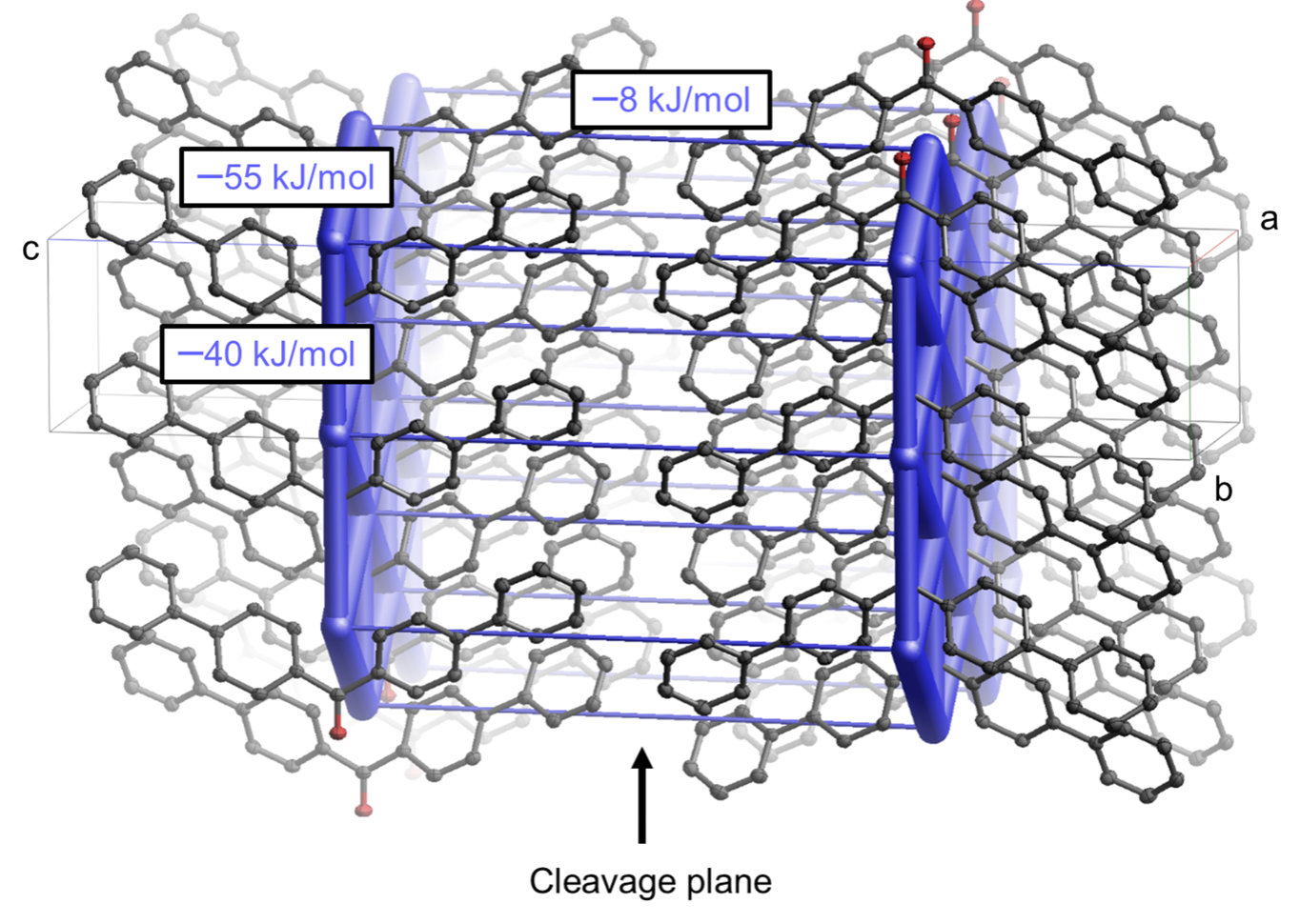}
	\caption{Depiction of the energy networks obtained via \textit{CrystalExplorer} program\cite{spackmanCrystalExplorerProgramHirshfeld2021} to illustrate the laminar arrangement of the crystal structure of \textbf{S1}.}
	\label{Fig.EnergyNetworks}
\end{figure}

We analyzed the energy networks of intermolecular interactions using the \textit{CrystalExplorer} package\cite{spackmanCrystalExplorerProgramHirshfeld2021} at the B3LYP/6-31(d,p) level of theory to illustrate the cohesion of molecules in different growth directions of the crystal (Fig. \ref{Fig.EnergyNetworks}). In agreement with our observations when mechanically manipulating crystals we find strong interactions along the crystallographic a and b directions originating from strong attractive C=O$\cdot\cdot\cdot$H–C interactions as well as multiple $\pi$$\cdot\cdot\cdot$H–C interactions, which in total lead to the edge-to-face arrangement of the biphenyl units as mentioned above. The individual layers of tightly packed matrix molecules in the \textit{ab} plane are held together by multiple weak dispersive van der Waals interaction, which denotes the cleavage plane.

\newpage
\section{Optical spectroscopy of bulk materials and frozen solutions}

\subsection{Steady-state spectroscopy}
\label{sec:Steady}
Fluorescence spectroscopy was performed using an inhouse-built setup with a LP450-SF25 laserdiode (Thorlabs, 452\,nm) and a QEPro spectrometer (OceanOptics) with a 475 nm long pass filter to remove the excitation light from the emission. For that, the corresponding diaryldiazomethane precursor \textbf{S3} was dissolved in the respective solvent (optical density OD $< 0.1$) in a 5\,mm NMR tube, containing ca. 0.2\,mL solution and was rapidly frozen by plunging into liquid nitrogen (77\,K).
After temperature equilibration (ca. 1\,min), photoactivation was carried out using the laserdiode while recording the fluorescence signal.

\begin{figure}[htpb]
	\centering
	\includegraphics[scale=0.6]{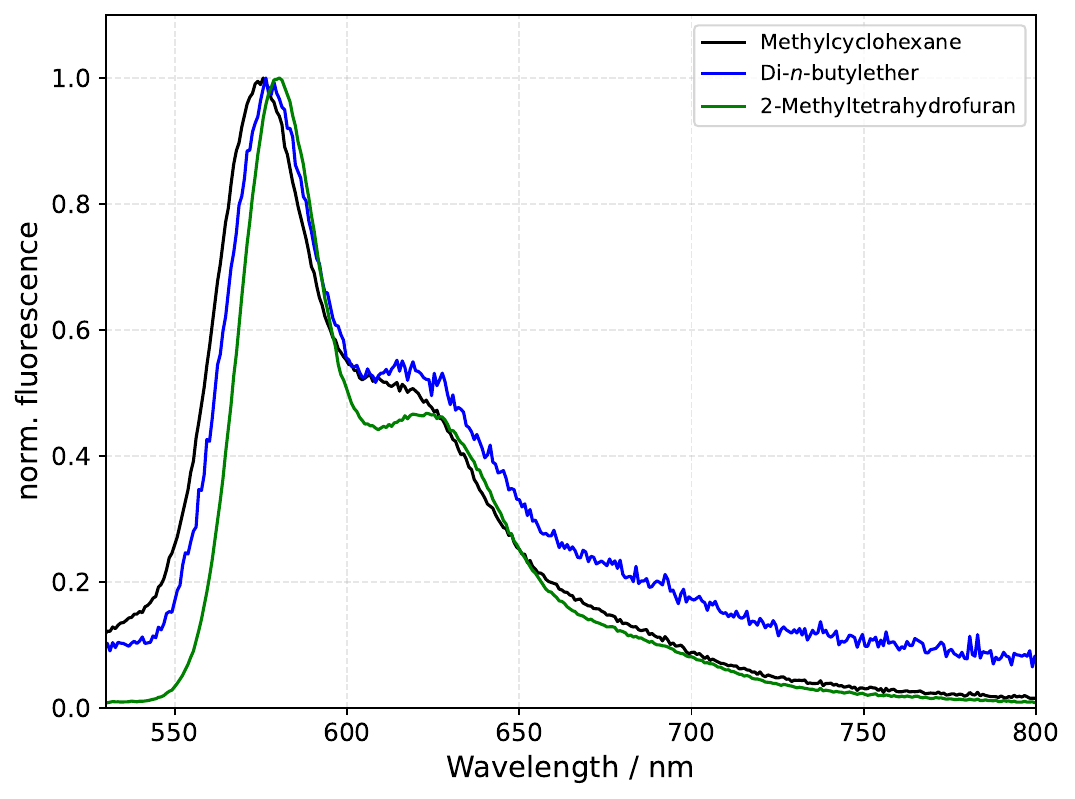}
	\caption{Fluorescence spectra of the carbene in frozen methylcyclohexane, di-\textit{n}-butylether and 2-methyltetrahydrofuran at 77\,K demonstrating emission from a locally excited state.}
	\label{fig:fluorescence_spectra_solvents}
\end{figure}

To investigate solvatochromic effects, three solvents with varying permittivities were selected (methylcyclohexane, di-\textit{n}-butylether, and 2-methyltetrahydrofuran). Experimental results revealed only a negligible bathochromic shift across the series (575--580~nm, ca. 0.02~eV; see Fig.~\ref{fig:fluorescence_spectra_solvents}). Qualitatively, the use of a frozen solvent medium constrains the ability of the solvent shell to adapt to charge transfer in the excited state, thereby reducing the solvatochromic effect. For the sake of comparison, we refer to the values reported by Nowacki and coworkers,\cite{karpiukIntramolecularElectronTransfer2017} who studied the electron-transfer characteristics of a series of donor-acceptor type chromophores in similar frozen solvents (including methylcyclohexane and 2-methyltetrahydrofuran) and found charge-transfer band shifts of approximately 0.5~eV. This finding allows us to rule out an excited charge-transfer state, which is consistent with our theoretical findings.

\subsection{Photoluminescence quantum yields}
Absolute fluorescence quantum yield $\phi$ were determined in triplicate using a Quantaurus-QY (C11347, Hamamatsu) equipped with a cryo module after excitation at 460 nm. For that, the corresponding diaryldiazomethane precursor \textbf{S3} was dissolved in 2-methyltetrahydrofuran (optical density OD $< 0.1$), the tube containing ca. 0.2\,mL solution rapidly frozen by plunging into liquid nitrogen (77\,K). After temperature equilibration (ca. 1\,min), quantitative photoactivation was carried out using a UV LED curing lamp (Jupitertech, (12.0\,W/cm$^{2}$, distance ca. 3 cm) with a central wavelength of 395 nm within a 1 min. The carbene shows a photoluminescent quantum yield of 0.09(1).

\subsection{Time-correlated single photon counting}
\label{sec:TCSPC}
\begin{figure}[H]
	\centering
	\includegraphics[scale=0.65]{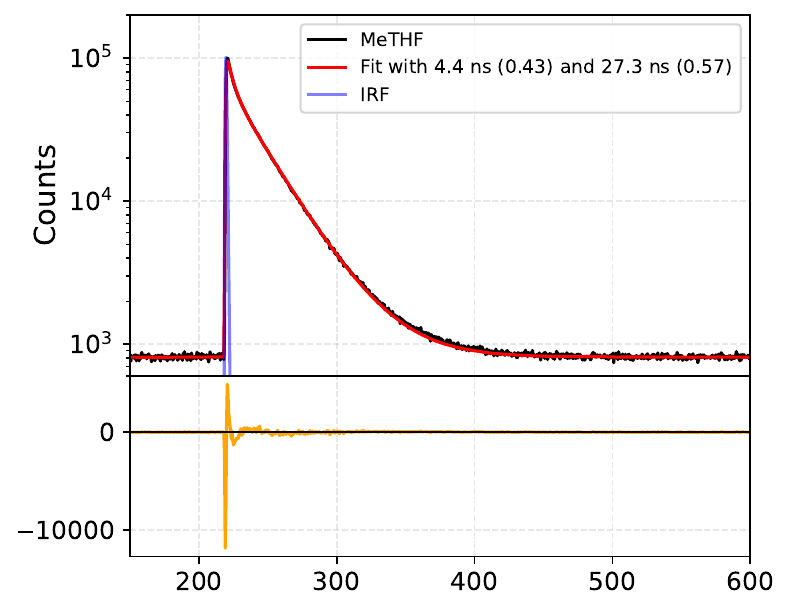}
	\caption{TCSPC measurement with reconvolution fit of activated carbene in 2-methyltetrahydrofuran at 77 K. The orange graph denotes the residuals of the fit.}
	\label{fig:tcspc_results_methf}
\end{figure}

\begin{figure}[H]
	\centering
	\includegraphics[width=0.9\textwidth]{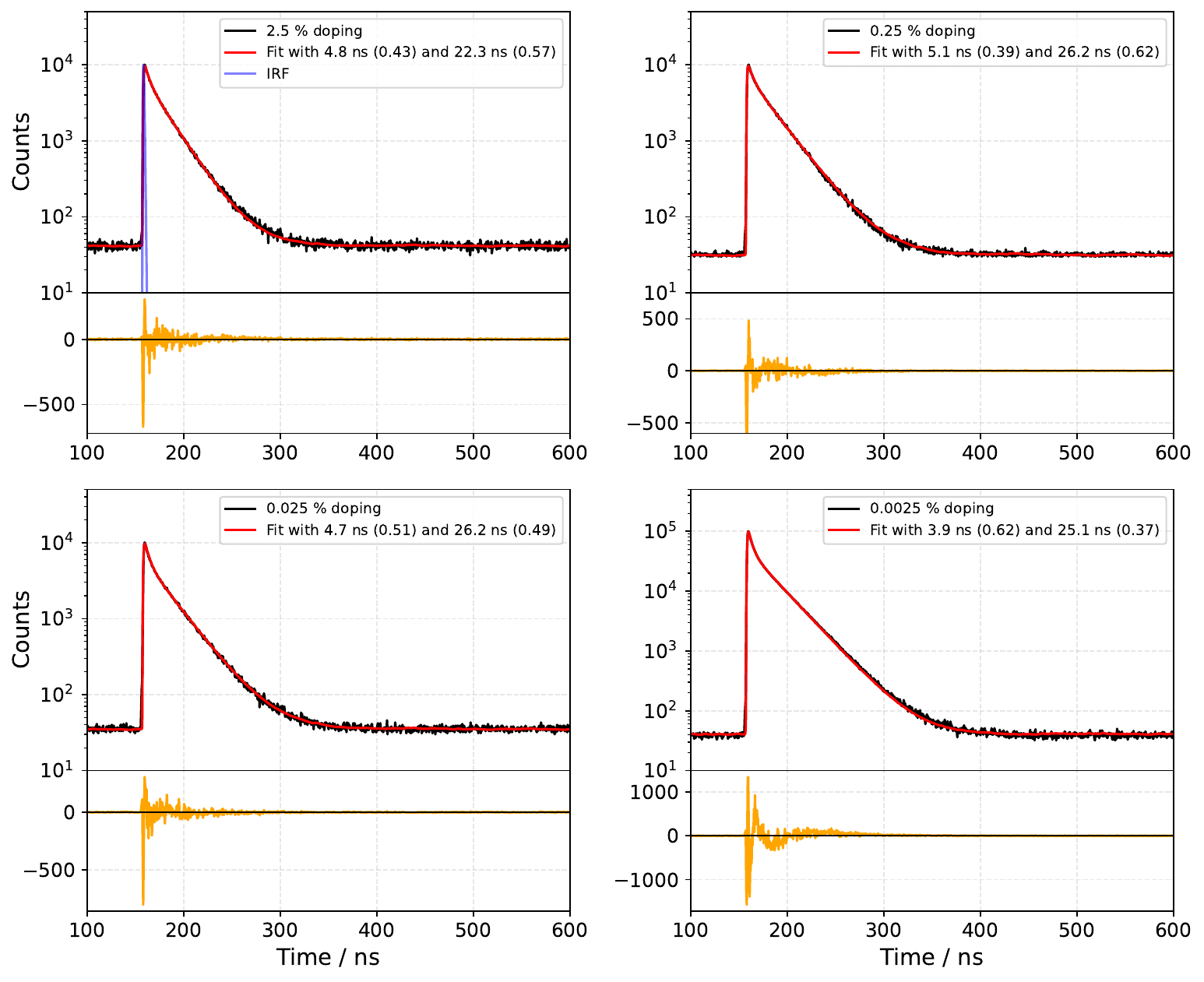}
	\caption{TCSPC measurements with reconvolution fits of activated doped crystals with different doping rates at 77 K. The orange graph denotes the residuals of the fit.}
	\label{fig:tcspc_results_crystal}
\end{figure}

Time correlated single photon counting (TCSPC) experiments were performed using an inhouse-built setup with a pulsed DLnsec laser (LABS electronics) with 520\,nm or 450\,nm wavelength excitation wavelength.
For filtering the excitation light from the emission, either a 550\,nm or a 475\,nm long pass filter was used in front of an avalanche photodiode SPCM-AQRH (Excelitas).
The pulse trigger and photon time tagging was done by using an OPX1000 (Quantum Machines).
The samples were placed in a 5\,mm NMR tube, cooled down to 77\,K and then photoactivated using a UV LED curing lamp (Jupitertech, (12.0\,W/cm$^{2}$, distance ca. 3 cm) with a central wavelength of 395\,nm within 30 seconds.
Measurements were performed on photoactivated \bifi{} in frozen 2-methyltetrahydrofuran, as well as on carbenes generated within \bike{} host crystals by photoactivating diazomethane precursors at varying dopant densities.
The 2-methyltetrahydrofuran sample was degassed by gently bubbling argon through the solution for 10 min prior freezing.
The results from the 2-methyltetrahydrofuran measurements show an overlap of two exponential decays with timescale 4.4\,ns and 27.3\,ns, and respective relative amplitudes of 0.43 and 0.57 (see Fig.~\ref{fig:tcspc_results_methf}). 
The fluorescence lifetime measurements on crystal samples is shown in Fig.~\ref{fig:tcspc_results_crystal}, and the fit results in Tab.~\ref{tab:lifetimes}.

The TCSPC data of the doped molecular crystals reveals a bi-exponential fluorescence decay. The long-lifetime component ($\tau_2 \approx 22.3 - 26.2$~ns) observed in the ensemble measurements aligns well with the excited-state lifetime of the $T_{1y}$ state measured at the single-molecule level via cryo-confocal microscopy ($24 \pm 2$~ns, see Sec.~\ref{sec:cryoconfocal_lifetime_measurements}).
Based on this, we tentatively assign the short-lifetime component ($\tau_1 \approx 3.9 - 5.1$~ns) to the $T_{1z}$ state with a dominant decay via ISC.
However, this assignment could not be definitively verified through single-molecule fluorescence lifetime measurements since spin-selective optical excitation for this state ($T_{0z} \rightarrow T_{1z}$) could not be observed for single molecules.

In systems where short-range intermolecular interactions are prevalent, elevated doping concentrations typically induce a pronounced shortening of the excited-state lifetime \cite{kimConcentrationQuenchingBehavior2017, liuPureRedDoubletEmission2020}. In contrast, the fitted lifetimes ($\tau_1$ and $\tau_2$) in our system remain remarkably constant across a broad range of dopant concentrations, spanning from 0.0025\% to 0.25\%. Only at the highest doping level investigated we observe a marginal decrease in lifetime, which likely signifies the onset of concentration quenching driven by reduced intermolecular distances.
This constant lifetimes at lower doping rates strongly indicate that the carbene molecules are well-isolated within the ketone host matrix, and the observed emission kinetics are driven purely by intrinsic monomolecular photophysics rather than intermolecular crosstalk.

Interestingly, while the lifetimes remain largely constant, the pre-exponential amplitudes exhibit a concentration dependence.
As the doping ratio decreases to 0.0025\%, the relative amplitude of the short-lived component ($A_1$) increases from 0.43 to 0.62, while the long-lived component ($A_2$) correspondingly decreases from 0.57 to 0.37 (see Tab.~\ref{tab:lifetimes}).
While we cannot explain this trend conclusively at this point, it may hint at a concentration-dependent inner filter effect (radiation trapping) in which the short timescale component is suppressed by repeated absorption/re-emission processes eventually funneling more photons into the more emissive long lived component at high doping rates.

\begin{table}[H]
	\centering
	\caption{ Fluorescence lifetimes ($\tau$) and relative amplitudes ($A$) for different doping concentrations from reconvolution fits of TCSPC data at 77 K.}
	\label{tab:lifetimes}
	\begin{tabular}{lcccc}
		\toprule
		\textbf{Concentration}  &  \boldmath{$\tau_1$} \textbf{(ns)}  &  \boldmath{$A_1$}  &  \boldmath{$\tau_2$} \textbf{(ns)}  &  \boldmath{$A_2$} \\ 
		\midrule
		2.5 \%                   &  4.8  &  0.43  &  22.3  &  0.57 \\
		0.25 \%                  &  5.1  &  0.39  &  26.2  &  0.62 \\
		0.025 \%                 &  4.7  &  0.51  &  26.2  &  0.49  \\
		0.0025 \%                &  3.9  &  0.62  &  25.1  &  0.37 \\ 
		\bottomrule
	\end{tabular}
\end{table}

\newpage
\section{Computational chemistry}
\subsection{Computational methods}

\begin{figure}[H]
	\centering
	\includegraphics[width=0.8\textwidth]{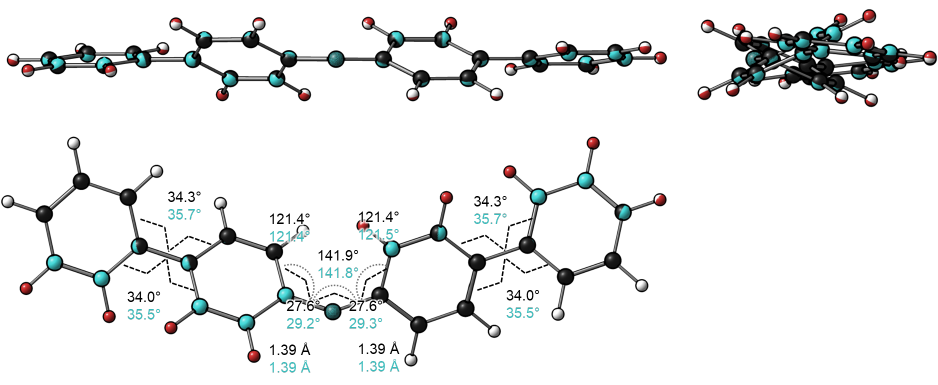}
	\caption{Merged geometries of DFT-optimized structures of \bifi{} for the triplet ground state configuration at the B3LYP/def2-TZVPP (black carbons, white hydrogens) and B97-3c/def2-mTZVP (turquoise carbons, red hydrogens) level of theory calculated in vacuum; root-mean-square-deviation of the atomic positions is 0.0321 \AA.}
	\label{fig:B97-3c_Benchmark}
\end{figure}

To validate the suitability of the B97-3c functional for geometry prediction within the ONIOM framework, we first optimized the structure of \bifi{} in vacuum in the triplet ground state and compared it to results obtained at the B3LYP/def2-TZVPP level. The latter has been shown, both in our experience and by others,\cite{shiraziPerformanceDensityFunctional2020a} to provide reliable geometries (Fig. \ref{fig:B97-3c_Benchmark}). The structures are nearly identical, with negligible deviations in overall shape or key geometric parameters, including bond lengths, dihedral angles, and the carbene angle.

\begin{figure}[H]
	\centering
	\includegraphics[width=0.93\textwidth]{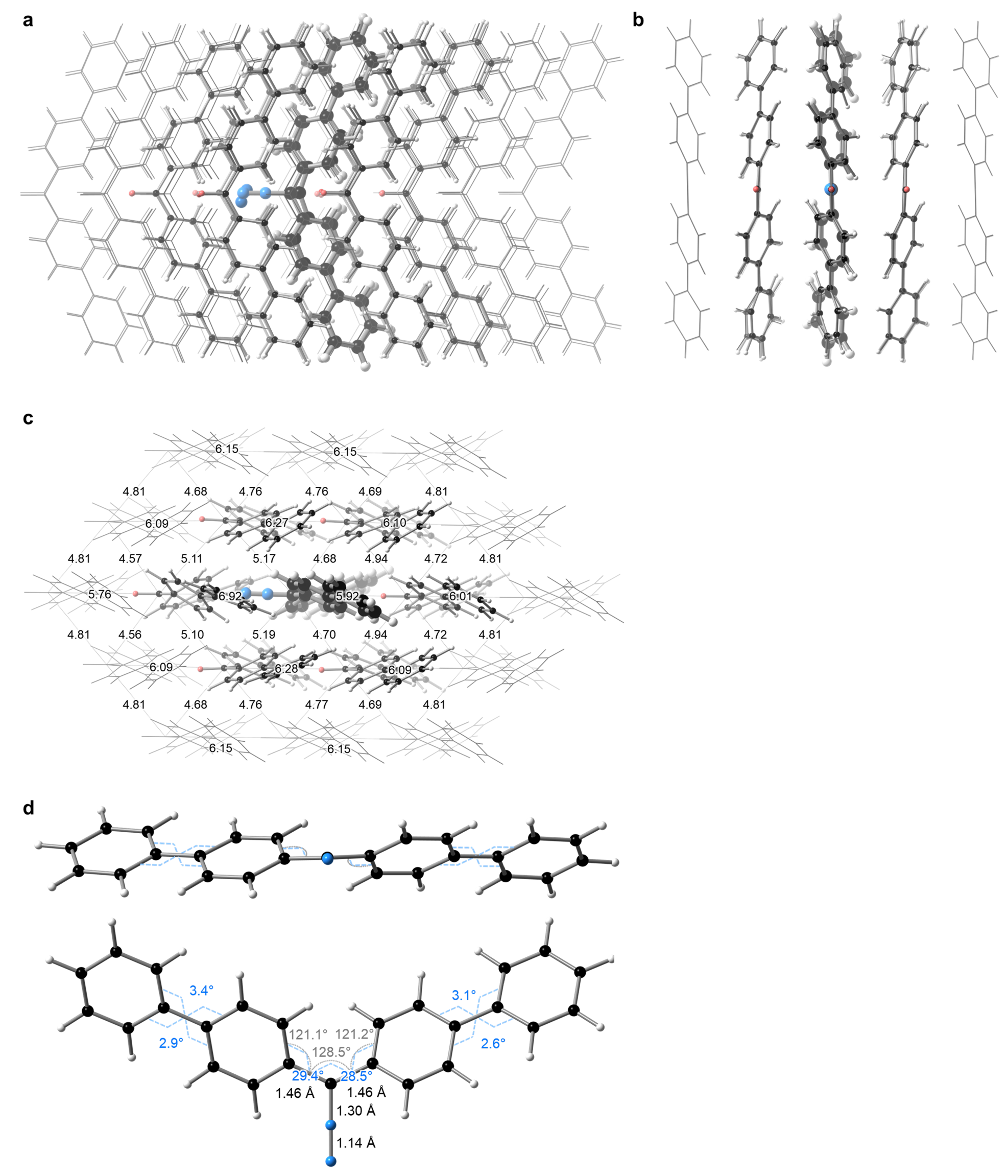}
	\caption{Illustration of the multiscale model for the carbene precursor consisting of 837 atoms; The high-level QM region including a diazomethane derivative (bold) and the surrounding shell of six matrix molecules treated at the B97-3c level of theory, and a outer shell of matrix molecules with fixed coordinates (12 molecules, wireframe) treated using semiempirical extended tight-binding method, GFN2-xTB (denoted as XTB2 \cite{Bannwarth_GFN2_2019}; Top (a), front (b), and side view with  distances (in Å) of the ketone and diazomethane carbons to illustrate the structural rearrangement in the first molecular shell around the carbene precursor (c); Molecular structure of diazomethane derivative isolated from the multiscale calculations together with relevant structural data).}
	\label{fig:Model_Diazo}
\end{figure}

Following the experimental sequence of doping and subsequent photoactivation to generate the activated carbene, the ONIOM calculations were analogously performed in two successive steps. First, the geometry of the closed-shell singlet diazo precursor was optimized  (Fig.~\ref{fig:Model_Diazo}).
In light of recent \textit{in-situ} X-ray diffraction studies on halogenated diarylcarbenes\cite{kawanoStructureDeterminationTriplet2007a} we then proceeded on the assumption that the nitrogen is trapped in the crystal lattice in close vicinity to the carbene center after photoactivation.
Hence, the relaxed structure of the diazomethane precursor in the matrix was used to model the system after photoactivation, \textit{i.e.} the resulting carbene in the triplet ground state and the closeby nitrogen molecule embedded in the matrix (T$_0$, Fig.~\ref{fig:Model_T0}a-c). For that, the nitrogen was cleaved off manually by mimicking the out-of plane distorted conical intersection for the photochemical nitrogen dissociation upon excitation to \textit{S}$_2$ as reported recently, and geometry optimization was performed~\cite{pitesaPhotoeliminationNitrogenDiazoalkanes2020c}.

The resulting relaxed cluster was used to extract the coordinates of the final qubit in the triplet ground state (see Fig.~\ref{fig:Model_T0}d). We note that this specific model setup limits geometric relaxation strictly to the qubit molecule and the first matrix shell and does not account for longe-range rearrangements within the matrix.

\newpage
\subsection{Precursor ground state geometry}

\subsection{Triplet ground state geometry ($T_0$)}
\begin{figure}[H]
	\centering
	\includegraphics[width=0.93\textwidth]{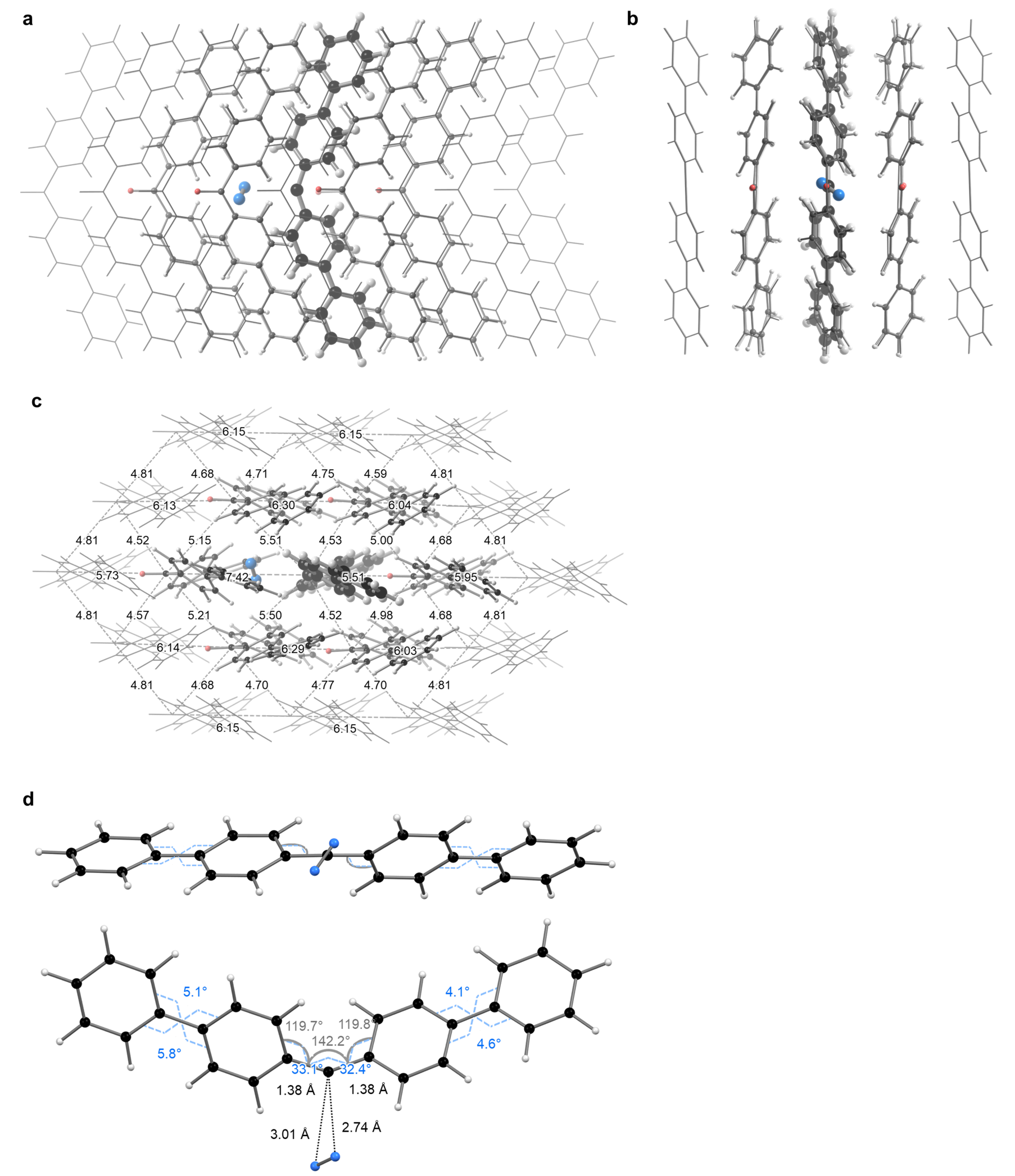}
	\caption{Illustration of the multiscale model of the qubit in $\Tg$ geometry; The high-level QM region (carbene and N$_{2}$, bold) and surrounding shell of six matrix molecules (ball-and-stick) treated at the B97-3c level of theory and a outer shell of 12 matrix molecules with fixed coordinates (wireframe) computed using semiempirical extended tight-binding method, GFN2-xTB (denoted as XTB2 \cite{Bannwarth_GFN2_2019}; Top (a), front (b), and side view with distances (in Å) of the central ketone and carbene carbons to illustrate the structural rearrangement in the first molecular shell around the qubit (c); Molecular structure of the carbene + N$_{2}$ in $\Tg$ geometry isolated from the multiscale calculations together with relevant structural data).}
	\label{fig:Model_T0}
\end{figure}

\newpage
\subsection{Excited state geometry in vacuuum ($T_1$)}
\begin{figure}[h]
	\centering
	\includegraphics[width=0.7\textwidth]{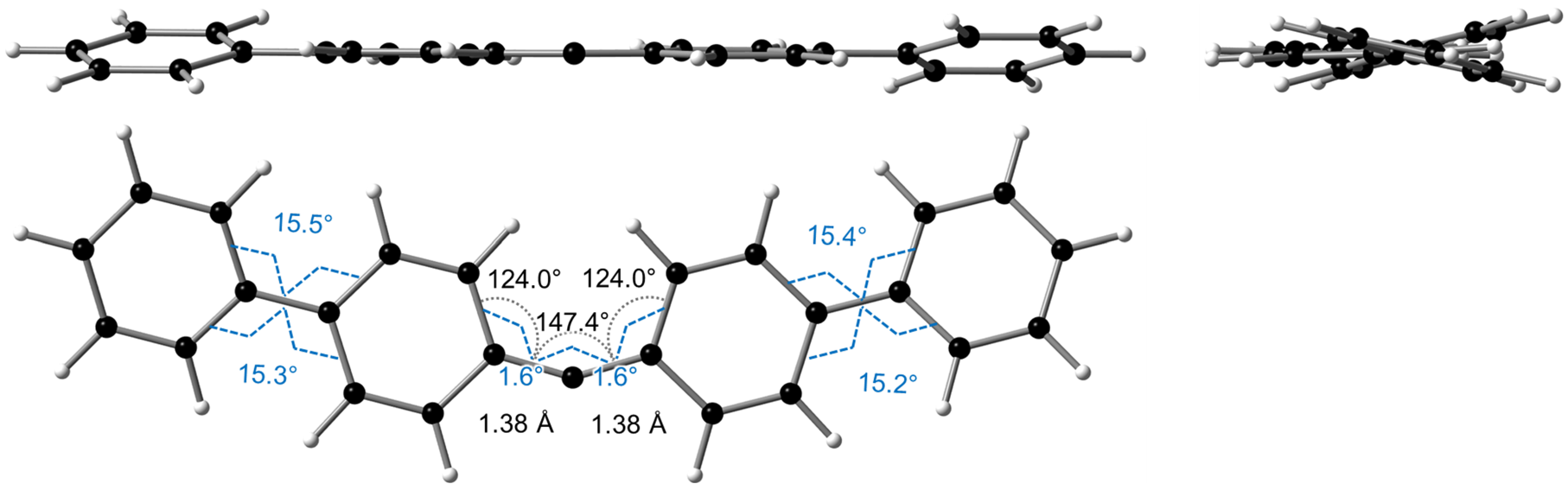}
	\caption{Optimized geometry for the first excited triplet state in vacuum obtained via TD-DFT calculations at the B3LYP/def2-TZVPP level of theory (unrestricted Kohns-Sham calculations).}
	\label{fig:T1_Vac}
\end{figure}

When optimized in vacuum, the carbene becomes considerably more coplanar with respect to both the central carbene and biphenyl dihedral angles (see Fig.~\ref{fig:T1_Vac}). Interestingly, the specific arrangement of our molecular matrix in the crystalline state pre-orients the biphenyl groups into a conformation closely resembling the excited-state geometry (see the dihedral angles in Fig.~\ref{fig:Model_T0}d). Inlight of the pronounced ZPL observed in the cryo confocal measurements, we hypothesize that the intrinsic tendency of the phenyl groups at the central carbene unit to also co-planarize is effectively counteracted by the rigid environment of the crystal lattice.

\subsection{Property prediction via TD-DFT}
\begin{figure}[b]
	\centering
	\includegraphics[width=0.75\textwidth]{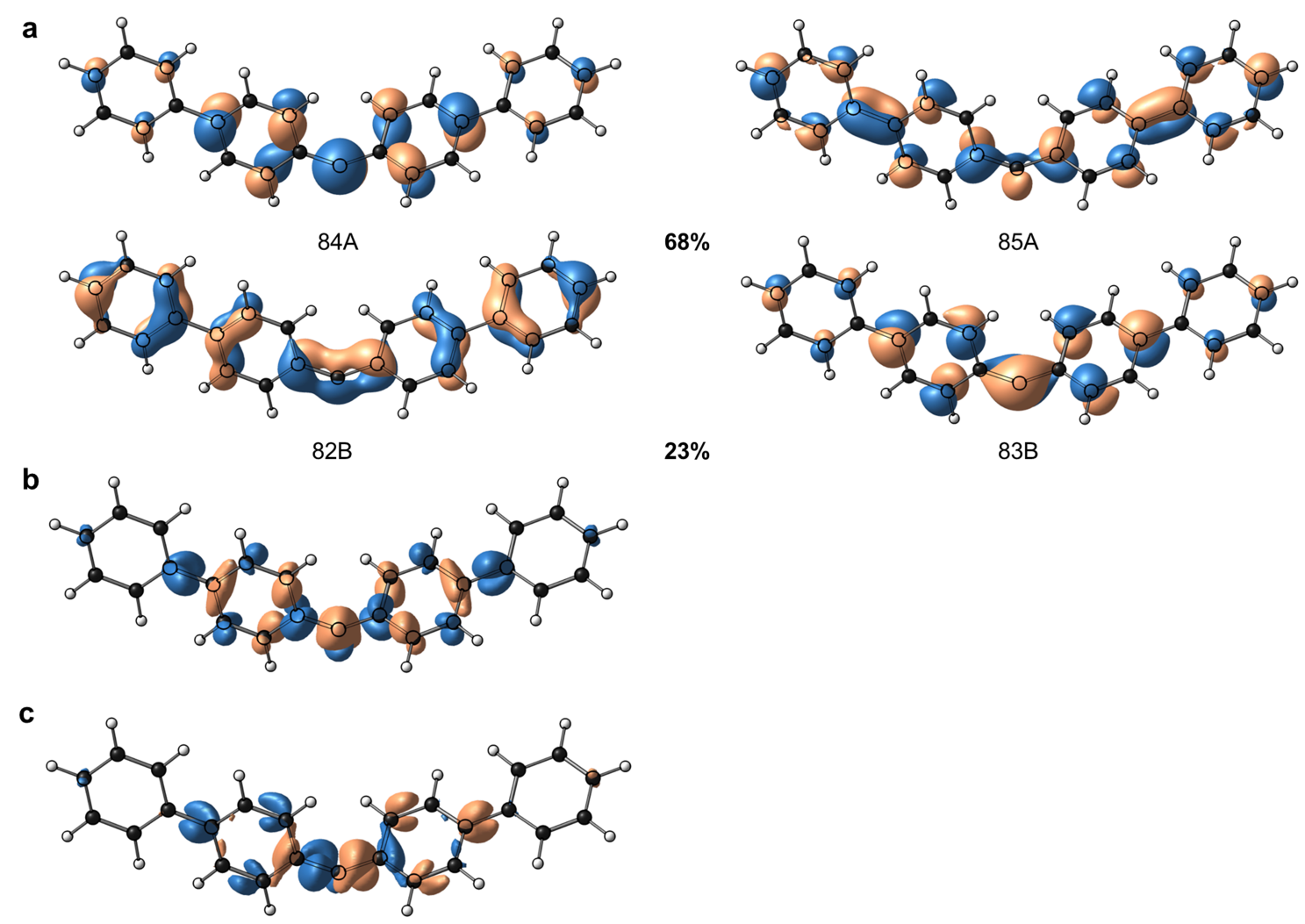}
	\caption{TD-DFT calculations at the B3LYP/def2-TZVPP level of theory (unrestricted Kohns-Sham calculations); Representation of natural transition orbitals (NTO) of the “particle” (right) and “hole” pairs (left) that is dominated by SOMO$\rightarrow$LUMO character (A-manifold, 68\%) and to a smaller extent by HOMO$\rightarrow$SOMO character (B-manifold, 23\%, a); Difference density with depletion (red) and accumulation (blue) zones showing a locally excited state (B3LYP/def2-TZVPP, isosurface=0.0015, b) and transition density plot showing regions that contribute to the transition dipole moment (TDM, isosurface=0.0015, b, compare to Fig. 1a in the manuscript).}
	\label{fig:TD-DFT}
\end{figure}
The TD-DFT calculation (B3LYP/def2-TZVPP, D4) of the carbene in its triplet ground state (pseudo-\textit{C}$_2$ geometry from ONIOM calculations) predicts the lowest energy excitation ($T_0 \rightarrow T_1$ transition) at 2.55~eV (487~nm) with an oscillator strength of $f = 0.084$. This lowest energy transition is multiconfigurational in character and is dominated by a SOMO$\rightarrow$LUMO (A-manifold, 68\%) and, to a smaller extent, by a HOMO$\rightarrow$SOMO (B-manifold, 23\%) transition. The difference density exhibits electron depletion and accumulation zone in which the contributions from the outer phenyl rings in the A- and B-manifold are mutually cancelled out and the remaining density is predominantly localized on the central carbene carbon and the next-neighbor phenyl rings. This spatial confinement indicates that the excitation can be characterized as a locally excited (LE) state within the central core of the molecule, with no relevant charge-transfer contribution, which is in line with the spectroscopic observation of a sharp and pronounced ZPL and negligible solvatochromic shift of the emission band. Furthermore, the transition density plot, which represents the mathematical product of the ground state and excited state wavefunctions also shows that the overlap region predominantly contributing to the transition dipole moment is mainly localized of the central carbene and the adjacent phenyls.

\subsection{Property prediction via CASSCF/NEVPT2}

\subsubsection{Active space deduction}

\begin{figure}[H]
	\centering
	\includegraphics[width=0.8\textwidth]{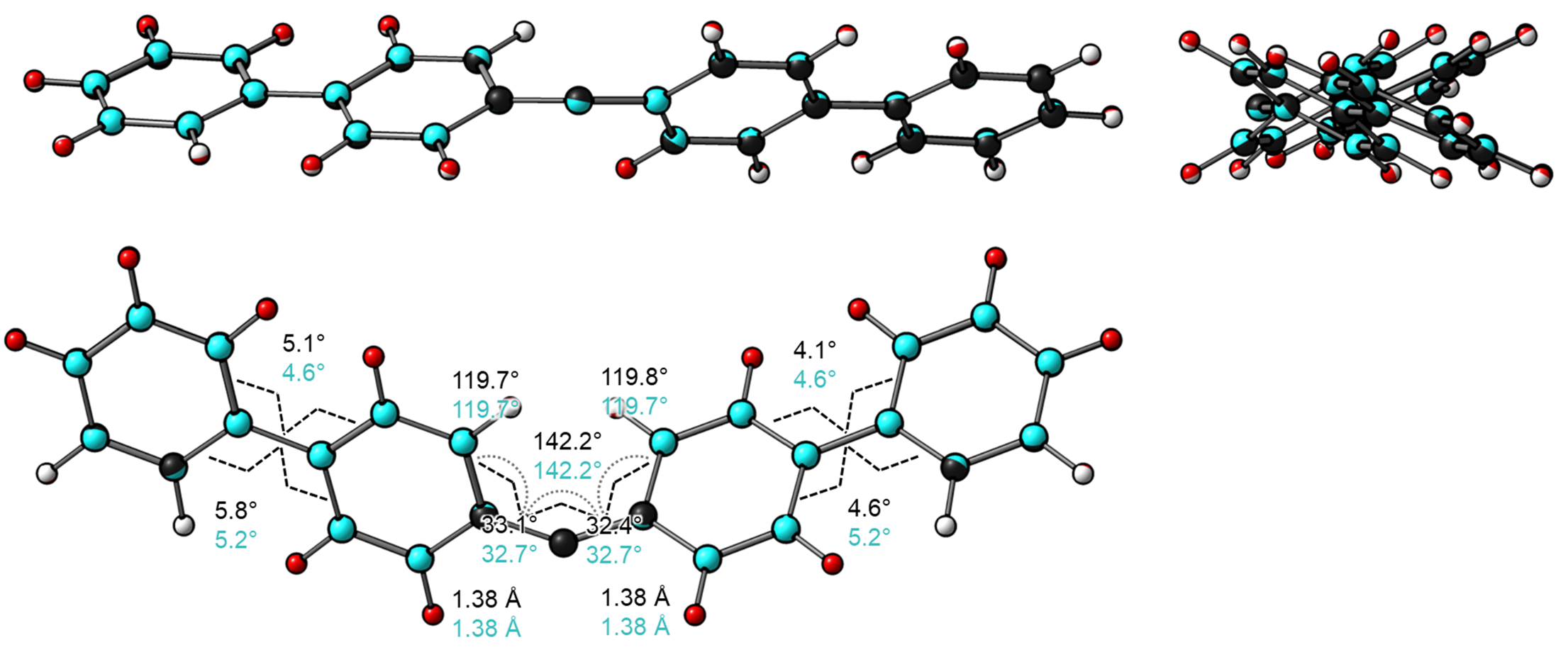}
	\caption{Merged geometries of pseudo-\textit{C}$_2$ symmetric \bifi{} isolated from ONIOM calcualtions (black carbons, white hydrogens) and the perfect \textit{C}$_2$ symmetric geometry (turquoise carbons, red hydrogens; root-mean-square deviation (RMSD) of 0.009~$\AA$).}
	\label{fig:Sym}
\end{figure}

The ONIOM-derived structure of \bifi{} exhibits a slight structural deviation from ideal \textit{C}$_2$-symmetry compared to the fully \textit{C}$_2$-symmetrized geometry, with a root-mean-square deviation (RMSD) of 0.009~\AA~(Fig. \ref{fig:Sym}). Noteworthy in this regard is the slightly more co-planarized biphenyl on the right hand side (avg. difference to the other biphenyl ca 1.1°). Further we note that the final geometry obtained in the ONIOM calculations is largely independent of the starting orientation of the cleaved-off nitrogen, yielding virtually identical structures if e.g. the nitrogen was cleaved off in an in-plane fashion. This leads us to the conclusion that the observed pseudo-\textit{C}$_2$ symmetry is not an artefact from ONIOM calcualtions but results from the asymmetric relaxation of the expelled nitrogen within the crystal lattice (c.f. Fig. \ref{fig:Model_T0}).

In light of this small geometric distortion, we note that multireference wavefunction methods are prone to artifactual symmetry breaking in such nearly symmetric frameworks~\cite{roosCompleteActiveSpace1987}. This phenomenon occurs when the active space artificially mixes orbitals of different physical symmetries to lower the absolute energy, causing the wavefunction to collapse into unphysical, localized minimum-energy solutions which can reduce the physical meaning~\cite{shuReducingPropensityUnphysical2013, marieExcitedStatesSymmetry2023}.
To circumvent these artifacts without modification of any nuclear coordinates, we employed the restricted-step second-order trust-region converger (TRAH) implemented in \textit{ORCA 6.1.0}~\cite{helmich-parisTrustregionAugmentedHessian2022}.
The robust orbital optimization provided by the TRAH algorithm effectively prevents root-flipping, ensuring that the resulting wavefunction largely maintains approximate \textit{C}2-symmetry and yields physically meaningful electronic states.

\begin{figure}[t]
	\centering
	\includegraphics[width=0.75\textwidth]{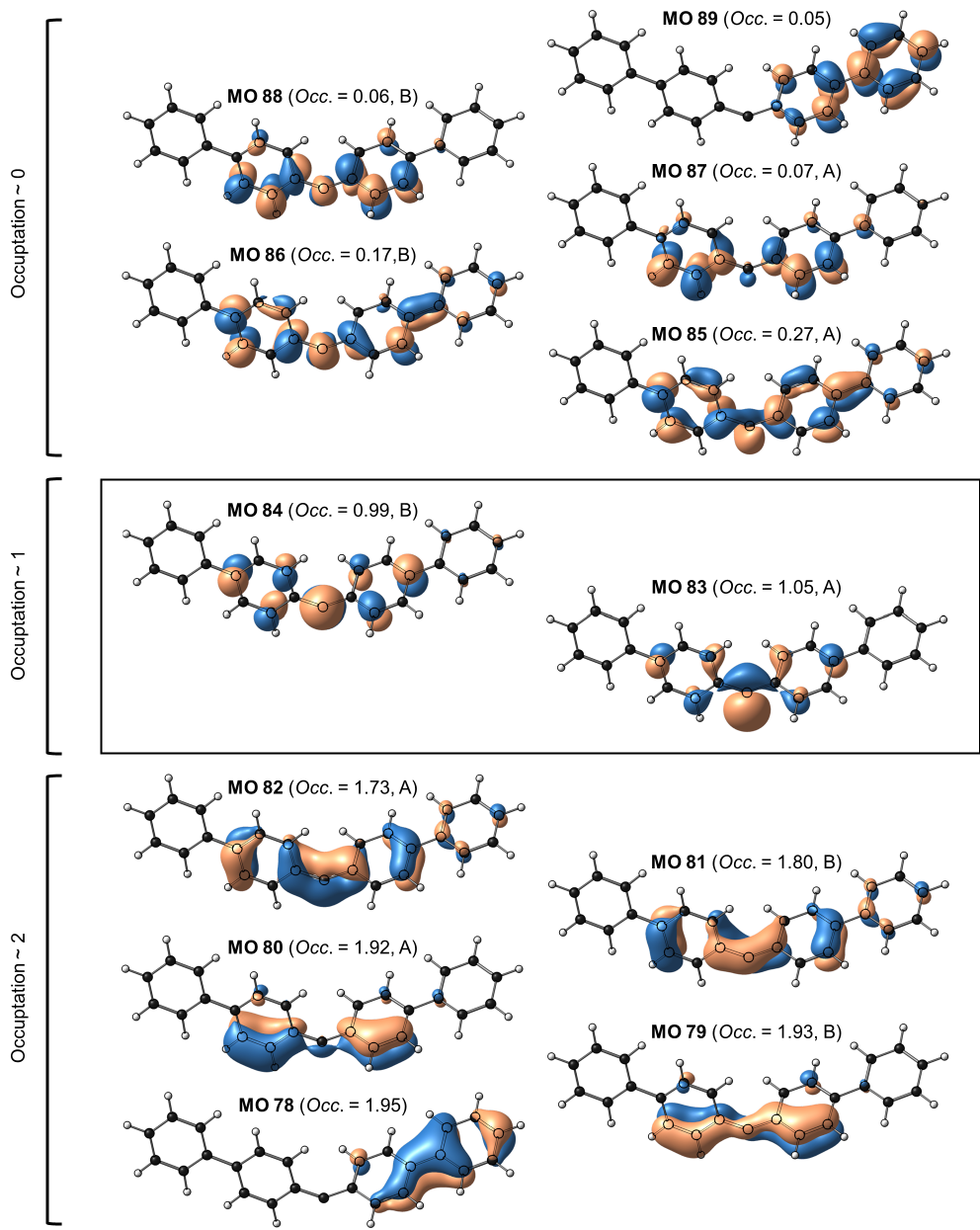}
	\caption{Molecular orbitals of the qubit within the chosen (12,12) active space together with the energies and occupancies for the ground state triplet (isovalue= 0.04) and the tentativley assigned irreducible representations.}
	\label{fig:CAS}
\end{figure}

Using the ONIOM pseudo-\textit{C}$_2$ symmetric geometry of our qubit, a minimum active space of 12 electrons in 12 orbitals was necessary to arrive at the minimum requirement for a sufficiently complete and balanced active space (0.05 $<$ occupancy $<$ 1.95). Multiple attempts to use a larger (14,14) active space were ultimately abandoned due to excessive computational demands when calculating the 4-particle reduced density matrix (4-RDM) required for strongly contracted N-electron valence state perturbation theory. When employing a (10,10) active space the occupation numbers at the outermost were 1.93 and 0.07 and the MO's shape and symmetry were essentially identical to MO's 79-88 depicted in Fig. \ref{tab:state_structure}. 

Exploratory SA8-CASSCF(10,10)/QD-SC-NEVPT2/def2-TZVPP calculations were performed to evaluate the influence of the neighboring nitrogen on the electronic structure of the carbene. We find no significant alterations to the predicted properties and, hence,  omitted the nitrogen from all further calculations for the sake of clarity.

\subsubsection{Complete active space}
The active orbitals obtained from above mentioned SA8-CASSCF(12,12)/QD-SC-NEVPT2/def2-TZVPP calcualtions are plotted in Fig. \ref{fig:CAS} using the Chemcraft software~\cite{zhurkoChemcraftGraphicalProgram2005}.
In total, six triplet and six singlet roots were computed, and a state-averaged CASSCF (SA-CASSCF) optimization was performed over the lowest three triplet and five singlet states, as these correspond to the optical range of interest (see Tab. \ref{tab:state_structure}).
Visual inspection of the active orbitals reveals that the CAS is predominantly delocalized over the central carbene carbon and the adjacent phenyl rings. Notably, the outermost active molecular orbitals, MO 78 (Occ. = 1.95) and MO 89 (Occ. = 0.05), exhibit an asymmetrically distorted wavefunction; their symmetrical counterparts reside just outside the active space as the frontier inactive MOs. Because expanding to a (14,14) active space was computationally prohibitive (see above), we proceeded under the assumption that these weakly correlated orbitals have only a marginal impact on the predicted properties (see below).

\subsubsection{Singlet triplet gap}
The vertical singlet-triplet energy gap ($\Delta E_{ST}$) was estimated from the abovementioned SA8-CASSCF(12,12)/QD-SC-NEVPT2/def2-TZVPP calculations. In this framework the vertical singlet-triplet energy gap ($\Delta E_{ST}$) is then provided by the Eigenvalue of state 3 with $\sim$0.47 eV, which can best be described as an closed shell singlet state with minor open shell singlet admixture. This value is slightly smaller than the previously reported $\Delta E_{ST}$ $\sim$0.63 eV of a familiar carbene doped into a molecular matrix\cite{roggorsOpticallyDetectedMagnetic2025a}.

\subsubsection{State structure}
\begin{table}[h]
	\centering
	\begin{tabular}{ l @{\hspace{2cm}} c c }
		\toprule
		\multicolumn{3}{l}{\textbf{CASSCF triplet states}} \\ 
		\midrule
		\multicolumn{3}{l}{ROOT 0: E = -957.5461485074 Eh} \\
		0.76728 & [0]   & 222221100000 \\
		0.03004 & [28]  & 222211110000 \\
		0.02085 & [246] & 222121101000 \\
		\midrule
		\multicolumn{3}{l}{ROOT 1: E = -957.4423817119 Eh (2.824 eV, 22774.2 cm$^{-1}$)} \\
		0.25860 & [1]   & 222221010000 \\
		0.20677 & [27]  & 222211200000 \\
		\bottomrule
		&&\\
		&&\\
		\midrule        
		\multicolumn{3}{l}{\textbf{CASSCF singlet states}} \\
		\midrule
		\multicolumn{3}{l}{ROOT 0: E = -957.5199886716 Eh} \\
		0.55517 & [0]   & 222222000000 \\
		0.17764 & [7]   & 222220200000 \\
		\midrule
		\multicolumn{3}{l}{ROOT 1: E = -957.4950588995 Eh (0.678 eV, 5471.5 cm$^{-1}$)} \\
		0.66064 & [1]   & 222221100000 \\
		\midrule
		\multicolumn{3}{l}{ROOT 2: E = -957.4465168185 Eh (1.999 eV, 16125.2 cm$^{-1}$)} \\
		0.30204 & [7]   & 222220200000 \\
		0.21209 & [0]   & 222222000000 \\
		\bottomrule
	\end{tabular}
	\caption{CASSCF-calculated multiconfigurational electronic wavefunctions of ground (root 0) and first excited states (root 1,2,...) of the triplet and singlet manifold. Shown here are the energies and relative contributions of single configurations (weights and electron occupations of the CAS, using 12 orbitals). Except for root 0 of the triplet states, configurations that contribute with $\le10\%$ are not shown but can be found in the Orca output file.}
	\label{tab:state_structure} 
\end{table}

\newpage
\subsubsection{Energy level structure}
\begin{table}[H]
	\centering
	\small
	\begin{tabular}{ c @{\hspace{0.8cm}} c @{\hspace{0.8cm}} c @{\hspace{0.8cm}} c @{\hspace{0.8cm}} c @{\hspace{0.8cm}} c @{\hspace{0.8cm}} c c c r }
		\toprule
		\textbf{State} & \textbf{Energy} & \textbf{Weight} & \textbf{Real} & \textbf{Imaginary} & \textbf{Block} & \textbf{Root} & \textbf{Spin} & \boldmath$M_s$ \\
		& \textbf{(cm$^{-1}$)} & & & & & & & \\
		\midrule
		0 & 0.0000 & 0.999924 & -0.000000 & -0.999962 & 0 & 0 & 1 & 0 \\
		\midrule
		\multirow{2}{*}{1} & \multirow{2}{*}{0.3763} & 0.499998 & 0.706999 & 0.012286 & 0 & 0 & 1 & 1 \\
		& & 0.499991 & 0.706993 & -0.012285 & 0 & 0 & 1 & -1 \\
		\midrule
		\multirow{2}{*}{2} & \multirow{2}{*}{0.4107} & 0.499966 & 0.012274 & -0.706976 & 0 & 0 & 1 & 1 \\
		& & 0.499973 & 0.012259 & 0.706981 & 0 & 0 & 1 & -1 \\
		\midrule
		3 & 3797.0085 & 0.999997 & 0.999999 & 0.000000 & 1 & 0 & 0 & 0 \\
		\midrule
		4 & 6976.3732 & 1.000000 & -1.000000 & 0.000000 & 1 & 1 & 0 & 0 \\
		\midrule
		5 & 14425.3414 & 0.999998 & 0.999999 & 0.000000 & 1 & 2 & 0 & 0 \\
		\midrule
		\multirow{3}{*}{6} & \multirow{3}{*}{21312.7452} & 0.497992 & 0.028947 & 0.705092 & 0 & 1 & 1 & 1 \\
		& & 0.004009 & 0.000001 & 0.063320 & 0 & 1 & 1 & 0 \\
		& & 0.497998 & 0.028956 & -0.705096 & 0 & 1 & 1 & -1 \\
		\midrule
		\multirow{3}{*}{7} & \multirow{3}{*}{21312.7636} & 0.490888 & -0.699769 & 0.034794 & 0 & 1 & 1 & 1 \\
		& & 0.018231 & 0.000000 & -0.135021 & 0 & 1 & 1 & 0 \\
		& & 0.490881 & -0.699765 & -0.034798 & 0 & 1 & 1 & -1 \\
		\midrule
		\multirow{3}{*}{8} & \multirow{3}{*}{21312.9599} & 0.011120 & 0.097405 & 0.040400 & 0 & 1 & 1 & 1 \\
		& & 0.977759 & 0.000000 & -0.988817 & 0 & 1 & 1 & 0 \\
		& & 0.011120 & 0.097407 & -0.040400 & 0 & 1 & 1 & -1 \\
		\midrule
		9 & 23243.8134 & 0.999999 & -1.000000 & -0.000000 & 1 & 3 & 0 & 0 \\
		\midrule
		\multirow{3}{*}{10} & \multirow{3}{*}{24099.8534} & 0.408159 & 0.636622 & -0.053588 & 0 & 2 & 1 & 1 \\
		& & 0.183667 & 0.000002 & -0.428563 & 0 & 2 & 1 & 0 \\
		& & 0.408174 & 0.636633 & 0.053592 & 0 & 2 & 1 & -1 \\
		\midrule
		\multirow{3}{*}{11} & \multirow{3}{*}{24099.8697} & 0.498989 & 0.044131 & 0.705011 & 0 & 2 & 1 & 1 \\
		& & 0.002041 & 0.000000 & -0.045173 & 0 & 2 & 1 & 0 \\
		& & 0.498971 & 0.044151 & -0.704997 & 0 & 2 & 1 & -1 \\
		\midrule
		\multirow{3}{*}{12} & \multirow{3}{*}{24099.9353} & 0.092852 & 0.304557 & 0.009841 & 0 & 2 & 1 & 1 \\
		& & 0.814292 & 0.000002 & 0.902381 & 0 & 2 & 1 & 0 \\
		& & 0.092856 & 0.304563 & -0.009842 & 0 & 2 & 1 & -1 \\
		\midrule
		13 & 26120.1734 & 1.000000 & -1.000000 & 0.000000 & 1 & 4 & 0 & 0 \\
		\bottomrule
	\end{tabular}
	\caption{Spinstates in $T_0$ geometry (block 0 = triplet states, block 1 = singlet states) included into state averaging, based on the multiconfigurational CASSCF-calculated wavefunctions (root 0, 1,...), employing QDPT with SOC and SSC contributions.}
	\label{fig:energy_levels}
\end{table}

\newpage
\subsubsection{Spin densities}
\begin{figure}[b]
	\centering
	\includegraphics[width=0.7\textwidth]{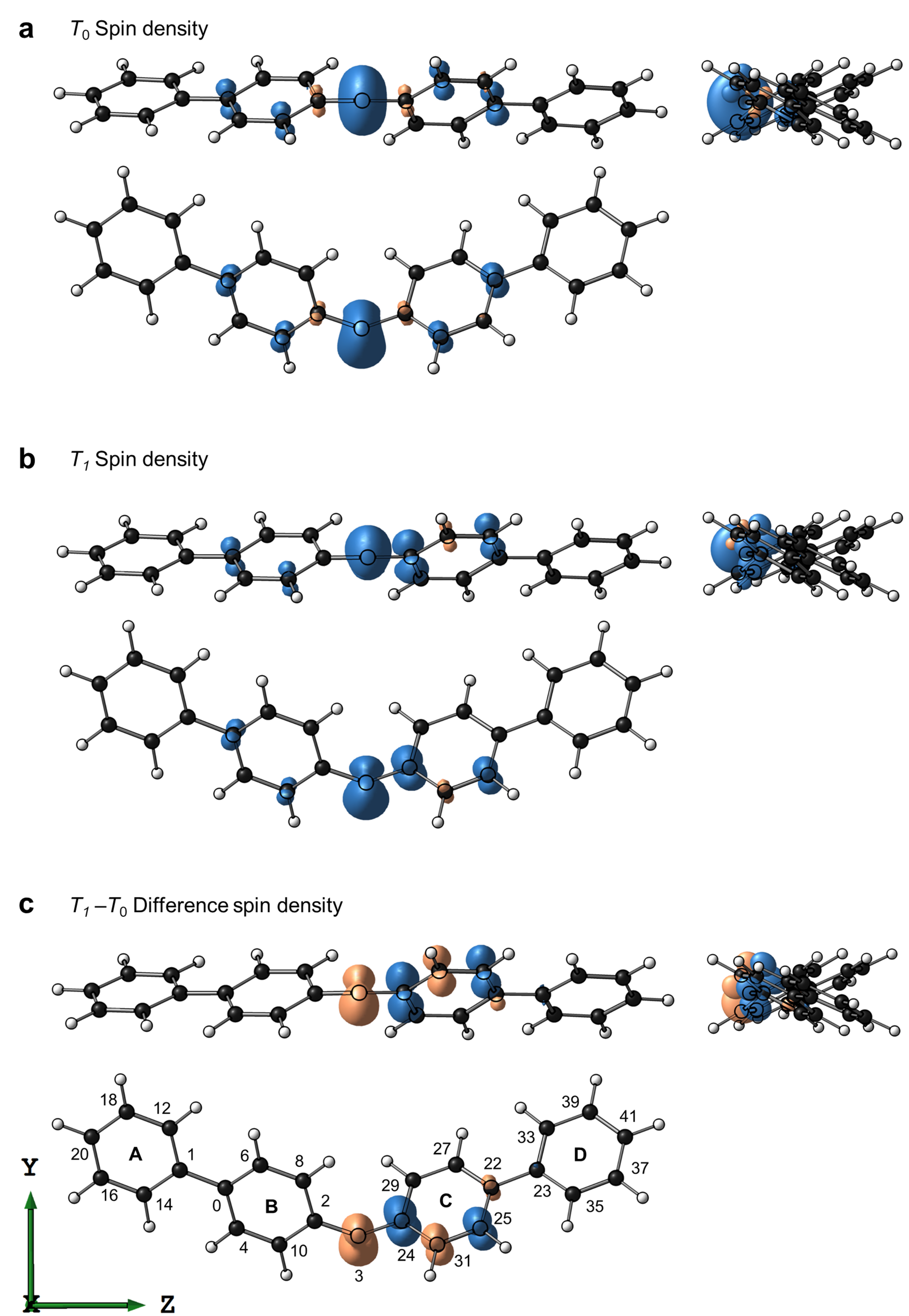}
	\caption{Spin density plots from state specific CASSCF/NEVPT2 calcualtion (CAS(12,12)), in the (a) $T_0$ and (b) $T_1$ electronic configuration and (c) difference spin density obtained by mathematically subtracting $T_1$ from $T_0$ density together with atom numbering (orange denotes spin depletion and blue accumulation zones when going from $T_0$ to $T_1$, isosurface=0.015, see also Tab. \ref{tab:spin_densities}.}
	\label{fig:SpinDensities}
\end{figure}
Changes in the spin distribution were analyzed based on tstate specific CASSCF(12,12)/QD-SC-NEVPT2/def2-TZVPPcalculations for the $T_0$ and $T_1$ electronic configuration. The transition from the $T_0$ to the $T_1$ state is accompanied by a stark change in the magnitude and sign of the axial and the transversal ZFS component \textit{D} and \textit{E}, respectively.
\begin{table}[H]
	\centering
	\caption{Loewdin spin densities and orbital breakdowns for selected carbon atoms in $T_0$ and $T_1$ electronic configuration (see Fig. \ref{fig:SpinDensities} for atom numbering).}
	\label{tab:spin_densities}
	\renewcommand{\arraystretch}{1.3}
	\begin{tabular}{lccc}
		\toprule
		\textbf{Atom} & \textbf{Spin density ($T_0$)} & \textbf{Spin density ($T_1$)} & \textbf{Orbital spin ($s$, $p_x$, $p_y$, $p_z$)} \\
		\midrule
		0 C   &  0.1251 &  0.1224 &  \begin{tabular}{@{}l@{}}$T_0$: $s$=0.0000, $p_x$=0.0939, $p_y$=0.0331, $p_z$=0.0047 \\ $T_1$: $s$=0.0000, $p_x$=0.0910, $p_y$=0.0321, $p_z$=0.0046\end{tabular} \\
		1 C   &  0.0181 &  0.0178 &  -- \\
		2 C   &  0.0804 &  0.0488 &  -- \\
		3 C   &  1.0787 &  0.6655 &  \begin{tabular}{@{}l@{}}$T_0$: $s$=0.0629, $p_x$=0.4565, $p_y$=0.5526, $p_z$=0.0001 \\ $T_1$: $s$=0.0443, $p_x$=0.1724, $p_y$=0.4024, $p_z$=0.0002\end{tabular} \\
		4 C   & -0.0214 & -0.0143 &  -- \\
		6 C   &  0.0156 &  0.0187 &  -- \\
		8 C   &  0.0683 &  0.0590 &  -- \\
		10 C  &  0.1165 &  0.0994 &  \begin{tabular}{@{}l@{}}$T_0$: $s$=0.0018, $p_x$=0.0844, $p_y$=0.0341, $p_z$=0.0106 \\ $T_1$: $s$=0.0007, $p_x$=0.0740, $p_y$=0.0289, $p_z$=0.0056\end{tabular} \\
		12 C  &  0.0080 &  0.0081 &  -- \\
		14 C  &  0.0076 &  0.0077 &  -- \\
		16 C  &  0.0022 &  0.0022 &  -- \\
		18 C  &  0.0022 &  0.0023 &  -- \\
		20 C  &  0.0061 &  0.0062 &  -- \\
		22 C  &  0.1455 &  0.0903 &  \begin{tabular}{@{}l@{}}$T_0$: $s$=0.0000, $p_x$=0.1124, $p_y$=0.0376, $p_z$=0.0054 \\ $T_1$: $s$=0.0000, $p_x$=0.0453, $p_y$=0.0150, $p_z$=0.0022\end{tabular} \\
		23 C  & -0.0013 &  0.0525 &  -- \\
		24 C  &  0.0790 &  0.3271 &  \begin{tabular}{@{}l@{}}$T_0$: $s$=0.0035, $p_x$=-0.0626, $p_y$=-0.0133, $p_z$=0.0080 \\ $T_1$: $s$=0.0026, $p_x$=0.1849, $p_y$=0.0715, $p_z$=0.0110\end{tabular} \\
		25 C  & -0.0242 &  0.1808 &  \begin{tabular}{@{}l@{}}$T_0$: $s$=0.0001, $p_x$=-0.0439, $p_y$=-0.0152, $p_z$=-0.0017 \\ $T_1$: $s$=0.0001, $p_x$=0.1307, $p_y$=0.0460, $p_z$=0.0063\end{tabular} \\
		27 C  &  0.0122 &  0.0482 &  -- \\
		29 C  &  0.0617 &  0.0586 &  -- \\
		31 C  &  0.1224 & -0.0254 &  \begin{tabular}{@{}l@{}}$T_0$: $s$=0.0018, $p_x$=0.0913, $p_y$=0.0335, $p_z$=0.0111 \\ $T_1$: $s$=0.0016, $p_x$=-0.0640, $p_y$=-0.0222, $p_z$=0.0039\end{tabular} \\
		33 C  &  0.0177 &  0.0399 &  -- \\
		35 C  &  0.0184 &  0.0410 &  -- \\
		37 C  &  0.0056 &  0.0172 &  -- \\
		39 C  &  0.0053 &  0.0176 &  -- \\
		41 C  &  0.0183 &  0.0643 &  -- \\
		\bottomrule
	\end{tabular}
\end{table}
In the $T_0$ state, radical character is highly localized at the C3 center ($\rho$=1.079), dominated by the $p_y$ (0.553, mainly $\sigma$-orbital) and $p_x$ (0.456, mainly $p$-orbital) orbitals and a weak contribution from the $s$-orbital (0.063) due to hybridization (Tab. \ref{tab:spin_densities}, for spin density plots and coordinate system see Fig. \ref{fig:SpinDensities}a). This spatial confinement is characteristic for carbenes and results in a strong dipolar spin-spin interaction and an oblate ZFS tensor with a large positive magnitude ($D_{T_0}$ = 11797\,MHz, $E_{T_0}$=-516\,MHz, see Sec.~\ref{sec:zfs_tensor}). Upon excitation to $T_1$, the central carbene carbon C3 donates significant spin density to the adjacent $\pi$-system; its total spin decreases to 0.665 as the localized $p_x$ contribution drops to 0.172 (Fig. \ref{fig:SpinDensities}b). Notably, this transition drives a pronounced shift in the spin distribution. While the spin density on C3 and C31 decreases substantially, there is a corresponding accumulation of spin density on C24 and C25, which increase to 0.079 and 0.327 in $T_1$, respectively (see Fig. \ref{fig:SpinDensities}c). The resulting elongated and more diffuse spin distribution in the $T_1$ state fundamentally alters the dipolar spin-spin interaction, thereby inverting the ZFS tensor to a prolate geometry with a reduced, negative magnitude in the excited state ($D_{T_1}$=-6161\,MHz, $E_{T_1}$=276\,MHz).

\subsubsection{Zero-field splitting: Tensor parameters and spin state overlap}
\label{sec:zfs_tensor}

The zero-field splitting (ZFS) parameters for the lowest-energy triplet state and the first excited triplet state were computed incorporating both spin-orbit coupling (SOC) and spin-spin coupling (SSC) contributions. The resulting ZFS tensors are then given by a coordinate frame (x,y,z) for the ground-state triplet and (x',y',z') for the excited-state triplet (see \ref{sec:zfs_tensor}) which are nearly identical (see tab. \ref{tab:spin_state_overlap}).

From Tab. \ref{fig:energy_levels} we calculate the zero-field splitting (ZFS) parameters $D$ and $E$ (see Tab. \ref{tab:zfs_parameters}) with contribution from both spin-spin and spin-orbit interactions (here, the ground state $T_0$ is given by states 0, 1 and 2 while the first excited triplet state $T_1$ consists of states 6, 7 and 8) using the Hamiltonian of the form
$$H_{\text{ZFS}} = D_x S_x^2 + D_y S_y^2 +D_z S_z^2 = D (S_z^2-1/3 \cdot S^2) + E (S_x^2 -S_y^2)$$
assuming a traceless tensor \textbf{D} and the convention \cite{pooleStandardizationConventionZero1974} $ |D_z| > |D_x| \geq |D_y|$, where the ZFS energies are $E(T_x)=1/3 \cdot D-E$, $E(T_y)=1/3 \cdot D+E$ and $E(T_z)=-2/3 \cdot D$ \cite{weilElectronParamagneticResonance2006}.
\begin{table}[h]
	\centering
	\begin{tabular}{  |l @{\hspace{0.5cm}} | c @{\hspace{0.5cm}} | c | }
		\toprule
		\textbf{ZFS Parameter}  &  \textbf{GS (MHz)}  &  \textbf{ES (MHz)} \\
		\midrule
		$D$  &  11797  &  -6161 \\
		$E$  &  -516  &  276 \\
		\bottomrule
	\end{tabular}
	\caption{ZFS parameters of \bifi{} in the triplet ground state (GS) and first excited state (ES).}
	\label{tab:zfs_parameters}
\end{table}

In order to estimate to which degree the optical transitions are spin-conserving between the ground and excited triplet states, we further calculate the overlap of the spin state as the vectorial product (see Tab. \ref{tab:spin_state_overlap}). As a result, the spin state conservation is greater than 0.97 for each state.

\begin{table}[h]
	\centering
	\begin{tabular}{ | l  @{\hspace{0.5cm}} | c  @{\hspace{0.5cm}} | c | }
		\toprule
		\textbf{GS}  &  \textbf{ES}  &  \textbf{overlap} \\
		\midrule
		x  &  z'  &  0.003 \\
		&  y'  &  0.004 \\
		&  x'  &  \textbf{0.993} \\
		\midrule
		y  &  z'  &  0.019 \\
		&  y'  &  \textbf{0.977} \\
		&  x'  &  0.003 \\
		\midrule
		z  &  z'  &  \textbf{0.978} \\
		&  y'  &  0.018 \\
		&  x'  &  0.004 \\
		\bottomrule
	\end{tabular}
	\caption{Spin state overlap between GS and ES.}
	\label{tab:spin_state_overlap}
\end{table}

\subsubsection{ISC pathways due to SOC}
Referring to our previous work \cite{roggorsOpticallyDetectedMagnetic2025a}, we only consider ISC pathways between states that are closest in energy and we report relative rates due to the lack of a calibration of the spin orbit coupling matrix elements (SOCMEs) to absolute relaxation rates (see Tab. \ref{tab:ISC_relative_rates}). As a result, both excited and ground state relaxation mostly couples from (to) the $T_{1z'(0z)}$ state.

\begin{table}[h]
	\centering
	\begin{tabular}{ | l @{\hspace{0.5cm}} | c @{\hspace{0.5cm}} | c | }
		\toprule
		\textbf{Pathway}  &  \textbf{spin state}  &  \textbf{relative rate} \\
		\midrule
		$T_1 \rightarrow{S_2}$    &  z'  &  \textbf{0.949} \\
		&  y'  &  0.042 \\
		&  x'  &  0.009 \\
		\midrule
		$S_0 \rightarrow{T_0}$  &  x  &  $<$ 0.001 \\
		&  y  &  $<$ 0.001\\
		&  \textbf{z}  &  \textbf{$>$ 0.999} \\
		\bottomrule
	\end{tabular}
	\caption{Selectivity of ISC given by the relative rates of the triplet-singlet states that are closest in energy.}
	\label{tab:ISC_relative_rates}
\end{table}

Our SA8-CASSCF(12,12)/QD-SC-NEVPT2/def2-TZVPP calculations indicate strong and almost exclusive $z$-selective ISC for both the $T_1 \rightarrow S_2$ and $S_0 \rightarrow T_0$ transitions. The latter is consistent with an insightful theoretical study on bent carbenes by Yuen-Zhou and co-workers \cite{pohQuantifyingSpinOpticalProperties2026a}, in that we find that the pseudo-$C_2$ symmetry suppresses $xy$-selective ground-state ISC while opening the $z$-selective, El-Sayed-allowed channel by favoring population of the non-bonding $\sigma$ orbital (see the state structure in Tab. \ref{tab:state_structure}).

On the other hand, the strongly $z$-selective ISC observed in the excited state aligns with our experimental findings and warrants closer inspection. We further analyze this phenomenon by applying the theoretical framework established by Yuen-Zhou and co-workers \cite{pohQuantifyingSpinOpticalProperties2026a}. The two states under consideration are predominantly built from contributions of the inner six molecular orbitals (MOs 81-86) comprising ca. 79\% of the mutliconfigurational electronic space for $T_1$ and ca. 84\% for $S_2$ with only a 1.5\% and 0.5\% contribution originating from the outermost asymmetrically distorted MOs (see Fig. \ref{fig:CAS}. Consequently, these outermost MOs are not considered relevant for the subsequent discussion.

In our bent, pseudo-$C_2$-symmetric carbene, the departure from a linear $D_2$ geometry as previously discussed\cite{pohElectronSextetsOptically2025e} breaks the strict spatial symmetry that typically restricts excited-state ISC to the $xy$-plane ($\Delta M_S = \pm 1$). This structural bending unlocks a $z$-selective ($\Delta M_S = 0$) ISC channel mediated by $B$-irrep triplet states (see Fig. \ref{fig:CAS} for irrep assignment). Crucially, because our $T_0 \rightarrow T_1$ excitation is predominantly SOMO-to-LUMO in character (68\% $A$-manifold, see Fig. \ref{fig:TD-DFT}), the standard charge recombination pathway and the newly activated charge separation pathway superimpose (not to be confused with the charge transfer (CT) state discussed in Sec.\ref{sec:Steady}). While it has been found that a HOMO-to-SOMO transition is able - due to a general destructive interference of the involved pathways - to allow for a suppression of the $\Delta\,M_S=0$ channel, in the case of a SOMO-to-LUMO transition, the opposite is true and the pathways interfere constructively, rendering the $\Delta\,M_S=0$ channel to be stronger than the $\Delta\,M_S=\pm1$ channel and as observed in our case, dominate the relaxation pathway. 

\newpage
\section{Microscopy and refractive index tensor of undoped crystals}
\label{section:microscopy}

\begin{figure}[htpb]
	\centering
	\includegraphics[width=0.95\textwidth]{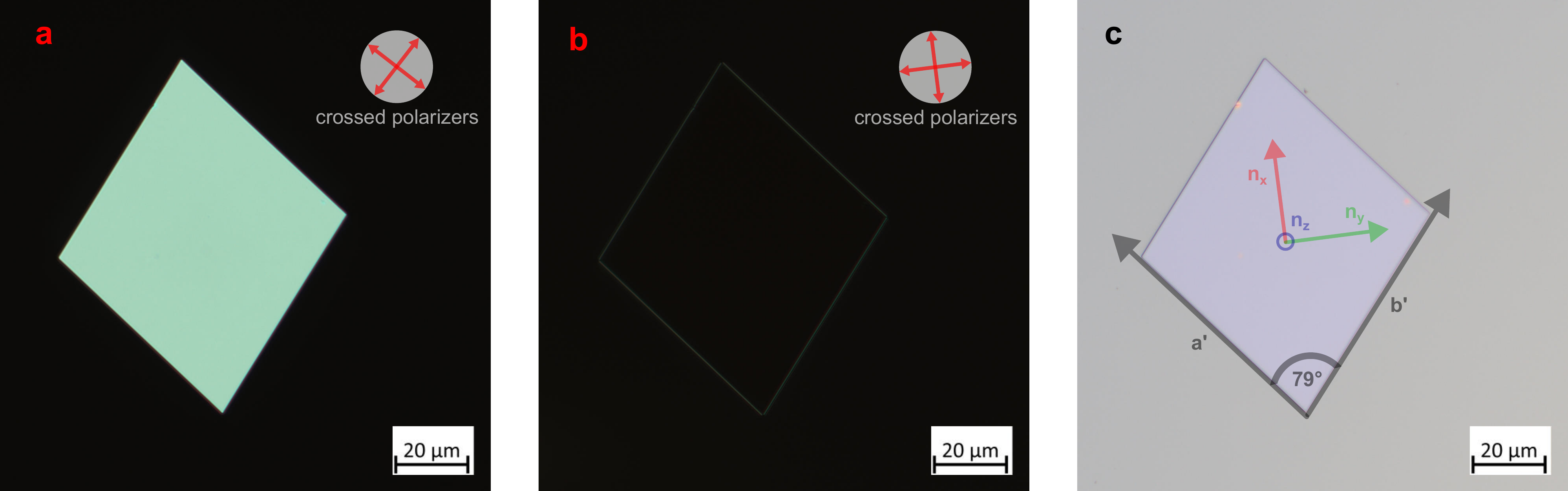}
	\caption{Microscopy images of a submicron thick crystal. From left to right: Image under crossed linear polarizers (indicated by insets) with maximum brightness of crystal appearance (a). Polarizers turned by 45° in respect to bright appearance, matching the in-plane refractive index axes (b). As a result, the crystal is rendered completely dark except for scattered light on the edges. A brightfield image of the crystal (c) with refractive index axes ($n_x$, $n_y$, $n_z$) and axes ($a'$, $b'$) drawn onto the  parallelogramic outline.}
	\label{fig:microscopy_images}
\end{figure}

\begin{figure}[htpb]
	\centering
	\includegraphics[width=0.95\textwidth]{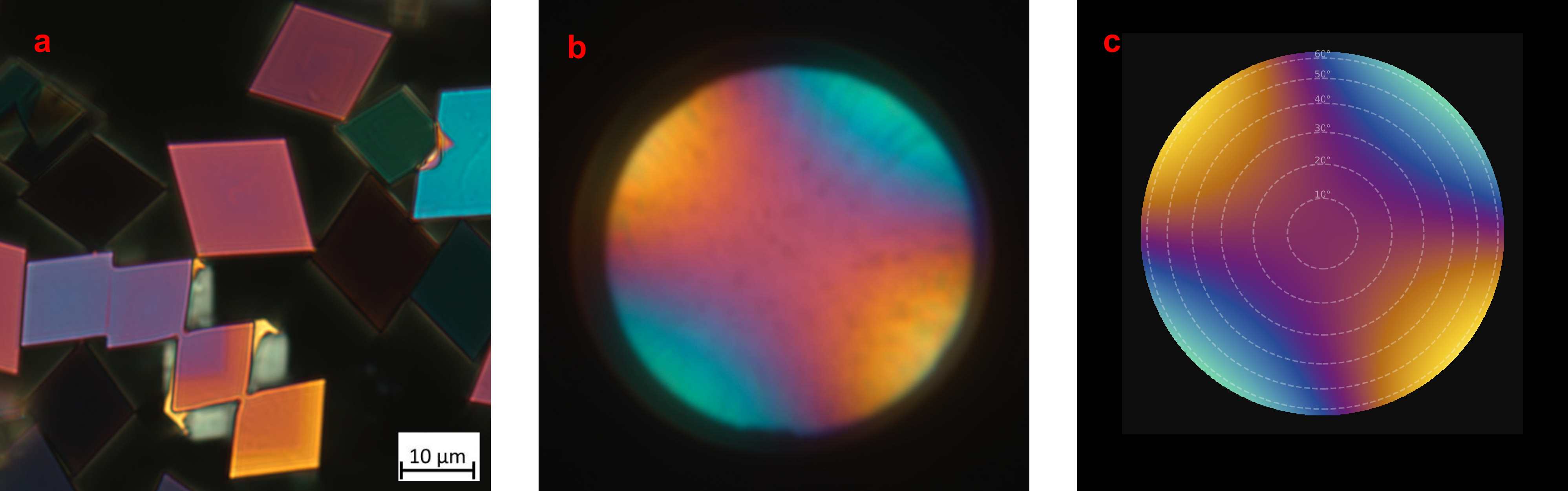}
	\caption{To evaluate the refractive index tensor via conoscopy, the central 2.7\,$\mathrm{\mu}$m thick crystal in the polarized image of (a) was chosen. (b) Experimental conoscopy pattern and (c) corresponding simulation, with an overlaid polar grid indicating the inclination angle collected by the high NA of the microscope. Simulation parameters and results are detailed in the main text.}
	\label{fig:conoscopy}
\end{figure}

Thin-film crystals of the undoped matrix with hundreds of nanometer to few micrometer thickness were slowly grown with several hours by partially sealing a drop casted, several 100\,nL of a 15\,mM solution of \bike{} (\textbf{S1}) in 1:1 tetrahydrofuran:mesitylene between a microscopy glass slide and a cover glass. Under these growth conditions, crystals typically exhibit top / bottom surfaces of parallelogramic outline (see figures \ref{fig:microscopy_images} and \ref{fig:conoscopy}) that have an acute angle of (79.0 $\pm$ 1)° and coincide with the crystal cleavage plane. Crystals were selected by their homogeneous appearance exhibiting uniformity in thickness, and studied with a Zeiss Axio Imager microscope equipped with a conoscopy unit.

Using polarized light microscopy, the refractive index tensor axes of the biaxial crystal can be identified to be the normal to the plane of the parallelogram and the two bisectors, as depicted in (c) of Fig. \ref{fig:microscopy_images}. The correct orientation of the refractive index axes in respect to the crystal outline can be identified by its conoscopy pattern (see Fig. \ref{fig:conoscopy}). Identifying the refractive index along the acute bisector with the immersion method ~\cite{stoiberImmersionMethod1994} under polarized light to be $\sim$\,1.55, one can thus deduce the full tensor by simulating the conoscopy pattern ~\cite{vanhornConoscopicMeasurementBirefringence2003a} ~\cite{sorensenRevisedMichelLevyInterference2013} . In the example shown, the crystal thickness was measured to be 2.7 $\mu$m. A nominal maximum numerical aperture of 0.9 for both the condenser and objective was employed as a simulation input, allowing the crystal's refractive index tensor to be estimated as $(n_x \sim1.55, n_y \sim1.74, n_z \gtrsim 1.85)$.

\newpage
\section{EPR Spectroscopy: Ensemble GS ZFS tensor}

The ground state zero-field splitting (GS ZFS) tensor of the carbene spin ensemble and its eigenframe compared to the crystal frame (and hence refractive index tensor frame, see \ref{section:microscopy}) were characterized using EPR spectroscopy on a Bruker Elexsys II X-band spectrometer, operated with a dielectric ring resonator (Bruker ER 4118X-MD5) and equipped with an Helium-flow cryostat (Oxford Instruments CF935) to allow temperature-dependent measurements and in-situ annealing. A small, selected crystal of \bifi{} doped into \bike{} (0.25 mol\,$\%$) with flawless optical appearance and $\sim$ 0.1\,mm$^3$ volume (2500 ppm doping ratio) is inserted into a quartz glass tube and fixed with a PVA coating. Crystal samples are rotated by a goniometer along the axis of the sample stick, perpendicular to $B_0$. For photoactivation of the carbene ensemble we illuminate the sample with a 395\,nm UV LED lamp (Jupitertech NSP1, maximum intensity of 12\,W/cm$^2$) held in front of the cryostat optical side-port window at cryogenic temperature. The photoactivation was followed by reading the EPR signal amplitude and was found to be completed after a duration of typically few ten seconds of illumination.

\subsection{Rotation-dispersion spectra: ground state (GS) ZFS frame}
\begin{figure}[b]
	\centering
	\includegraphics[width=0.95\textwidth]{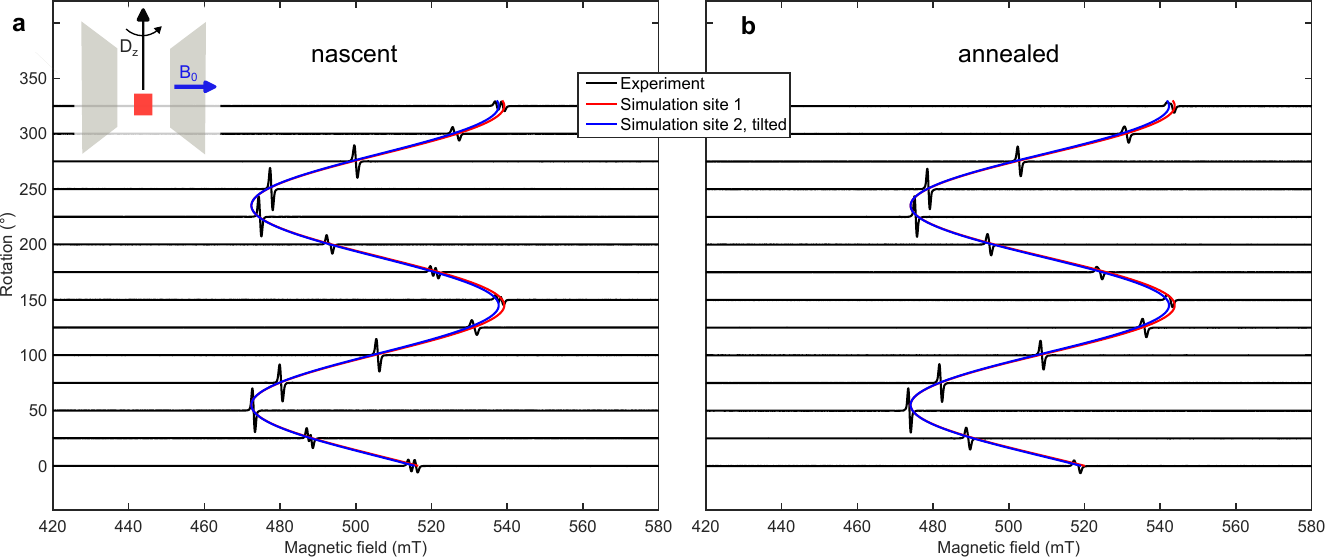}
	\caption{Rotation-dispersion spectra with simulation of the resonance line positions, of a nascent (a) and annealed (b) crystal with its cleavage plane normal (corresponding to the ZFS $D_z$ axis by analysis) orientated along the rotation axis as depicted in the inset of (a).}
	\label{fig:EPR-s2-Dz}
\end{figure}

\begin{figure}[t]
	\centering
	\includegraphics[width=0.95\textwidth]{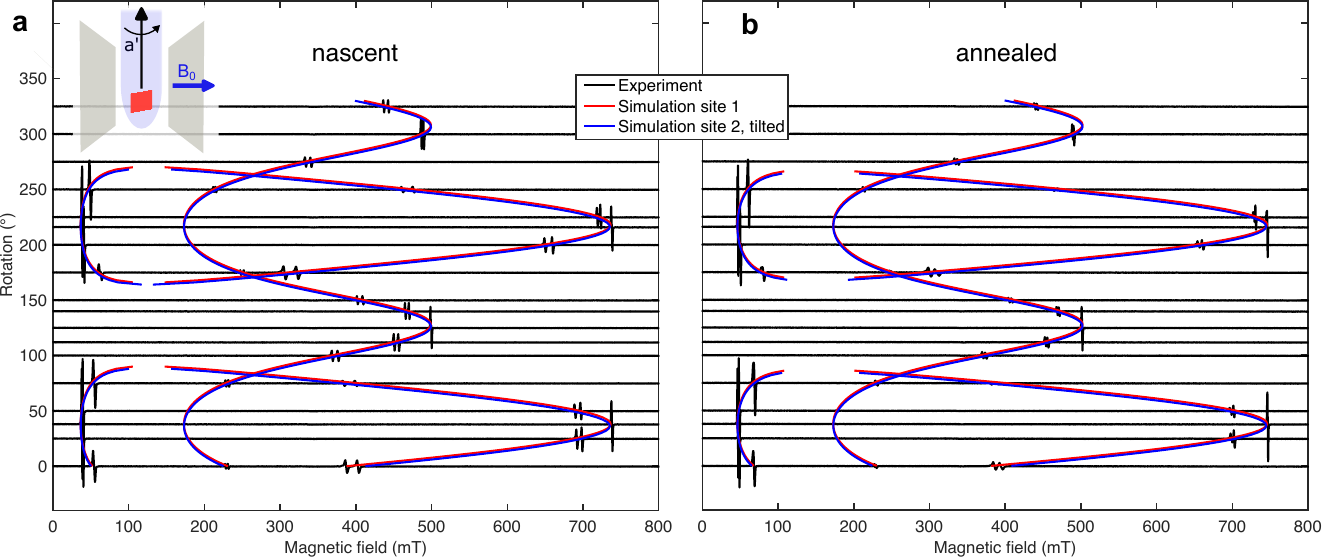}
	\caption{Rotation-dispersion spectra with simulation of the resonance line positions, of a nascent (a) and annealed (b) crystal with its crystal $a'$ axis (refer to Fig. \ref{fig:microscopy_images} c) orientated along the rotation axis as depicted in the inset of (a).}
	\label{fig:EPR-s3-a}
\end{figure}

\begin{figure}[t]
	\centering
	\includegraphics[width=0.4\textwidth]{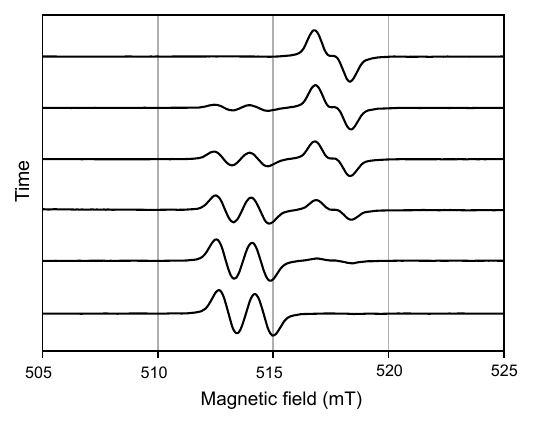}
	\caption{Annealing at $\sim$ 120\,K of the crystal of Fig. \ref{fig:EPR-s2-Dz} using the rotation position of the goniometer at 0°, and recording the field-swept EPR spectrum about every 10 minutes.}
	\label{fig:EPR_annealing}
\end{figure}

After photoactivation at temperatures of $\sim$ 50\,K, the continuous-wave field-swept EPR spectrum typically contains two close-lying resonances of the same order of magnitude, which can be identified by a few degree tilt between otherwise magnetically identical sites of spin 1. To achieve a complete characterization of the zero-field splitting (ZFS) tensor, rotation-dispersion spectra were recorded for two orthogonally oriented crystals. In the first configuration, the rotation axis was aligned with the normal to the cleavage plane; in the second, it coincided with the crystallographic $a'$-axis (see Fig. \ref{fig:microscopy_images}c). In both cases, the spectra are recorded for the nascent and annealed state of the carbene after photoactivation (see Fig. \ref{fig:EPR_annealing} for illustration of the effect of annealing on the signals). We used the following experimental and spectrometer settings: Recordings of nascent and annealed spectra were all done at 50\,K, annealing was performed between 120 and 130\,K, further, the carbene spins where excited with 15\,$\mu$W microwave power before the cavity with a quality factor Q $\sim$ 10'000, magnetic field sweeps were done with 20\,ms conversion time and 200 mT (800 mT) sweep width with 4000 (8000) data points and 2.5\,G (5\,G) amplitude modulation at 100\,kHz for Fig. \ref{fig:EPR-s2-Dz} (Fig. \ref{fig:EPR-s3-a}). 

The spectral positions of the rotation-dispersion spectra are then simulated with EasySpin \cite{stollEasySpinComprehensiveSoftware2006}. For the nascent case, we find a good agreement using $(D, E)_{\text{nascent}}$ = (10900, -520)\,MHz. For the annealed case we use the values of $(D, E)_{\text{annealed}}$ = (11160, -541)\,MHz found from the cryo-confocal measurements. Further, we introduce a tilt of 2.5° between the paramagnetic sites along the $D_y$ axis. In both cases, we assume an isotropic g-factor of 2.0023. It is important to note the lack of significant deviation in ZFS values between the nascent and annealed states. This may indicate that the crystalline matrix is rigid to an extent that it precludes significant structural changes of the carbenes.

\subsection{Comparison: Crystal, ground state ZFS and refractive index frame}

\begin{figure}[htpb]
	\centering
	\includegraphics[width=0.65\textwidth]{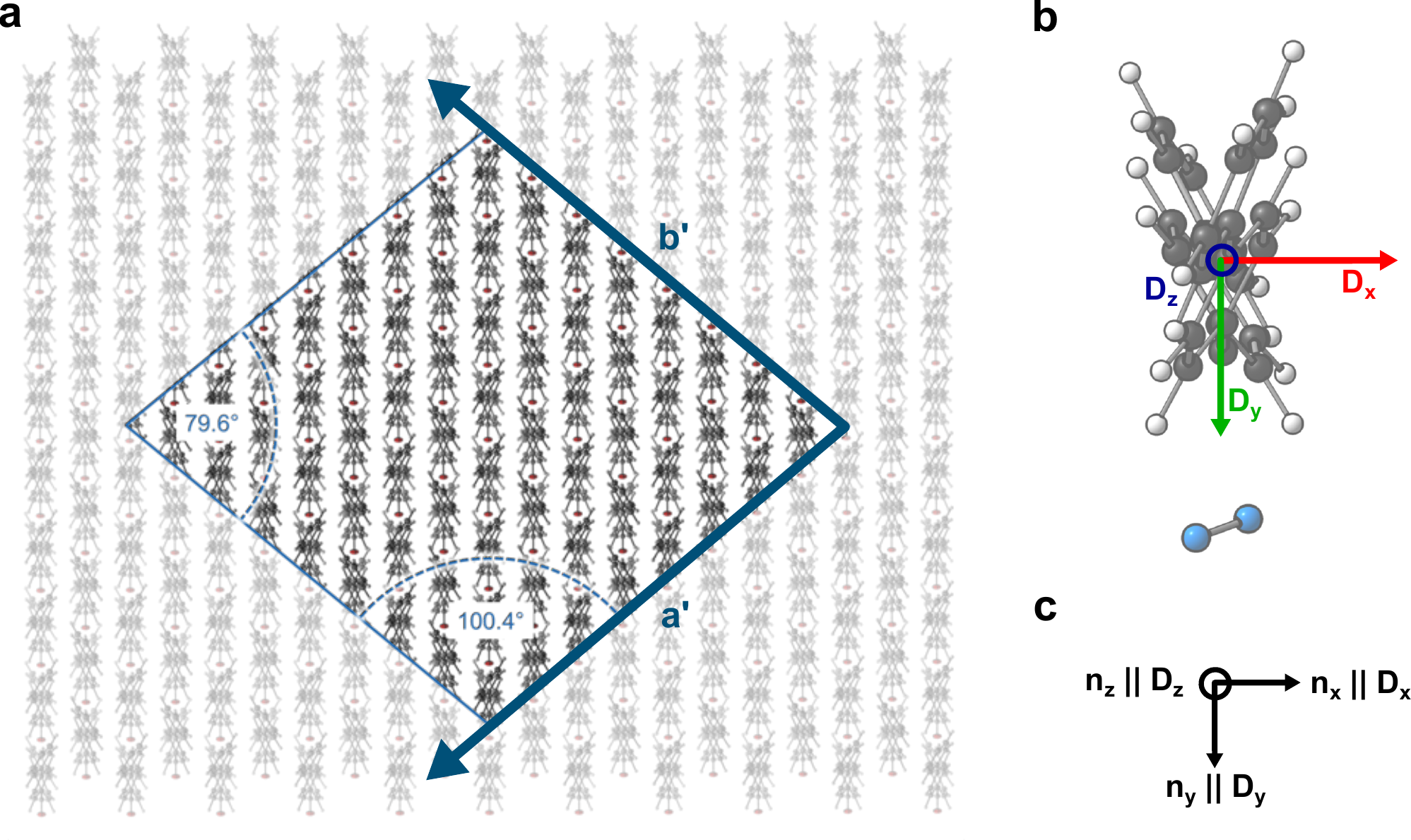}
	\caption{Crystal image from X-ray structure data of the matrix (see section X-ray crystallographic analysis, a), view is normal to the cleavage plane, highlighting a monoclinic cell which matches well to the parallelogramic outline of grown crystals (see Fig. \ref{fig:microscopy_images}, therein the axes definition of $a'$ and $b'$ which are repeated here). The predicted ZFS axes from quantum chemistry calculations of the carbene (see section Computational chemistry), adapted to the crystal frame by assuming that a doped carbene molecule substitutionally replaces a ketone molecule of the matrix, fits well to the ZFS axes referenced to the crystal frame from EPR measurements (b). All refractive index and ZFS axes are summarized in (c), compare to Tab. \ref{tab:frame_orientations}.} 
	\label{fig:morphology}
\end{figure}

Combining the polarization microscopy results on undoped thin-film crystals and the measurement of the ground state ZFS tensor with EPR on doped bulk samples (0.25 mol\,$\%$), we arrive at the conclusion that ZFS and refractive index tensor axes are co-aligned (up to a few degree of alignment, measurement and simulation uncertainty) and can be referred to characteristic crystal axes (see Tab. \ref{tab:frame_orientations}). This result matches very well the predicted orientations from the X-ray structure together with theoretically derived ZFS tensor axes (see Fig. \ref{fig:morphology}).

\subsection{Thermal Stability}
When doped into the matrix and photoactivated the carbenes are thermally stable at least up to 250\,K with no signal loss within the measurement error over several hours. At around 300\,K the carbenes become unstable with slow signal decrease over hours.

\begin{table}[H]
	\centering
	\begin{tabular}{| l | r | r |}
		\toprule
		\textbf{Crystal frame}  &  \textbf{GS ZFS Frame}  &  \textbf{Refractive index frame} \\
		\midrule
		Acute bisector of CP  &  $D_x$  &  $n_x$ \\
		Obtuse bisector of CP  &  $D_y$  &  $n_y$ \\
		Normal to CP  &  $D_z$  &  $n_z$ \\
		\bottomrule
	\end{tabular}
	\caption{Orientation of the ground state ZFS frame in respect to the crystal frame and refractive index frame (see Fig. \ref{fig:microscopy_images}). CP: cleavage plane.}
	\label{tab:frame_orientations}
\end{table}

\newpage
\section{Cryo-confocal measurements}

\subsection{Comparison of annealed and nascent carbenes}

\begin{figure}[H]
	\centering
	\includegraphics{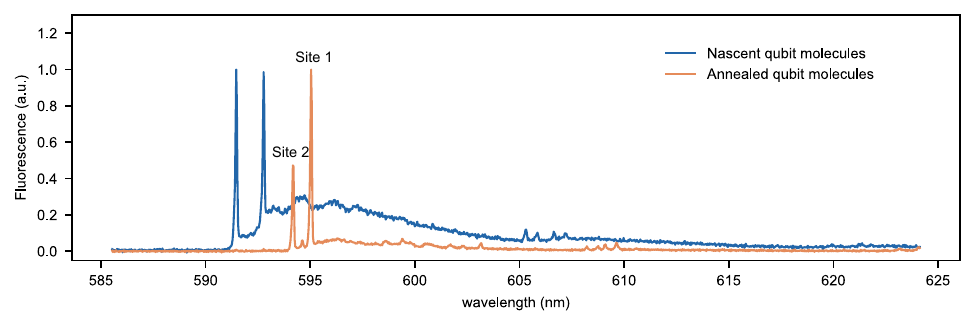}
	\caption{Fluorescence spectra for both nascent and annealed qubit molecules in the matrix. Both spectra are normalized to their respective maximum, thus, total fluorescence levels cannot be compared.}
	\label{fig:fluo_annealed_activated}
\end{figure}

\begin{figure}[H]
	\centering
	\includegraphics[width=0.55\textwidth]{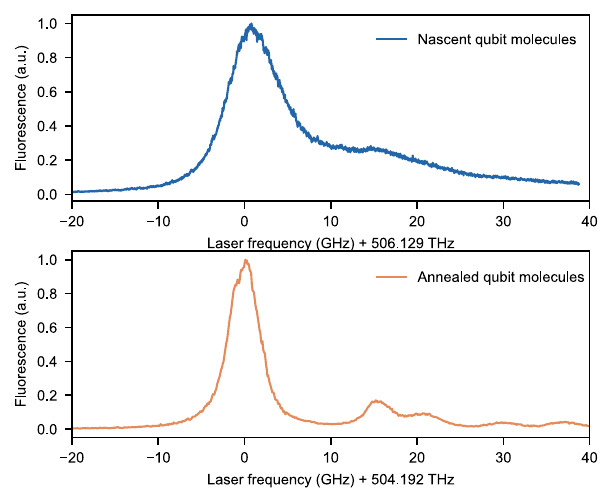}
	\caption{Laser excitation spectra for the higher wavelength molecular site for both nascent and annealed \bifi{} embedded in \bike{}. For both measurements, microwave driving has been applied on the \Tgz{} to \Tgx{} and \Tgz{} to \Tgy{} transition. In both cases, the laser frequency was offset to the respective overlap of the \Tgx{} to \Tex{} and \Tgy{} and \Tey{} transition (as in the main text).}
	\label{fig:exc_annealed_activated}
\end{figure}

\begin{figure}[H]
	\centering
	\includegraphics[width=0.68\textwidth]{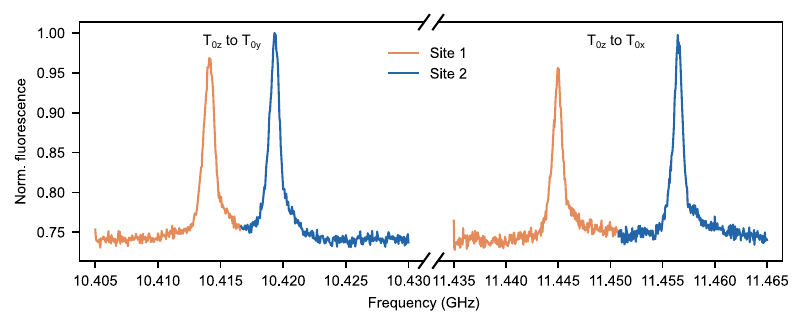}
	\caption{Full ODMR spectrum, recorded using off-resonant laser excitation at 580\,nm and laser power of 270\,nW. The signal stems from both molecular sites of \textbf{nascent} \bifi{} molecules, and shows two pairs of ODMR lines corresponding to the \Tgz{} to \Tgx{}/\Tgy{} transitions of both molecule classes.}
	\label{fig:odmr_annealed_activated}
\end{figure}

When photo-activated at temperatures below 120\,K, \bifi{} embedded in \bike{} appears spectrally blue shifted as compared to the annealed qubit molecules investigated in-depth in the main text (heated up to 200\,K for about one hour, see Fig.~\ref{fig:fluo_annealed_activated}). In addition, the nascent carbenes incorporate a much more dominant phonon side-band. Similarly, laser excitation spectra of nascent carbenes clearly show an increased inhomogeneous broadening in which, however, the fine structure is still visible (Fig.~\ref{fig:exc_annealed_activated}).
By taking into account the transition frequencies in Fig.~\ref{fig:odmr_annealed_activated}, the ground-state ZFS parameters for the nascent molecules of both molecular site 1 and 2 were deduced to be $|D_{T_0,\text{site 1}}| = 10.9300(2)$\,GHz, $|E_{T_0,\text{site 1}}|=516.0(2)$\,MHz, $|D_{T_0,\text{site 2}}| = 10.9380(2)$\,GHz and $|E_{T_0,\text{site 2}}|=519.0(2)$\,MHz.

\subsection{Off-resonant ODMR on both annealed molecular sites}
\begin{figure}[H]
	\centering
	\includegraphics[width=0.68\textwidth]{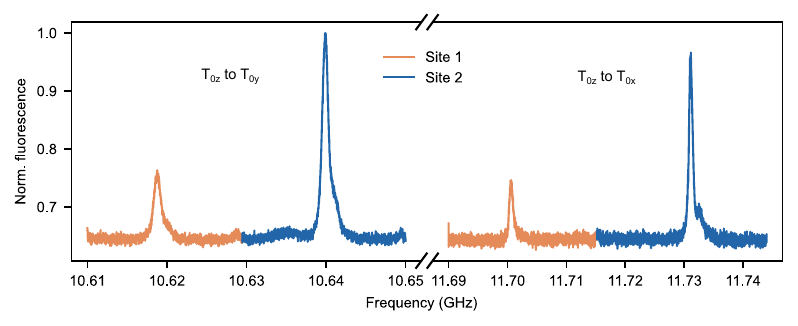}
	\caption{Full ODMR spectrum, recorded using off-resonant laser excitation at 580\,nm and laser power of 270\,nW. The signal stems from both \textbf{annealed} molecular sites (compare Fig.~2a in the main text), and shows two pairs of ODMR lines corresponding to the \Tgz{} to \Tgx{}/\Tgy{} transitions of both molecule classes.}
	\label{fig:ODMR_both_sites}
\end{figure}

In order to gain more insight into the nature of site 2 (shorter wavelength emission, see Fig.~2a), ODMR spectra were recorded using off-resonant laser excitation at a wavelength of 580\,nm (see Fig.~\ref{fig:ODMR_both_sites}).
\label{sec:ODMR_sites}
\begin{figure}[H]
	\centering
	\includegraphics[width=0.68\textwidth]{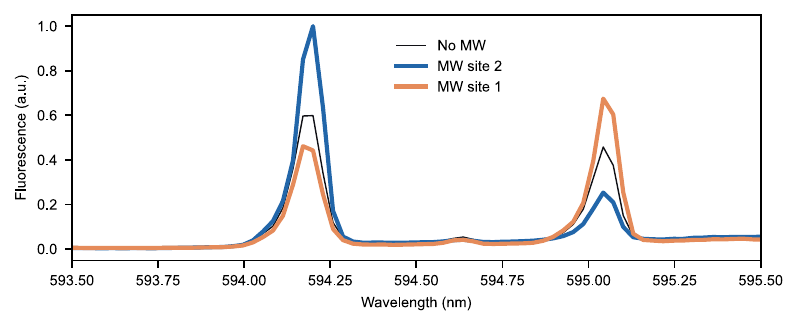}
	\caption{Fluorescence Spectrum, zoomed-in to the ZPLs of the two molecular sites. Excitation was performed using laser wavelength of 580\,nm and laser power of 270\,nW. In addition, microwave radiation was applied to manipulate the fluorescence intensity from both sites separately.}
	\label{fig:fluorescence_MW}
\end{figure}

To further strengthen the claim, that the two ZPLs in the emission spectrum correspond to the two respective pairs of ODMR lines and assign the individual lines, fluorescence spectra are recorded while applying microwave signals on one or the other pair of ODMR resonances.
Since the applied microwaves modulate the fluorescence intensity of the corresponding site, assignment of ODMR frequency to molecular site can be performed.
By taking into account the transition frequencies $\Tgz{} \leftrightarrow \Tgy{}$ at 10.640\,GHz or $\Tgz{} \leftrightarrow \Tgx{}$ at 11.731\,GHz for molecular site 2 (Fig.~\ref{fig:ODMR_both_sites}), the ground-state ZFS parameters for molecular site 2 were deduced to be $|D_{T_0,\text{site 2}}| = 11.1860(2)$\,GHz and $|E_{T_0,\text{site 2}}|=546.0(2)$\,MHz.

\subsection{Determination of the Debye-Waller factor}

\begin{figure}[H]
	\centering
	\includegraphics[]{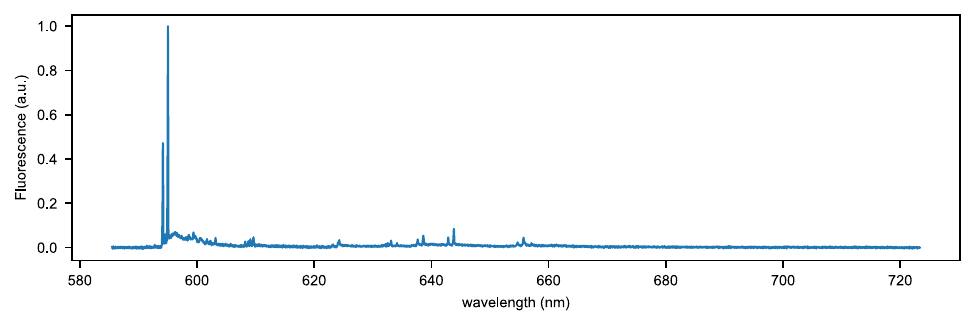}
	\caption{Full fluorescence emission spectrum, showing two molecular sites. Excitation was performed using laser wavelength of 580\,nm and laser power of 270\,nW.}
	\label{fig:fluorescence_full}
\end{figure}

\begin{figure}[H]
	\centering
	\includegraphics[]{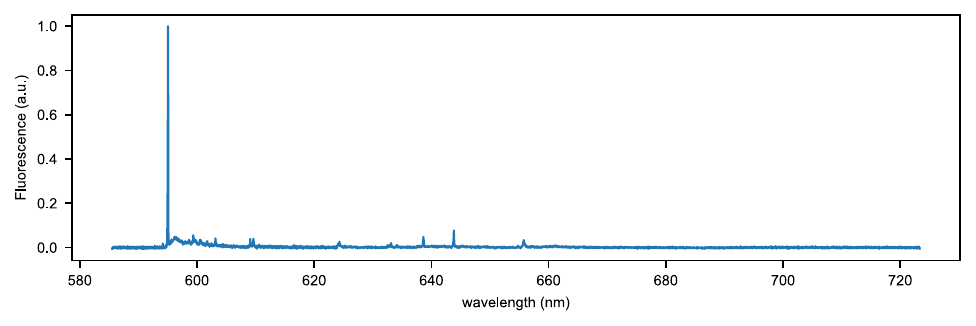}
	\caption{Full fluorescence emission spectrum, showing only molecular site 1. Excitation was performed using laser wavelength of 580\,nm and laser power of 270\,nW. In order to remove site 2, the full emission spectrum was modulated using site selective microwave driving, and subtraction.}
	\label{fig:fluorescence_site1}
\end{figure}

Fig.~\ref{fig:fluorescence_full} shows a full emission spectrum of annealed \bifi{} in \bike{}. In order to derive the Debye-Waller factor, especially for site 1, the modulating property of microwave driving site-selective spin transitions is used to create a difference spectrum, that only contains contributions from molecular site 1 (see Fig.~\ref{fig:fluorescence_MW} for the underlying effect, and Fig.~\ref{fig:fluorescence_site1} for the resulting single site spectrum).
Fluorescence background is mostly removed by offsetting the spectral region $\le$590\,nm to an average of 0.
From this, the Debye-Waller Factor can be determined as the ratio of intensity within the ZPL divided by the intensity in the spectral window between ZPL and 720\,nm, resulting in 19.47\%.

\subsection{Ensemble excitation spectra and \ciso{} isotope effects}
\begin{figure}[htbp]
	\centering
	\includegraphics[width=0.8\textwidth]{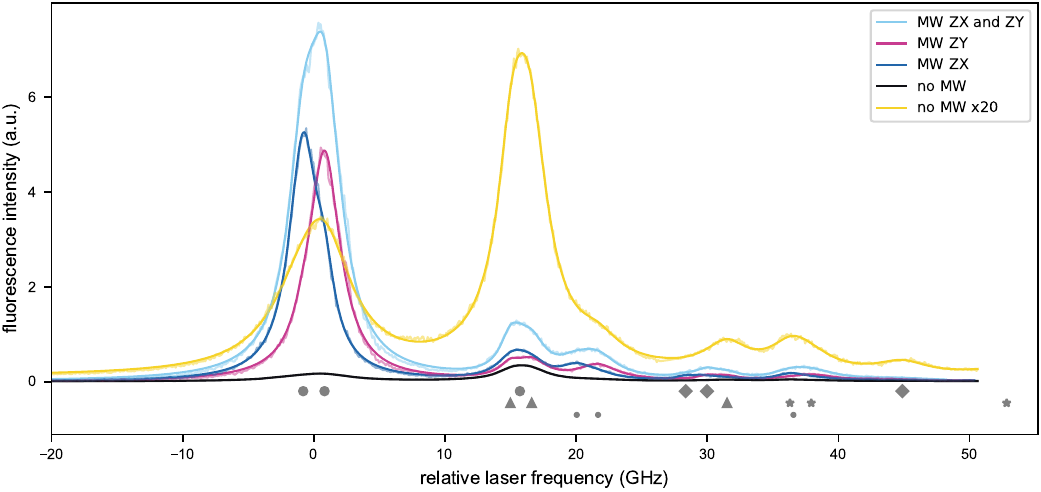}
	\caption{Fluorescence excitation spectra of an ensemble of qubit molecules in crystal site 1 for four cases of microwave fields: MW resonant to $\Tgz{} \leftrightarrow \Tgx{}$, MW resonant to $\Tgz{} \leftrightarrow \Tgy{}$, MW resonant to both transitions, all MW fields off (see legend).
		Five sets of each three Lorentzians are fit to the data.
		The spectral positions of the three Lorentzians of each set are marked by a symbol underneath; and each of the sets has a different symbol (circle, triangle, dot, diamond and star).
		The frequency separation within each triplet are identical.}
	\label{fig:PLE_many}
\end{figure}
The excitation spectra of ensembles of qubit molecules depicted in Fig.~\ref{fig:PLE_many} show multiple resonances with varying amplitudes depending on the type of MW application.
We attribute the strongest peak of each of the curves labeled ``mw ZX'', ``mw ZY'' and ``no mw'' (marked by a circle below) to optical transitions $\Tgx{} \rightarrow \Tex{}$, $\Tgy{} \rightarrow \Tey{}$ and $\Tgz{} \rightarrow \Tez{}$, respectively.
As described in the text, MW radiation that is resonant to spin transitions $\Tgz{} \leftrightarrow \Tgx{}$ and $\Tgz{} \leftrightarrow \Tgy{}$ recycles molecules into optical excitation and emission that have previously been shelved into the dark spin state \Tgz{}.
The latter process mainly describes the behavior of the two strongest optical resonances around $0\,$GHz.
The optical transition $\Tgz{} \rightarrow \Tez{}$ around $15\,$GHz is almost unaffected.
All weaker optical resonances in Fig.~\ref{fig:PLE_many} that are marked with triangles, dots, diamonds and stars cannot be explained as optical transitions of the same molecules that lead to the strongest resonances; they have to belong to molecules that differ to some extent.
Our hypothesis is the existence of at least five optically distinguishable classes of molecules (five different symbols in the figure), each with triples of optical resonances that have the same intra-triple separation but each triple is shifted with respect to each other.
Therefore, we have fit $5\times3$ Lorentzian lines to each of the four spectra in Fig.~\ref{fig:PLE_many} with same frequencies across spectra but varying widths and amplitudes.
Furthermore, three Lorentzians of a triple always have the same mutual separation.

The main triple of resonances (marked by a circle) in the gray curve (i.e. ``mw ZX and ZY'') has a relative area of 0.789 compared to the total area.
Interestingly, the probability for all 25 carbon atoms of our qubit molecule to be \cis{} is 0.758 for a natural abundance of 0.989.
This similarity motivated the consideration of isotope effects to slightly shift optical resonances of otherwise identically packed molecules inside the ketone crystal.
Indeed, such isotope effects have been investigated before for other optically narrow band molecular emitters like pentacene and terrylene in p-terphenyl crystals and for singlet to triplet excitation spectra in anthracene and naphthalene crystals \cite{kummerAbsorptionExcitationEmission1997,bascheOpticalSpectroscopySingle1994a,brouwer13CIsotopeEffects1996,doberer13CIsotopeShifts1983}.
In the latter reference targeted isotopic enrichment was performed to correlate different strengths of isotopic shifts with specific \ciso{} locations on the molecules.

Summarizing, our hypothesis for interpreting the spectra in Fig.~\ref{fig:PLE_many} is the existence of several subensembles of \bifi{} molecules that belong to the same crystal site (i.e. site 1, see Sec.~\ref{sec:ODMR_sites}) but differ in isotopic composition.
Most of the molecules are supposed to have a pure \cis{} backbone while all other molecules might have one or more \ciso{} isotopes spread over the molecule.
Depending on the number and position of the \ciso{} isotopes on the molecules, the influence on the ZPL shift might be different and thus leads to several additional resonances in the fluorescence excitation spectra.
Although, we have fit 5 molecular subensembles to the spectra, there might be more than that.
Some might be hard to resolve, others might be too weak to be identified due to probability, and others might just not be sufficiently bright because the necessary microwave transition frequency might be different than for the molecules comprising the main resonances.
\begin{figure}[htbp]
	\centering
	\includegraphics[width=1.0\textwidth]{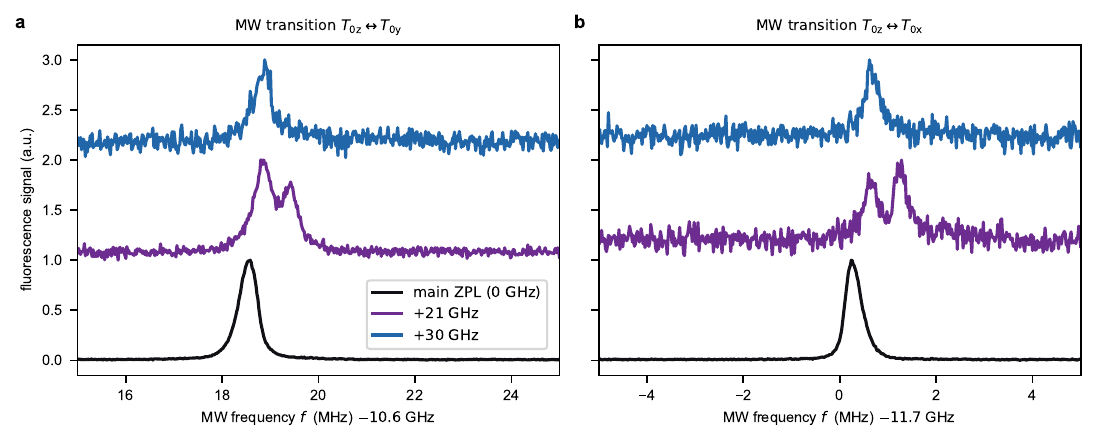}
	\caption{
		ODMR spectra on spin transitions
		(a), $\Tgz{} \leftrightarrow \Tgy{}$ and
		(b), $\Tgz{} \leftrightarrow \Tgx{}$, each for different laser excitation frequencies.
		Different laser excitation frequencies might excite molecular sub-ensembles that differ in isotopic composition of their carbon backbone (see text).
	}
	\label{fig:ODMR_isotope}
\end{figure}
Figure~\ref{fig:ODMR_isotope} shows ODMR resonances for three different optical excitation frequencies potentially belonging to at least three different isotopic compositions.
The resonances are slightly shifted with respect to each other and are split in one case.
The addition of a single \ciso{} nuclear spin to a molecule that consists of \cis{} isotopes otherwise doubles the amount of energy levels within the electron spin triplet ground state.
However, at zero magnetic field the \ciso{} spin shifts the energy depending on hyperfine interaction but does not lift the degeneracy of the new spin level pairs.
Therefore, we do not expect to necessarily find more ODMR resonances than without any \ciso{} spin.
One of the spectra in Fig.~\ref{fig:ODMR_isotope} exhibits two resonances instead of one.
Reasons might be for example more than one \ciso{} spin per molecule or the optical excitation of two sub-ensembles of molecules with similar optical frequency but different location of a single \ciso{} spin on the molecule.

\begin{figure}[htbp]
	\centering
	\includegraphics[width=0.8\textwidth]{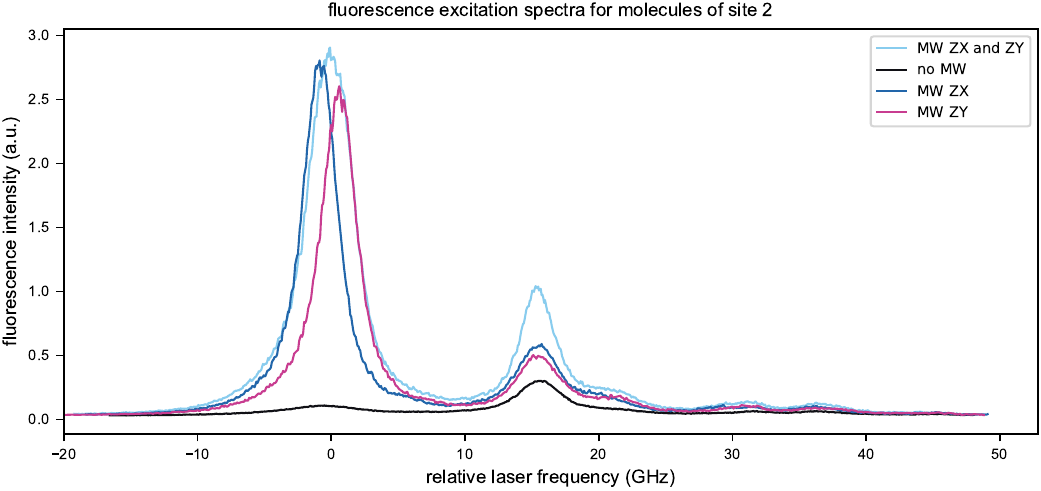}
	\caption{Fluorescence excitation spectra of an ensemble of qubit molecules in crystal site 2 for four cases of microwave fields: MW resonant to $\Tgz{} \leftrightarrow \Tgx{}$, MW resonant to $\Tgz{} \leftrightarrow \Tgy{}$, MW resonant to both transitions, all MW fields off (see legend).
	}
	\label{fig:PLE_ensemble_site_2}
\end{figure}
Finally, we show the excitation spectra of molecule ensembles in crystal site 2 (see Fig.~\ref{fig:PLE_ensemble_site_2}).
The absolute position of the ZPL of molecules in site 1 and site 2 can be taken from Fig.~\ref{fig:fluorescence_site1}.
The relative laser frequency given in Figs.~\ref{fig:PLE_many},\ref{fig:PLE_ensemble_site_2} helps comparing both spectra.
Both exhibit similar separations of spin selective transitions and isotope shifts.

\subsection{Single molecule spectra and spin state assignment}
\begin{figure}[htbp]
	\centering
	\includegraphics[width=1.0\textwidth]{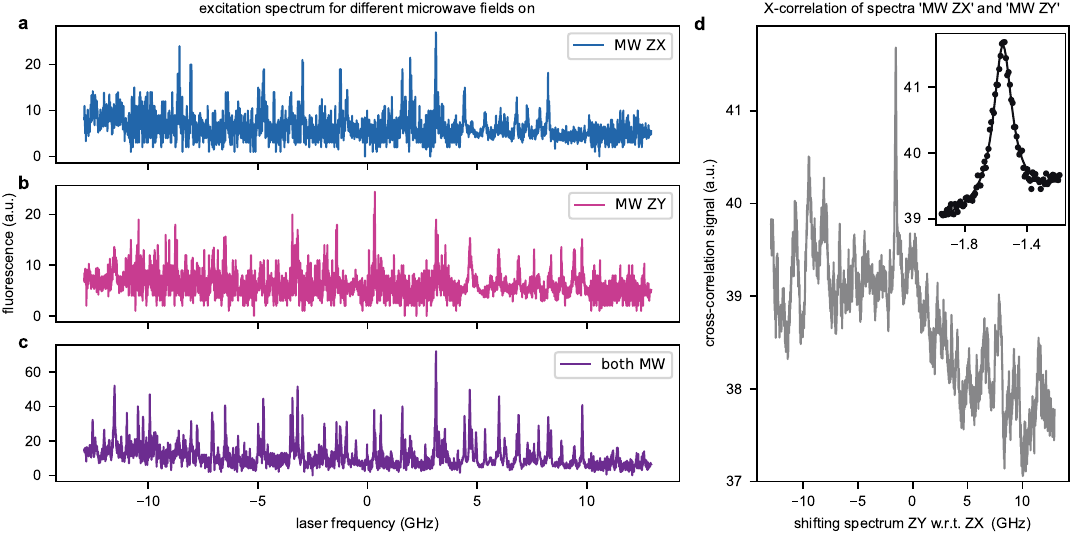}
	\caption{Fluorescence excitation spectra of few qubit molecules.
		(a), An additional MW field is applied resonant to spin transition $\Tgz{} \leftrightarrow \Tgx{}$, ((b), to $\Tgz{} \leftrightarrow \Tgy{}$ and (c) to both transitions.
		(d) Cross-correlation of data in \textbf{a} and \textbf{b} according to
		$d_k = \sum_{n=N}^{M-1} a_{n+k} \cdot b_n / (M-N)$.
		The inset zooms into the region of the maximum correlation value with a local fit of a Lorentzian on a linear slope to the apparent peak.
		According to the fit, the optical transition $\Tgx{} \rightarrow \Tex{}$ is shifted with respect to $\Tgy{} \rightarrow \Tey{}$ by $1.555(1)\,$GHz to lower frequencies.
	}
	\label{fig:PLE_few}
\end{figure}
The inhomogeneously broadened ensemble excitation spectra in Fig.~\ref{fig:PLE_many} exhibit separated spin-selective transitions.
For optical transitions $\Tgx{} \rightarrow \Tex{}$ and $\Tgy{} \rightarrow \Tey{}$, however, the separation is smaller than the inhomogeneous linewidth.
The excitation spectra of several individual qubit molecules in Fig.~\ref{fig:PLE_few} shows the same shift but for much narrower optical lines.
The depicted cross-correlation of spectra of qubits excited on their $\Tgx{} \rightarrow \Tex{}$ transition ($f(XX')$, Fig.~\ref{fig:PLE_few}a) or on their $\Tgy{} \rightarrow \Tey{}$ transition ($f(YY')$, Fig.~\ref{fig:PLE_few}b)  reveals an average splitting $\Delta f_{XY}$ of these optical transition frequencies $f$ of
\begin{align}
	\Delta f_{XY}  & = f\left(YY'\right) - f\left(XX'\right) = 1.555(1)\,\mathrm{GHz}\\
	& = 2 \Ees{} - 2 \Egs{} \nonumber \\
	& = 2 \Ees{} + 1.0818(2)\,\mathrm{GHz} \nonumber 
\end{align}
with a width of the correlation peak of $143(4)\,$MHz, which is about three times larger than the single emitter linewidth (Fig.~\ref{fig:PLE_few}d).

From spin-selective optical line separation $\Delta f_{XY}$ and the ground state ZFS parameter \Egs{}, the excited state ZFS parameter \Ees{} can be estimated to be $\Ees{} \approx 237(1)\,\mathrm{MHz}$.
The last piece of information to fully estimate the energy levels of the ground and excited triplet states is the excited state ZFS parameter \Des{}.
This requires information, for example, from the excited state ODMR frequencies $f(Z'Y') = 3.958(1)\,$GHz and $f(Z'X') = 4.423(1)\,$GHz of spin transition $\Tez{} \leftrightarrow \Tey{}$ and $\Tez{} \leftrightarrow \Tex{}$, respectively (see Fig.~\ref{fig:excited_state_ODMR}).
The latter measurements also give another estimate for \Ees{}, which differs by 5\,MHz from the previous estimate.
\begin{align}
	\Des{}  & = -\left[ f\left(Z'X'\right) + f\left(Z'Y'\right)\right] / 2\\
	& = -4.1905(10)\,\mathrm{GHz} \nonumber \\
	\Ees{}  & = \left[f\left(Z'X'\right) - f\left(Z'Y'\right)\right] / 2\\
	& = 0.2325(10)\,\mathrm{GHz} \nonumber
\end{align}

The spin-selective optical transition $\Tgz{} \rightarrow \Tez{}$ has not been observed for single molecules, yet.
This might be due to the high probability of ISC in the excited state compared to a fluorescent decay into the ground state.
From Fig.~\ref{fig:PLE_many} we estimate that the expected fluorescence intensity is more than ten times smaller for excitation on transition $\Tgz{} \rightarrow \Tez{}$ than for transitions $\Tgx{} \rightarrow \Tex{}$ or $\Tgy{} \rightarrow \Tey{}$.

The separation of energy levels with the qubit molecules are summarized in Tab.~\ref{tab:Elvls_exp}.
\begin{table}[htbp]
	\centering
	\begin{tabular}{ccc}
		\toprule
		&  Ground state triplet  &  Excited state triplet \\
		\midrule
		ZFS $D$  &  $11.1597(2)\,$GHz  &  $-4.1905(10)\,$GHz\\
		ZFS $E$  &  $-0.5409(2)\,$GHz  &  $0.2325(10)\,$GHz \\
		\midrule
		&  separation of optical transitions  &  \\
		$\Delta f_{XY}$  &  $1.55\,$GHz  &  \\
		$\Delta f_{XZ}$  &  $16.12\,$GHz  &  \\
		$\Delta f_{ZY}$  &  $14.58\,$GHz &  \\
		\bottomrule
	\end{tabular}
	\caption{Summary of experimentally obtained energy level separations of \bifi{} molecules in a \bike{} matrix at around 4.5\,K.}
	\label{tab:Elvls_exp}
\end{table}

\subsection{Optically detected magnetic resonance of the excited state}
\begin{figure}[htbp]
	\centering
	\includegraphics[width=0.9\textwidth]{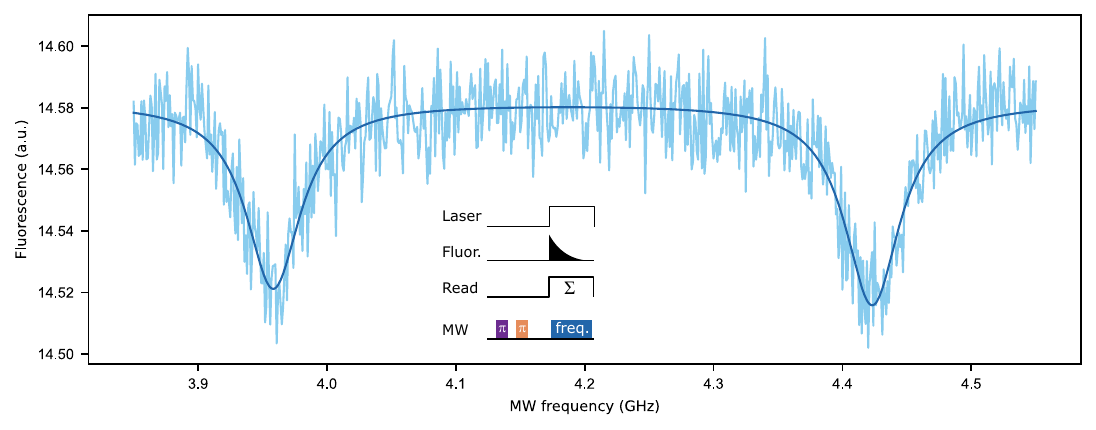}
	\caption{
		Optically detected magnetic resonance spectra of the excited state.
		To this end we work with ensembles of molecules and apply MW $\pi$-pulses on the $\Tgz{} \leftrightarrow \Tgx{}$ and $\Tgz{} \leftrightarrow \Tgy{}$ transitions in the ground state (i.e. without laser illumination, see inset).
		Hence, the molecules are prepared in their bright fluorescent state for optical excitation with a laser around 0\,GHz (c.f. Fig.~\ref{fig:PLE_many}).
		During this laser pulse the molecules transition either $\Tgx{} \rightarrow \Tex{}$ or $\Tgy{} \rightarrow \Tey{}$.
		In addition an MW pulse with variable frequency is applied during the laser pulse.
		The latter pulse affects the electron spin in the excited triplet state.
		If on resonance the MW pulse flips the excited state spin from either \Tex{} or \Tey{} into \Tez{}, which then more quickly decays to the ground state \Tgz{} yielding less fluorescence.
		The spectrum shows the two resonances $\Tey{} \leftrightarrow \Tez{}$ ($3.958(1)\,$GHz) and $\Tex{} \leftrightarrow \Tez{}$ ($4.423(1)\,$GHz) as exactly such dips in fluorescence.
		The latter assignment is performed by taking into account the optical transition frequencies corresponding to \Tgy{} to \Tey{} and \Tgx{} to \Tex{}.
	}
	\label{fig:excited_state_ODMR}
\end{figure}

For efficient ODMR in the excited state a sufficiently strong resonant microwave field is required to flip the spin with the excited state lifetime.
Here, such fields should correspond to Rabi frequencies of a few 10\,MHz.
We have reduced the maximum MW field because of too strong heating and therefore sacrificed contrast in these measurements.
Fig.~\ref{fig:excited_state_ODMR} shows the excited state ODMR spectrum and describes the pulse sequence utilized for ensembles of molecules.

\subsection{Fluorescence lifetime measurements}
\label{sec:cryoconfocal_lifetime_measurements}
\begin{figure}[htbp]
	\centering
	\includegraphics[width=0.9\textwidth]{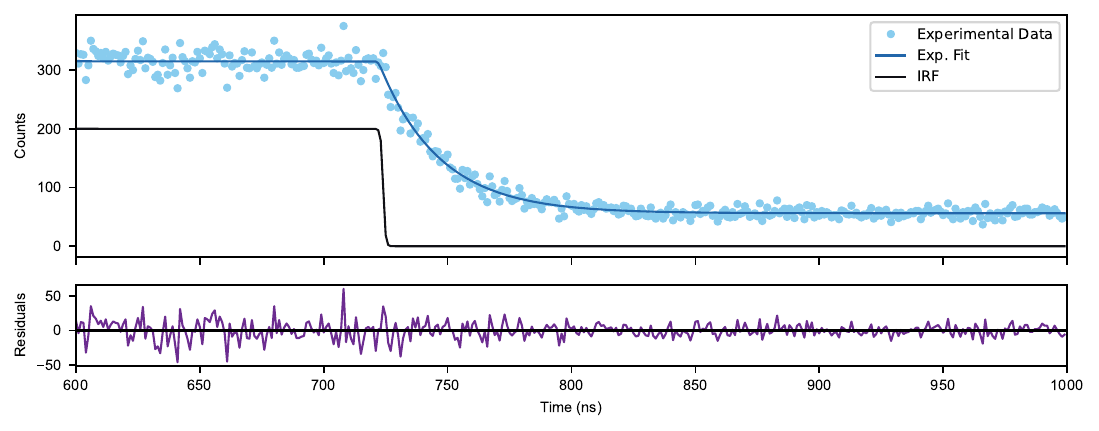}
	\caption{
		Single molecule excited state lifetime measurement.
		(Upper graph) The fluorescence response of a single molecule excited on its $\Tgy{} \leftrightarrow \Tey{}$ is shown for when the laser is switched off (around 720\,ns).
		The instrument response function (IRF) to the laser pulse's falling edge is vertically scaled and plotted for comparison.
		A single exponential decay convolved with IRF is fit to the data.
		(Lower graph) Residuals of fit to data.
	}
	\label{fig:excited_state_lifetime_single}
\end{figure}

In the cryo-confocal setup excited state lifetime measurements were performed on single molecules and for resonant optical excitation compared to off-resonant excitation of molecules ensembles as described in \ref{sec:TCSPC}.
To this end, the dye laser light was set on resonance with transition $\Tgy{} \leftrightarrow \Tey{}$ and was chopped by an AOM and an EOM.
While the AOM achieves a higher on/off contrast ($\approx 10^4$), switching it is slower than the EOM.
Therefore we cascaded AOM and EOM laser switching.
The total pulse sequence starts with a MW $\pi$-pulse on the $\Tgz{} \leftrightarrow \Tgy{}$ transition, preparing the ground state spin in its \Tgy{} state.
Then the AOM turns on and shortly after the EOM.
After 500\,ns the EOM turns off and the AOM remains on for another 450\,ns.
The excited state lifetime is extracted from the fluorescence decay after the EOM is switched off (see Fig.~\ref{fig:excited_state_lifetime_single}).
To estimate the instrument response function (IRF) to the falling edge of the laser pulse we have recorded the bare laser light on the single photon detector (see IRF in Fig.~\ref{fig:excited_state_lifetime_single}).
The fit to the molecule fluorescence response is a convolution of a single exponential decay with the IRF.
It yields an excited state lifetime of
\begin{align}
	\tau_\mathrm{exc} = 24(2)\,\mathrm{ns}.
\end{align}
The lifetime of the \Tez{} of a single molecule could not be determined because we did not find a single molecule $\Tgz{} \leftrightarrow \Tez{}$ transition in excitation spectroscopy.
The lifetime of the \Tey{} state, however, is comparable to the long lifetime component in the ensemble measurement results presented in Fig.~\ref{fig:tcspc_results_crystal}.

\clearpage
\bibliographystyle{naturemag}
\bibliography{references}
	
\end{document}